\documentclass[prd, showpacs, 
               nofootinbib, preprintnumbers, floatfix]{revtex4}%
\usepackage{amssymb, hyperref, amsmath}
\usepackage{graphicx}

\begin{document}
\newcommand{\ignore}[1]{}
\newcommand{\vdot}{\!\cdot\!}
\bibliographystyle{apsrev}
\newcommand{\bvec}[1]{\ensuremath{\mbox{\boldmath $\mathrm{#1}$}}}

\unitlength=1.2mm \preprint{NT@UW-11-09}

\title{Review of Baryon Spectroscopy in Lattice QCD}
\author{Huey-Wen Lin}
\email{hwlin@phys.washington.edu}
\affiliation{Department of Physics, University of Washington, Seattle, WA 98195}

\date{June 7, 2011}

\begin{abstract}
The complex patterns of the hadronic spectrum have puzzled physicists since the early discovery of the ``particle zoo'' in the 1960s. Today, the properties of these myriad particles are understood to be the result of quantum chromodynamics (QCD) with some modification by the electroweak interactions. Despite the discovery of this fundamental theory, the description of the hadronic spectrum has long been dominated by phenomenological models, due to the difficulties of addressing QCD in the strong-coupling regime, where nonperturbative effects are essential. By making numerical calculations in discretized spacetime, lattice gauge theory enables the ab~initio study of many low-energy properties of QCD. Significant efforts are underway internationally to use lattice QCD to directly compute properties of ground and excited-state baryons. Detailed knowledge of the hadronic spectrum will provide insight into the character of these states beyond what can be extracted from models.

In this review, I will focus on the latest progress in lattice calculations of the $P_{11}(1440)$, the poorly known hyperon spectrum and the energies of highly-excited states of the nucleon, Delta and other light-flavor baryons. In the heavy-flavor sector, I will concentrate on recent lattice-QCD calculations of baryon masses, particularly those that make predictions concerning yet-to-be-discovered baryons, such as 
$\Omega_{cc}$, $\Xi^\prime_b$ or triply-heavy baryons.
\end{abstract}

\maketitle

\section{Introduction}\label{sec:intro}

Hadron spectroscopy plays an important role in understanding the theory of quantum chromodynamics (QCD), which has been successful in describing many aspects of the strong interaction. At the most basic level, the SU(3) nature of the color interaction predicts the existence of certain kinds of color-neutral hadrons: quarks and antiquarks can bind into mesons, and three quarks can form baryons. There have been many experimental observations of these simple hadrons with all kinds of flavor content and a variety of excitations. QCD also suggests the possibility of bound states formed only by gluons, called glueballs, or more complex hadrons, such as exotic mesons, tetraquarks, pentaquarks, dibaryons and hadronic molecules; many hadronic spectroscopy experiments are looking hard to find evidence of such states.

In the regime of simple meson and baryon spectroscopy, there are complicated patterns of experimentally observed states, and although many of them are understood by theory, there remains much work to be done. Even for light-quark hadrons, made of up, down or strange quarks only, there are many excited states, especially in the baryon sector, whose physical properties are poorly understood. One famous example is the nature of the Roper resonance, $N(1440)\ P_{11}$, which has been the subject of interest since its discovery in the 1960s. It is quite surprising that the rest energy of the first excited state of the nucleon is less than the ground-state energy of nucleon's negative-parity partner, the $N(1535)\ S_{11}$~\cite{Yao:2006px}, a phenomenon never observed in meson systems. There are several alternative interpretations of the Roper state, for example, as the hybrid state that couples predominantly to QCD currents with some gluonic contribution~\cite{Carlson:1991tg} or as a five-quark (meson-baryon) state~\cite{Krehl:1999km}.

There are also baryonic excited states predicted by the quark model, based on three constituent quarks but not yet discovered in experiments, the so-called ``missing resonances'', which also provide challenges for the experimentalists and also for the phenomenologists. Many experiments at, for example, Jefferson Lab, MIT-Bates,
LEGS, MAMI, ELSA, and GRAAL, have been collecting data from one and two-meson production to broaden our understanding of excited states. These processes are so complicated that many theoretical centers at, for example, Jefferson Laboratory, George Washington University, Bonn, and Mainz, have built models to be able to analyze the experimental data. These joint efforts will provide us detailed understand of the QCD theory and paint a more complete picture of hadronic resonances.

In the heavy-quark sector, even many of the ground states, such as doubly bottom or triply charmed baryons are yet to be discovered. 
Experimental and theoretical studies of charmed and bottom hadrons have been the focus of vigorous research over the last several years~\cite{Barberio:2008fa,Voloshin:2007dx,:2007rw,:2007ub}, such as recent experimental discoveries of new charmed baryons at SELEX~\cite{Mattson:2002vu,Ocherashvili:2004hi}. 
In the summer of 2007, CDF\cite{:2007rw} reported the first observation of the heavy baryons $\Sigma_b$ and $\Sigma_b^*$, and then both D0\cite{:2007ub} and CDF\cite{:2007un} observed the bottom baryon $\Xi_b^-$, breaking a long period of silence following the observation of the $\Lambda_b$ in 1991. 
The bottom-baryon spectrum has become somewhat controversial due to recent results from D0 and CDF. Last summer, D0\cite{Abazov:2008qm} reported a first observation of the doubly strange bottom baryon $\Omega_b^-$ at 6.165(10)(13)~GeV. However, a recent CDF work puts the $\Omega_b^-$ mass at 6.0544(68)(9)~GeV, a difference of 111(12)(14)~MeV. With a discrepancy of 6.2 standard deviations, it appears that the two collaborations cannot both be observing the $\Omega_b$. 
A theoretical understanding of the bottom baryon spectroscopy from first-principles QCD is crucial and can help disentangle such discrepancies in experiment.

The Beijing Spectrometer (BES-III), a detector at the recently upgraded Beijing Electron Positron Collider (BEPCII), has great potential for accumulating large numbers of events to help us understand more about charmed hadrons. The antiProton ANnihilation at DArmstadt (PANDA) experiment, a GSI future project, and the LHC is also expected to provide new results to help experimentally map out the heavy-baryon sector. For these reasons, lattice quantum chromodynamics calculations of the spectrum of heavy baryons are now very timely and will play a significant role in providing theoretical first-principles input to the experimental program. 
At the Tevatron, much data remains to be fully analyzed; one could possibly expect more discoveries over the next couple years. 
Furthermore, it is anticipated that in the upcoming dedicated bottom-physics experiment LHCb at CERN or a future Super-B factory, there will be many more discoveries of bottom-hadron properties. Combining such experiments with improved theoretical understanding, our knowledge about these states will in the near future be significantly enhanced.

Although significant progress has been made in refining phenomenological models since the development of the original quark model in the 1960s, determining the properties of hadrons from the first principles of the Standard Model requires us to directly solve QCD in the strong-coupling regime. Unfortunately, when the coupling constant becomes large, perturbative techniques based on Feynman diagrams do not work well. Instead, we use a discretization of space and time in a finite volume to calculate the observables of QCD numerically, a technique known as lattice QCD.

Lattice quantum chromodynamics (LQCD) has been used successfully to compute many experimentally observable quantities from first-principles calculation of Euclidean-time hadron correlation functions, even occasionally predicting experimental results before they are measured. However, the early successes of LQCD were mostly restricted to computation of the physical properties of the lowest-energy states in each quantum number channel. This is due to the focus on the large-time behavior of correlation functions, where uncertainties due to excited-state contributions are exponentially suppressed. Given that signal-to-noise in correlation functions also falls exponentially at large times, success is often dictated by available computational resources.

With the most state-of-the-art supercomputers available, we are just beginning to be able to simulate at the physical pion mass. 
A recent work by the BMW collaboration\citep{Durr:2008zz} calculating multiple lattice spacings, volumes and pion masses as light as 180~MeV provided an excellent demonstration of how ground-state hadron masses with fully understood and controlled systematics are consistent with experiment.

In this review, selected channels of light- and heavy-flavor baryons efforts from lattice QCD calculations are summarized and compared. In Sec.~\ref{sec:theory}, a brief introduction to lattice gauge theory is given, along with some examples of operators and correlator construction. A few commonly used methods used to extract masses of hadrons are also presented. 
Lattice-QCD calculations of excited baryon spectroscopy involving light and strange quarks are presented in Sec.~\ref{sec:light-strange}. The status of Roper-resonance studies is summarized. We describe progress of the Hadron Spectrum Collaboration (HSC)'s effort on highly (both radially and orbitally) excited states, along with a few remarks on recent developments and future efforts. We also cover lattice-QCD calculations in hyperon channels with $2+1$ dynamical ensembles. 
In Sec.~\ref{sec:heavy}, charm and bottom-baryon spectroscopy from various lattice-QCD groups are summarized. The heavy flavor is simulated with relativistic heavy quark action, NRQCD and static quarks with various light-fermion actions for the light/strange quarks. 
In Sec.~\ref{sec:conclusion}, we summarize the current state of the field and discuss some aspects of future developments.

\section{Theoretical Background}\label{sec:theory}
\subsection{Lattice-QCD Setup}

Lattice QCD is a discrete version of the continuum QCD theory. The path integral over field strengths at infinitely many Minkowski space-time points from continuum QCD is approximated using only finitely many points in a discrete Euclidean space lattice with periodic boundary conditions. In momentum-space, this automatically provides both an ultraviolet cutoff at the inverse lattice spacing and an infrared cutoff at the inverse box size, and it is important to make sure that these cutoffs do not affect the physics. Since the real world is effectively continuous and infinitely large, a lattice calculation needs to take limits where the lattice spacing $a \rightarrow 0$ and the simulated spacetime volume $V \rightarrow \infty$ to eliminate the artifacts introduced in a discretized finite box.

At a fixing lattice spacing, the systematics due to the finite lattice spacing can be controlled by a systematic expansion called Symanzik improvement\cite{Symanzik:1983dc} for the lattice fields, actions and operators. The goal is to construct an effective continuum action using operators that are invariant under discrete rotations, parity-reversal and charge-conjugation transformations, that represent the long-distance limit of the lattice theory and zero the leading finite-$a$ errors. However, the breaking of continuous (Euclidean) SO(4) symmetry allows many new degrees of freedom, leading to various lattice actions that return to the same continuum action once the symmetry is restored. Thus, there exist many gauge and fermion actions for us to choose from.

For example, in constructing a relativistic heavy-quark action on the lattice, one starts with the Dirac operator and the gluon field tensor (and the presence of the heavy quark mass distinguishes between the time and space components of each). There are eight operators with dimensions up to five:
\begin{eqnarray*}
\label{eq:O-aOp}
&&{} \overline{\psi} \psi,\
 \overline{\psi} \gamma_0 D_0  \psi,
 \overline{\psi}\vec{\gamma} \vdot \vec{D} \psi,\
 \overline{\psi} \left(D_0\right)^2  \psi,\
 \overline{\psi}\big(\vec{D}\big)^2 \psi,\
 i\overline{\psi}  \sigma_{0i} \vdot F_{0i}\psi,\
 i\overline{\psi}  \sigma_{ij} \vdot F_{ij} \psi,\
 \overline{\psi}\big[\gamma_0 D_0,\vec{\gamma} \vdot \vec{D}\big] \psi.
\end{eqnarray*}
However, not all of the above operators are relevant. That is, the redundant continuum operators may have their coefficients arbitrarily set without affecting the resulting physics. The remaining coefficients of the relevant operators will then be tuned to give the correct physics in the continuum limit.

%
Today, most gauge actions used are $O(a^2)$-improved and leave small discretization effects ($O(a^3\Lambda_{\rm QCD}^3)$) due to gauge choices. On the other hand, most fermion actions are only $O(a)$-improved and have systematic errors of $O(a^2\Lambda_{\rm QCD}^2)$ that become dominant. For this reason, lattice calculations are generally distinguished according to the fermion action used. Differences among the actions are benign once all systematics are included, and the choice of fermion action is constrained by limits of computational and human power and by the main physics focus. The commonly used actions are: domain-wall fermions (DWF)\cite{Kaplan:1992bt,Kaplan:1992sg,Shamir:1993zy,Furman:1994ky}, overlap fermions\cite{Neuberger:1997fp}, clover (or Sheikholeslami-Wohlert) action\cite{Sheikholeslami:1985ij}, twisted-Wilson fermions\cite{Frezzotti:2000nk} and staggered fermions\cite{Naik:1986bn,Orginos:1998ue}.

For a calculation done at a typical fixed lattice spacing, 0.12~fm, the discretization effect for an $O(a)$-improved fermion action is on the order of $O(\Lambda_{\rm QCD}^2a^2) \approx 3\%$. However, one should keep in mind that it varies with fermion actions in the actual calculation and the physical quantities (for example, the charmonium hyperfine splitting has stronger lattice-spacing dependence than the proton mass). At a fixed lattice spacing, we may expect different amounts of discretization effect for different fermion actions; however, after continuum extrapolation, that is,  using multiple calculations at different lattice spacings to extrapolate to zero lattice spacing, we expects universal results for all quantities.

In order to make predictions using QCD on the lattice, we calculate observables corresponding to vacuum expectation values of operators $O$, taking the form
\begin{eqnarray}
\langle O \rangle & = &
  \frac{1}{Z} \int [dU][d\psi][d\overline{\psi}]
              e^{-S_F(U, \psi, \overline{\psi})-S_G(U)}O(U, \psi,\overline\psi) \nonumber \\
& = & \frac{1}{Z} \int [dU] {\rm det}M e^{-S_G(U)} O(U) \nonumber \\
& = & \frac{1}{Z} \int [dU] e^{-S_{\rm eff}(U)} O(U),
\end{eqnarray} 
where 
\begin{equation}\label{eq:Z}
 Z  =  \int [dU][d\psi][d\overline{\psi}] e^{-S_F-S_G},
\end{equation} 
$S_G$ is the gauge action and $S_F=\overline{\psi} M \psi$ is the fermion action with Dirac operator $M$. The bilinear structure of the fermion action allows the integration over the fermion fields to be done explicitly, bringing down a factor of ${\rm det}\ M$. 
This means that the anticommuting fermion fields (impossible to simulate on a computer) are integrated out, leaving an integrand that depends only the values of the gauge fields. In the early days of lattice QCD, the computational resources were insufficient to compute the fermionic determinant; instead, the determinant was approximated by a constant. This is equivalent to removing quark loops from the Feynman diagrams of a perturbative expansion, and this technique became known as the ``quenched approximation''. Althought quenching retains many of the important properties of QCD, such as asymptotic freedom and confinement, it introduces an uncontrollable systematic error, and modern calculations keep at least the up, down and strange quarks in the sea.

The discrete integral we have derived can be evaluated numerically using Monte Carlo methods. The Monte Carlo integration uses random points within the domain of gauge configurations to approximately evaluate the integral. The ``importance sampling'' technique is introduced to perform this task more efficiently: instead of choosing points from a uniform distribution, they are chosen from a distribution proportional to the Boltzmann factor $e^{-S_{\rm eff}(U)}$, which concentrates the points where the function being integrated is large.

Using this method, we accumulate an ensemble of gauge field configurations, generated using a Markov-chain technique. Based on the current state of the gauge configuration, a new configuration is selected. A transition probability $P([U^\prime] \leftarrow [U])$  is determined based solely on this new configuration and the current one. The new configuration is added to the ensemble or rejected, according to the resulting probability. If the ``detailed balance'' condition 
\begin{equation}
P([U^\prime] \leftarrow [U]) e^{-S_{\rm eff}(U)} =
    P([U] \leftarrow [U^\prime]) e^{-S_{\rm eff}(U^\prime)}
\end{equation} 
is satisfied by this update procedure, then the canonical ensemble is a fixed point of the transition probability matrix. Under this condition, repeated updating steps will bring the gauge-field distribution to the canonical ensemble.

\begin{figure}[t]
\begin{center}
\includegraphics[width=0.75\textwidth]{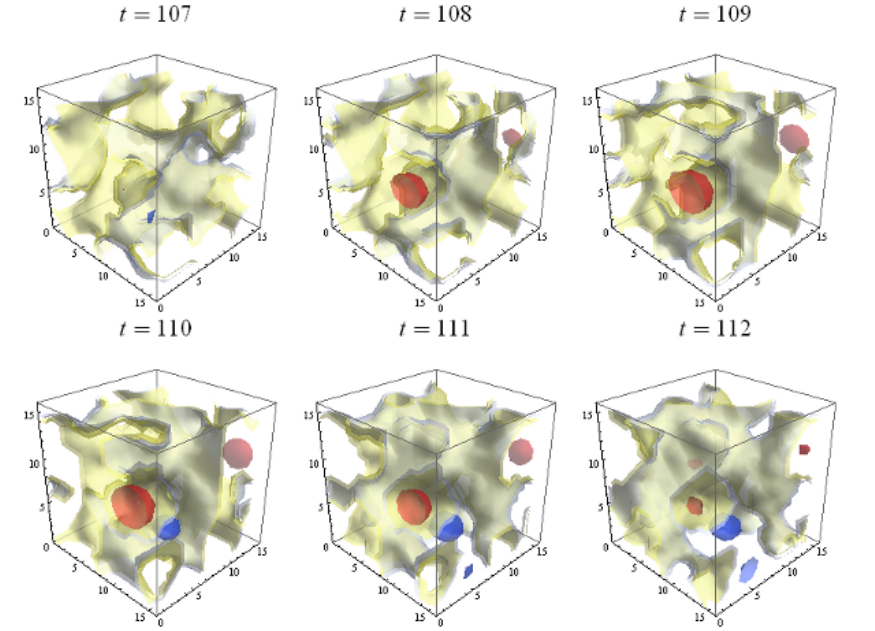}
\end{center}
\caption{\label{fig:instantons}
A selection of time-slices showing the topological charge of the vacuum calculated on one gauge-field configuration on a lattice with spacing approximately $0.1227 \times 10^{-15}$ meters with 16 lattice sites in each spatial direction. The evolution of the vacuum shown in this image occurs over $5.8 \times 10^{-25}$ seconds. The red regions correspond to instantons with positive topological winding number, while the blue correspond to instantons with negative winding number.
}
\end{figure}

Figure~\ref{fig:instantons} shows a few timeslices of the spatial distribution of the topological charge of a QCD vacuum ensemble taking from one of the lattices generated by HSC.

Once we have accumulated gauge-field configurations using this procedure, the expectation value of our observable $\langle O \rangle$ in the original field theory is numerically approximated by the sample average on our ensemble:
\begin{equation}
\overline{\langle O \rangle} \approx \frac{1}{N}\sum_n^N O(U^{{n}}_\mu (x)).
\end{equation}


Some of the primary observables we study on the lattice are two-point correlation functions. A particle of interest is created by an operator carrying the appropriate quantum numbers. 
For example, meson states can be created with local bilinear operators as shown in Table~\ref{tab:meson_op}. Baryons may be constructed in the same way; for example, for nucleon spin-1/2, we can write down 3 local-site (that is all 3 quarks resides within the same lattice point) operators: 
\begin{eqnarray}
\label{eq:proton-op}
\chi_N^{(1)}(x) &=& \epsilon^{abc}(u^{T,a}(x)C\gamma_5 d^b(x))u^c(x) \nonumber \\
\chi_N^{(2)}(x) &=& \epsilon^{abc}(u^{T,a}(x)C d^b(x))\gamma_5 u^c(x) \nonumber \\
\chi_N^{(3)}(x) &=& \epsilon^{abc}(u^{T,a}(x)C\gamma_5\gamma_4 d^b(x)) u^c(x)
\end{eqnarray} 
More complicated hadrons (such as pentaquarks) and multi-particle states simply increase the number of quarks and antiquarks involved.

\begin{table}
\caption{Meson states created by local operators of the form
$\bar{\psi}\Gamma\psi$, labeled in spectroscopic notation}
\label{tab:meson_op}
\begin{center}
\begin{tabular}{cccc}
\hline\hline
          $\Gamma$              & $^{2S+1}L_J$   & $J^{PC}$  & States \\
          \hline
          $\gamma_5$            & $^1S_0$        & $0^{-+}$  & $\pi, \eta_c$ \\
          $\gamma_i$            & $^3S_1$        & $1^{--}$  & $\rho, J/\Psi$ \\
          $\gamma_i\gamma_j$    & $^1P_1$        & $1^{+-}$  & $b_1, h_c$ \\
          $1$                   & $^3P_0$        & $0^{++}$  & $a_0, \chi_{c0}$ \\
          $\gamma_5\gamma_i$    & $^3P_1$        & $1^{++}$  & $a_1, \chi_{c1}$ \\
 \hline\hline
\end{tabular}
\end{center}
\end{table}

\begin{figure}[t]
\begin{center}
\includegraphics[width=0.75\textwidth]{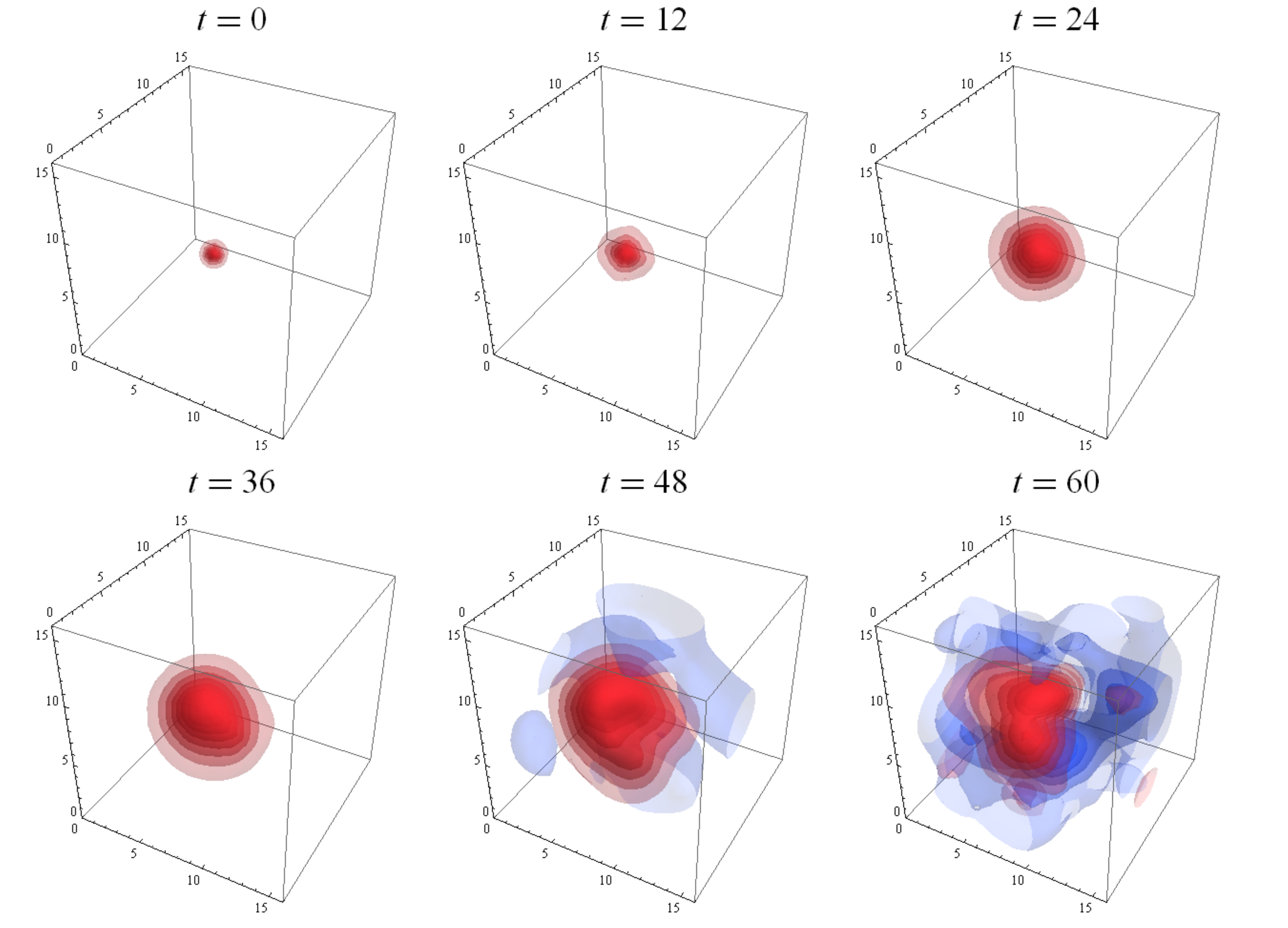}
\end{center}
\caption{\label{fig:proton}
A selection of timeslices showing the real part of the proton two-point correlation function calculated on one gauge-field configuration on a lattice with spacing approximately $0.1227 \times 10^{-15}$ meters, having 16 lattice sites in each spatial direction. The difference between adjacent timeslices corresponds to $\Delta t = 1.17 \times 10^{-25}$ seconds, and the evolution of the proton correlation function shown in this image occurs over $7.02 \times 10^{-24}$ seconds. To enhance detail, the colors (red being more positive and blue more negative) have been normalized on each timeslice. The proton source on timeslice $t = 0$ has a Gaussian distribution in space.
}
\end{figure}

For a meson created at a specific space-time $y=({\bf \vec y}, 0)$ (point source) and annihilated at $x=({\bf \vec x}, t)$ (point sink), the corresponding Euclidean two-point function is 
\begin{equation}\label{eq:mesoncorrelator}
C^{(2)}(x,y)=\langle {\overline O}(x) O(y) \rangle =
  \frac{1}{Z} \int [dU][d\psi][d\overline{\psi}]
              e^{-S_{\rm eff}(U, \psi, \overline{\psi})} (\overline \psi(x) \Gamma \psi(x))(\overline \psi(y) \Gamma \psi(y)).
\end{equation} 
Figure~\ref{fig:proton} shows a few timeslices of the spatial distribution of a proton two-point correlator on a QCD vacuum ensemble generated by HSC.

When $t$ is large enough, the lowest-energy (ground) state dominates, and its mass can be obtained from fitting the correlator to an exponential: 
\begin{equation}\label{eq:massform}
 C^{(2)}(t)= \left\langle \sum_{\vec x} C^{(2)}(({\vec x}, t),({\vec y}, 0))\right\rangle
\sim A_\Gamma e^{-m_\Gamma t}.
\end{equation} 
The so-called ``effective mass'' can be obtained from $\ln{C^{(2)}(t)}/\ln{C^{(2)}(t+1)}$; plots of the effective mass as a function of time are helpful in crosschecking the fitted values with different fitting methods and adjusting the fitting range away from excited-state contamination. However, such a character makes it difficult to study excited states on the lattice, since their signals will become exponentially dominated by lower states at large times; later, I will describe a few efforts to overcome this difficulty.

%

Very often, we impose a smeared source on the quark field using gauge-invariant smearing or gauge-fixed configurations to improve overlap of the creation and annihilation operators with the desired states:
\begin{equation}\label{eq:smear_source}
\psi^s(0) = \sum_{\vec y} F({\vec y}, 0) \, \psi({\vec y},0)
\end{equation} 
with smearing function $F$.

Jackknife analysis often adopted to analyze lattice data. It is is a method of regrouping a data sample (similar to bootstrap analysis) in order to make a reliable estimate of the error for a quantity computed from that sample. Consider an ensemble of $N$ gauge configurations: $E_1, E_2,...,E_N$. The $k^{\rm th}$ jackknife result for a quantity $A$ is obtained by
\begin{equation}\label{eq:jackknife}
A^{(J)}_k = \langle A \rangle_{E_i \neq E_k}
\end{equation}
and the variance of the jackknife estimators is 
\begin{equation}\label{eq:jackknife2}
(\sigma^{(J)})^2 = \frac{N-1}{N}
\sum_{i}^{N}(A^{(J)}_i-\overline{A^{(J)}})^2
\end{equation}
where $\overline{A^{(J)}}$ is the average of the jackknife blocks. The jackknife analysis is a better estimator when dealing with correlated measurements.

So where does lattice QCD stand in general?

Ensemble generation is still costly even with the supercomputers available today and the latest algorithmic improvements. The cost of an ensemble increases with smaller lattice spacing, larger volume and lighter pion mass composed by the sea quarks. 
Since light quark masses greatly increase the cost of an ensemble, lattice calculations often use unphysically heavy pions. Currently, more groups are generating pion masses around or below 200~MeV, and some have started to work at the physical pion mass. When using heavy pions, to get results that connect with the real world, we need to extrapolate the pion mass to its physical value. Since chiral perturbation theory is often used, this process is generally called chiral extrapolation. One side benefit of this process is that we can help to fix the low-energy parameters of chiral effective theory.

\begin{figure}
\includegraphics[width=0.52\textwidth]{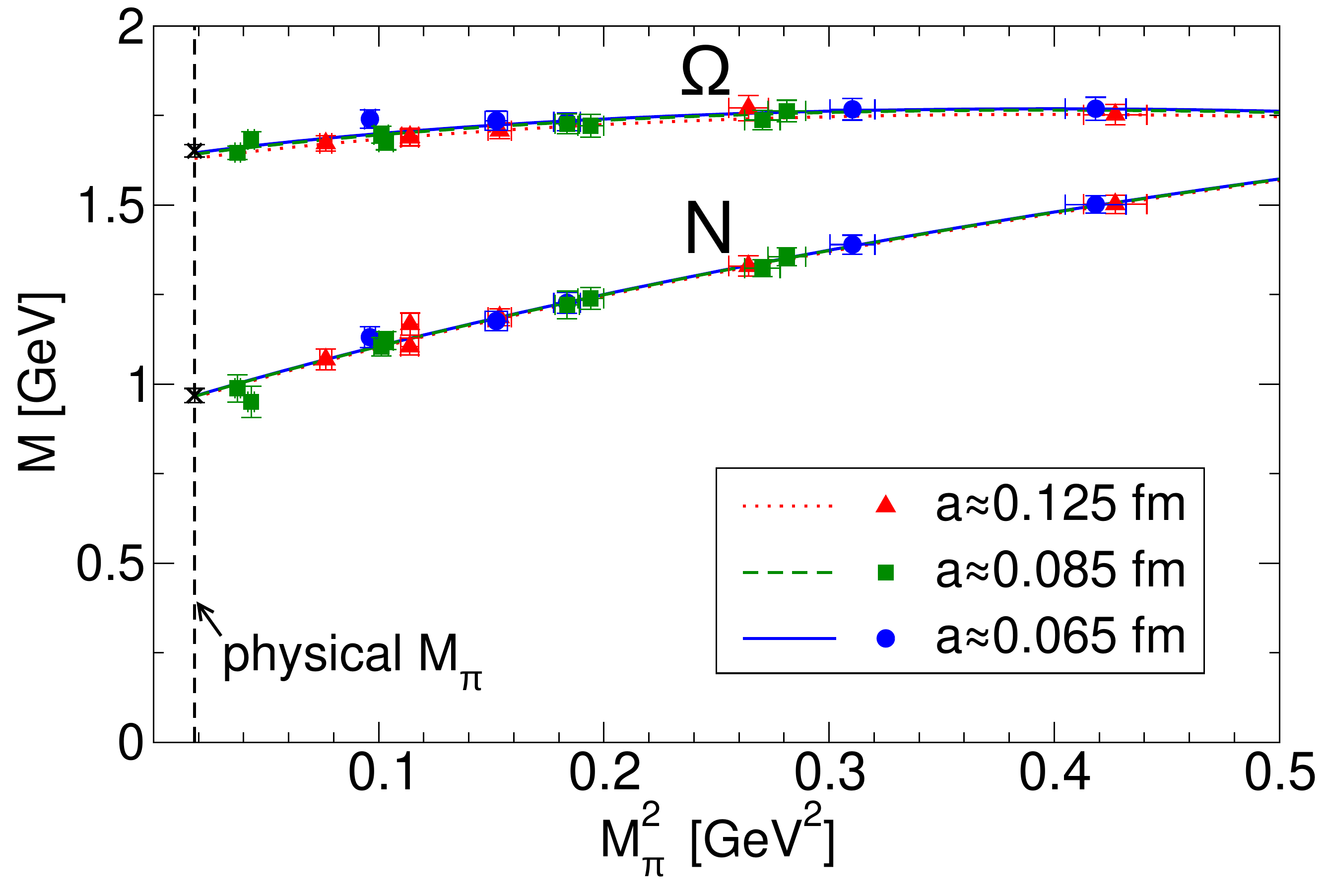}
\includegraphics[width=0.4\textwidth]{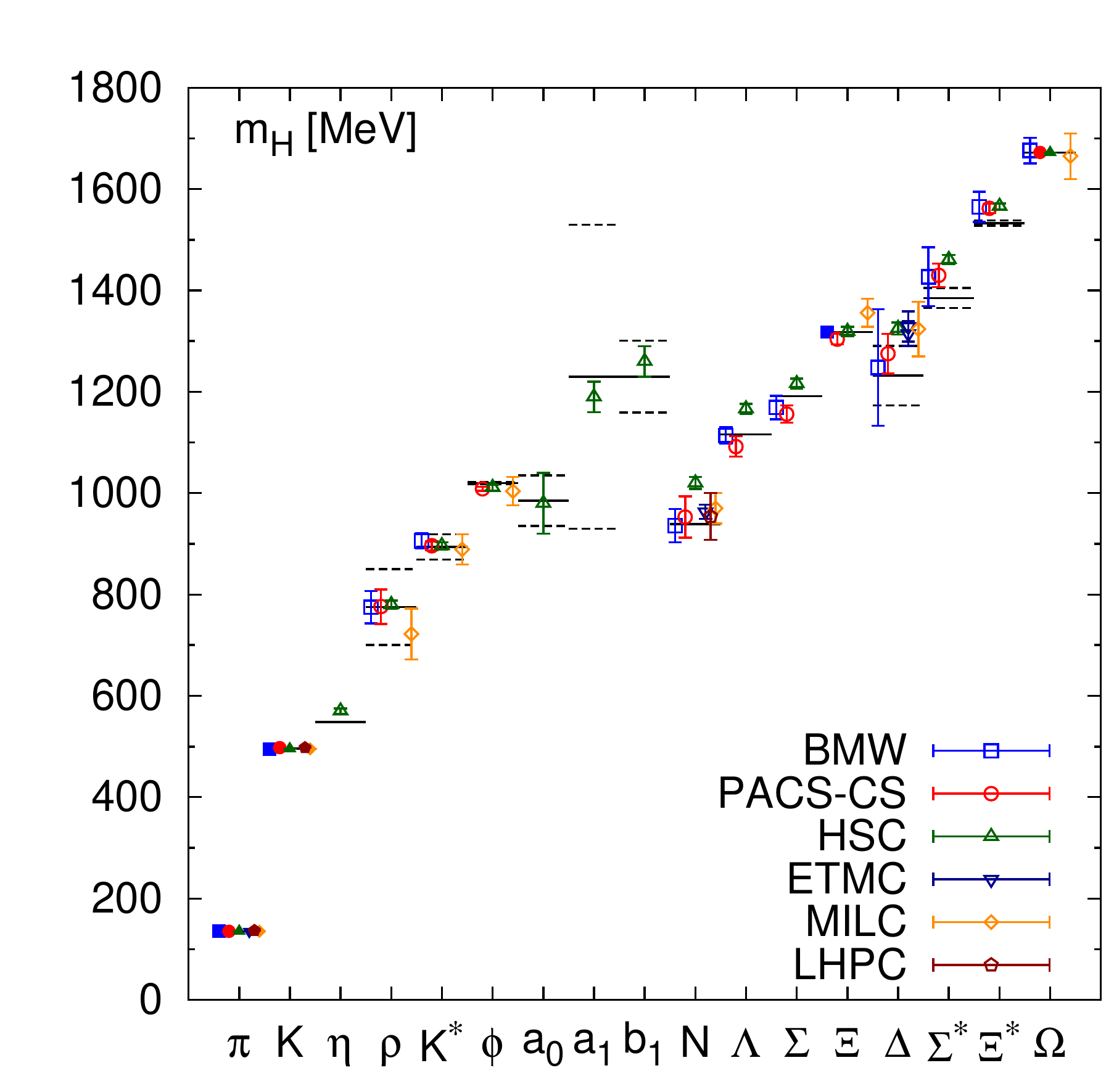}
\caption{\label{fig:basic-spec}(left)
The pion-mass dependence of the nucleon and $\Omega$ baryon masses at three lattice spacings: $a\approx 0.065$, 0.085, 0.125~fm. Figure taken from Ref.~\cite{Durr:2008zz}.
(right) The low-lying spectrum of light hadrons from dynamical lattice-QCD simulations from various collaborations around the world with different choices of fermion actions. The solid and dashed lines are the PDG values for the masses and widths, respectively. Figure taken from Ref.~\cite{Scholz:2009yz}.
}
\end{figure}

The left-hand side of Fig.~\ref{fig:basic-spec} shows typical masses (nucleon and $\Omega$) extracted from lattice QCD as functions of $m_\pi^2$. The different-color points indicate calculations done using different lattice spacings; the dashed line indicates the physical pion mass. With a lightest pion mass about 180~MeV and no visible lattice-spacing dependence, the chiral and continuum extrapolations are well controlled and the results agree nicely with experiment. 
The right-hand side of Fig.~\ref{fig:basic-spec} shows more hadron masses extracted by various collaborations with different types of discretization due to their choices of lattice fermion actions. The excellent agreement among different calculations and with experiment for these simple quantities gives us confidence to work on other calculations that are either difficult to measure or not-yet measured in experiment.

\subsection{Extracting Masses}

A variety of analysis techniques have been applied to extracting the hadron spectrum from correlation functions in lattice QCD. In this section, we discuss some of these approaches that have been used for extracting excited states, including multiple-correlator least-squares fitting, Bayesian fitting\cite{Michael:1993yj,Michael:1994sz,Lepage:2001ym}, black-box methods\cite{Fleming:2004hs,Lin:2008gv,Fleming:2009wb,Fleming:2010ebook},
and the variational method\cite{Michael:1985ne,Luscher:1990ck}.

\subsubsection{Multiple-Correlator Fits}

The most widely used method is nonlinear least-squares (NLLS) fitting to a model function, such as a sum over two or more exponentials. The brute-force way of extracting excited states is to fit using a form that includes them explicitly. Usually, one needs multiple correlators as input to make the fit algorithm converge. With multiple smeared correlators, one minimizes the quantity 
\begin{eqnarray}
\chi^2=\sum_{s,t} \frac{(C^{(2)}_s(t)-\sum_n a_{n,s}e^{-E_nt})^2}
                   {\sigma_s^2(t)},\label{eq:chi2}
\end{eqnarray}
where the correlators are each fit by the sum of $N$ exponentials, having independent amplitudes $a$ but the same masses (energies) $E$. The greater the number of distinguishable input correlators, the easier it will be to extract higher states.

Unfortunately, since the contribution of each state to the two-point correlation function declines exponentially over time with exponent equal to the mass, the excited-state contributions are swiftly dominated by the ground state and by noise. Indeed, much work where the ground states are of primary interest take pains to only fit the data at large times where the excited states have vanished. And since the excited states will have significant contributions only for the first few timeslices beyond the creation operator, little data is available to constrain the fits. Thus, this method is restricted to very small numbers of fitted masses.

\subsubsection{Bayesian Fitting}

One method for extracting additional masses from correlators is to apply more sophisticated fitting techniques. Using information we already know, we can introduce what are known as Bayesian priors to constrain our fits. These constraints may improve the precision with which difficult-to-determine masses are found by a fitting method.

The Bayesian technique applies known information to shift the estimated probabilities of correctness for parameter sets being considered by a fitting procedure. This prior information may be as simple as the assertion that all physical masses must be positive ($E_i>0$) or that each new excited state must be heavier than the states preceding it ($E_i > E_{i-1}$). We could even attempt to extract states from a correlator in an iterative fashion. Assuming we have somehow acquired a ground state for our data (say, by a non-Bayesian fit), we can make the prior assumption that the ground state is actually equal to that fitted value within the determined uncertainty.

Using this prior information, we wish to modify the likelihood estimate for all parameter sets considered by our Bayesian fitter. The probability of correctness for a particular parameter set is generally represented by the $\chi^2$ value, which for a non-Bayesian fit is
\begin{equation}
\chi^2_{\rm dof} = \frac{1}{N-n} \sum_{i=1}^N
       \left(\frac{y_i - f(\{a,E\}_{0..n-1},x_i)}{\delta y_i}\right)^2,
\end{equation}
where we fit $N$ data points $y\pm\delta y$ to some form $f$ parametrized by $n$ amplitudes $a$ and masses (energies) $E$. Rather than allowing the fitter to arbitrarily change the amplitudes and masses, we would like it to take into consideration the masses we have already determined. Therefore, we simply add a term that will increase $\chi^2$ (decrease the probability of correctness) if the fitter moves away from these values:
\begin{equation}
\Delta \chi^2 = \sum_{j=0}^{m-1}
            \left(\frac{E_j-\hat{E}_j}{\delta \hat{E}_j}\right)^2,
\end{equation}
where we have already determined the first $m$ masses to be $\hat{E}\pm\delta \hat{E}$.

\subsubsection{Black-Box Methods}

Operationally, even simple nonlinear fits can be fraught with difficulty, from establishing the range of Euclidean times included in the data set to stabilizing the convergence of minimization algorithms by careful choices of initial guesses or temporarily freezing selected fit parameters during the minimization process, all of which require manual intervention.

At such times, we often turn to black-box methods for guidance, because they do not require intervention to determine initial guesses and fitting ranges: estimates of correlation functions go into the black box and estimates of hadron energies come out. The main detraction of black-box methods is that the produced estimates are expected to have larger uncertainties, making them suboptimal relative to least-squares methods. Marrying the two approaches can effectively combine the best features of both, leading to a highly-automated analysis programs producing optimal estimates of energies.

%
%

The problem to solve is the nonlinear system of equations $\bvec{y} = \bvec{V}(x)
\ \bvec{a}$
\begin{equation}
  \label{eq:VandermondeQCD}
  \left[\begin{array}{c}
  y_1    \\
  y_2    \\
  y_3    \\
  y_4    \\
  \vdots \\
  y_{2M}
  \end{array}\right] = \left[\begin{array}{cccc}
  1          & 1          & \cdots & 1          \\
  x_1        & x_2        & \cdots & x_M        \\
  x_1^2      & x_2^2      & \cdots & x_M^2      \\
  x_1^3      & x_2^3      & \cdots & x_M^3      \\
  \vdots     & \vdots     & \ddots & \vdots     \\
  x_1^{2M-1} & x_2^{2M-1} & \cdots & x_M^{2M-1}
  \end{array}\right] \left[\begin{array}{c}
  a_1    \\
  \vdots \\
  a_M
  \end{array}\right]
\end{equation}
for $x_m = \exp\left[-a E_m\left(\vec{p}\right)\right]$ and $a_m = A_m\left(\vec{p}\right) \exp\left[ -t_0 E_m\left(\vec{p}\right)\right]$ where $y_n = C\left(\vec{p},t_n\right)$. $\bvec{V}(x)$ is known as $2M \times M$ rectangular Vandermonde matrix. 
The $M=1$ solution is simple to compute and widely known in the lattice QCD literature; it is the effective-mass solution. The $M=2$ solution was explicitly constructed in Refs.~\cite{Fleming:2004hs,Guadagnoli:2004wm,Lin:2008gv}. One can analytically solve up to $M=4$ systems.

However, there are ways to get even higher states. For the general approach for $M > 4$ (Prony-Yule-Walker method or just Prony method\cite{Prony:1795,Yule:1927,Walker:1931}) and solutions with multiple correlation functions, please refer to Ref.~\cite{Fleming:2009wb} for more details.

\subsubsection{Variational Method}

The variational method~\cite{Michael:1985ne,Luscher:1990ck} is a powerful tool for extracting multiple excited states using a matrix of two-point correlators.

We can often write down multiple independent operators that describe the same quantum numbers. In the simplest case, we can merely adjust the smearing function applied to the quarks in a single-site operator. If we include operators that involve different sites, there are more operators we can write down to generate even larger bases for the operators. We organize these by their behavior under the elements of the cubic group (the group describing the rotational symmetry of the lattice) into irreducible representations (irreps). Since the action of QCD respects the symmetry of the cubic group, an operator belonging to one irrep never create states having overlap with other irreps.

Using these operators, we construct an $r \times r$ correlator matrix, $C_{ij}(t)$, where each element of the matrix is a two-point correlator composed from different operators $O_i$ and $O_j$. Then we consider the generalized eigenvalue problem
\begin{equation}
C(t)\psi = \lambda(t,t_0)C(t_0)\psi,
\end{equation}
where the selection of $t_0$ depends on the range of validity of our approximation of the correlators by the lowest $r$ eigenstates. If $t_0$ is too large, the highest-lying states will have exponentially decreased too far to have good signal-to-noise ratio; if $t_0$ is too small, many states above the $r$ we can determine will contaminate our extraction. Over some intermediate range in $t_0$, we should find consistent results.

If the eigenvector for this system is $|\alpha\rangle$, and $\alpha$ goes from 1 to $r$, the correlation matrix can be approximated as
\begin{equation}
	C_{ij}=\sum_{n=1}^r v_i^{k*} v_j^n e^{-t E_n}
\end{equation}
with eigenvalues
\begin{equation}
	\lambda_n(t,t_0)=e^{-(t-t_0)E_n}
\end{equation}
(up to exponential corrections with exponents equal to the mass splitting with the next state not included in the approximation) by solving
\begin{equation}
	C(t_0)^{-1/2}C(t)C(t_0)^{-1/2}\psi=\lambda(t,t_0)\psi.
\end{equation}
The resulting eigenvalues $\lambda_n(t,t_0)$, called the principal correlators, are then further analyzed to extract the energy levels, $E_n$. Since they have been projected onto pure eigenstates of the Hamiltonian, each principal correlator should be fit well by a single exponential. The leading contamination due to higher-lying states is another exponential having higher energy; a two-state fit may help to remove this contamination.

\section{Light/Strange Flavor}\label{sec:light-strange}

\subsection{Roper Resonance}

In both meson and baryon spectroscopy there are many experimentally observed excited states whose physical properties, such as spin and parity, are poorly understood. In such cases, theoretical input from LQCD could help to solidify their identification. 
Among the excited nucleon states ($N^*$), the nature of the Roper resonance, $N(1440)\ P_{11}$, has been the subject of interest since its discovery in the 1960s. The reversal of the mass ordering of the positive-parity excited state (the Roper) and the negative-parity ground state ($N(1535)\ S_{11}$) is quite unusual, never being seen in mesons. This has led phenomenologists to postulate that the Roper might not simply be a radial excitation of the nucleon. Rather, it might be a hybrid state that couples predominantly to QCD currents with some gluonic contribution~\cite{Carlson:1991tg} or a five-quark (meson-baryon) state~\cite{Krehl:1999km}. Experiments at Jefferson Laboratory, MIT-Bates, LEGS, Mainz, Bonn, GRAAL, and Spring-8 offer new opportunities to understand in detail how nucleon resonance properties emerge from nonperturbative aspects of QCD.

\begin{figure}[t]
\begin{center}
\includegraphics[width=.65\textwidth]{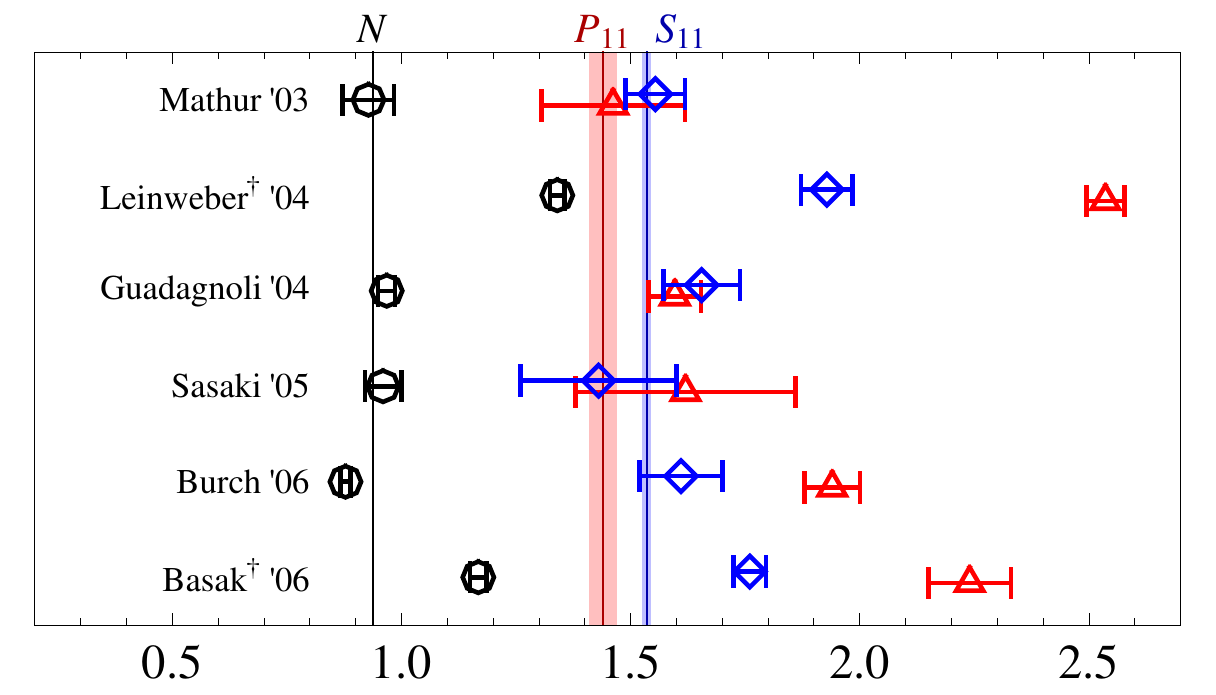}
\caption{\label{fig:all-roper}Summary of previous lattice calculations with
    extrapolation to the physical pion mass point or the lowest simulated pion
    point (labeled as ``$\dag$'').}
\end{center}
\end{figure}

Early LQCD calculations using the quenched approximation~\cite{Sasaki:2001nf,Mathur:2003zf,Guadagnoli:2004wm,Leinweber:2004it,Sasaki:2005ap,Sasaki:2005ug,Burch:2006cc}, found the computed spectrum inverted relative to experiment, with $P_{11}$ heavier than $S_{11}$. Fig.~\ref{fig:all-roper} and Table~\ref{tab:SP_summary} show a summary of parameters and analysis methods used in these works and the extrapolated masses at the physical pion mass. Only Ref.~\cite{Mathur:2003zf}, which uses lighter pion masses, seems to observe a potential mass reversal in their central values. There is agreement with experimental values within errors, since the masses of the $P_{11}$ and $S_{11}$ are overlapping within their statistical errorbar.

However, if we look at most of the published even-parity LQCD results ($N_f=0$)~\cite{Mathur:2003zf,Brommel:2003jm,Basak:2005ir,Sasaki:2005ug,Burch:2006cc,Lasscock:2007ce,Mahbub:2010jz} as functions of pion mass, we find big discrepancies in the calculated nucleon first-excited mass. The Roper calculation on the lattice is in a more chaotic situation than other states. 
One may suspect that the discrepancy is caused by how the lattice spacing is determined. In fact, there are different choices available for the scale-setting technique: Some prefer to use experimentally well-measured hadronic quantities, such as $f_\pi$ or $m_\Omega$ or bottomonium splittings to set the scale. Others prefer to calculate quantities that are not likely to depend on the lattice volume, such as the string tension in the heavy-quark potential in the static limit; this is matched up with phenomenological parameters at distance scale $r_0$ or $r_1$, which requires independent determination. However, if the different ways of setting the scale caused the discrepancy we find in Roper masses, we would expect the same amount of discrepancy to be reflected in the nucleon mass as well; however, the effects there are much smaller. 
Refs.~\cite{Mathur:2003zf,Brommel:2003jm} both use chiral fermion actions (overlap fermions and chirally improved fermions respectively), which preserves the chiral symmetry at finite lattice spacing. However, their Roper masses are no closer than any other set. 
Refs.~\cite{Burch:2006cc,Mahbub:2010jz} use local nucleon operators with different smearing parameters and analyze the data with the variational method; their results seems to be consistent within $2\sigma$. Ref.~\cite{Basak:2005ir} also uses the variational method and adopts more sophisticated operators organized according to their representations under the cubic symmetry group. These calculations using Wilson-type fermions are expected to have larger discretization error than other $O(a)$-improved fermion actions. Taking account the systematics, Refs.~\cite{Basak:2005ir,Burch:2006cc,Mahbub:2010jz} are rather consistent, but disagree with results from Ref.~\cite{Mathur:2003zf} which adopts Bayesian fitting method.

\begin{figure}[t]
\begin{center}
\includegraphics[height=.32\textheight]{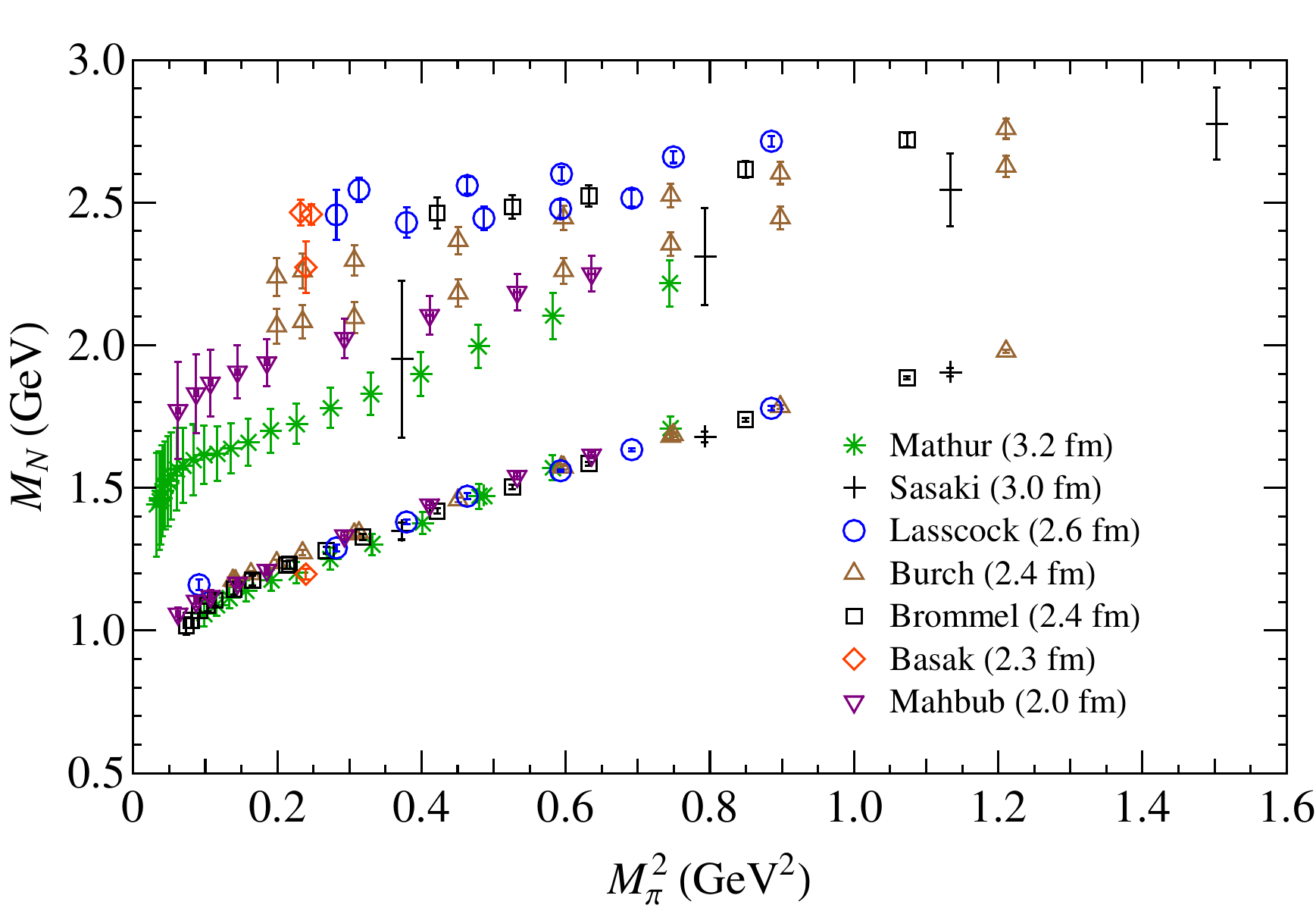}
\includegraphics[height=.32\textheight]{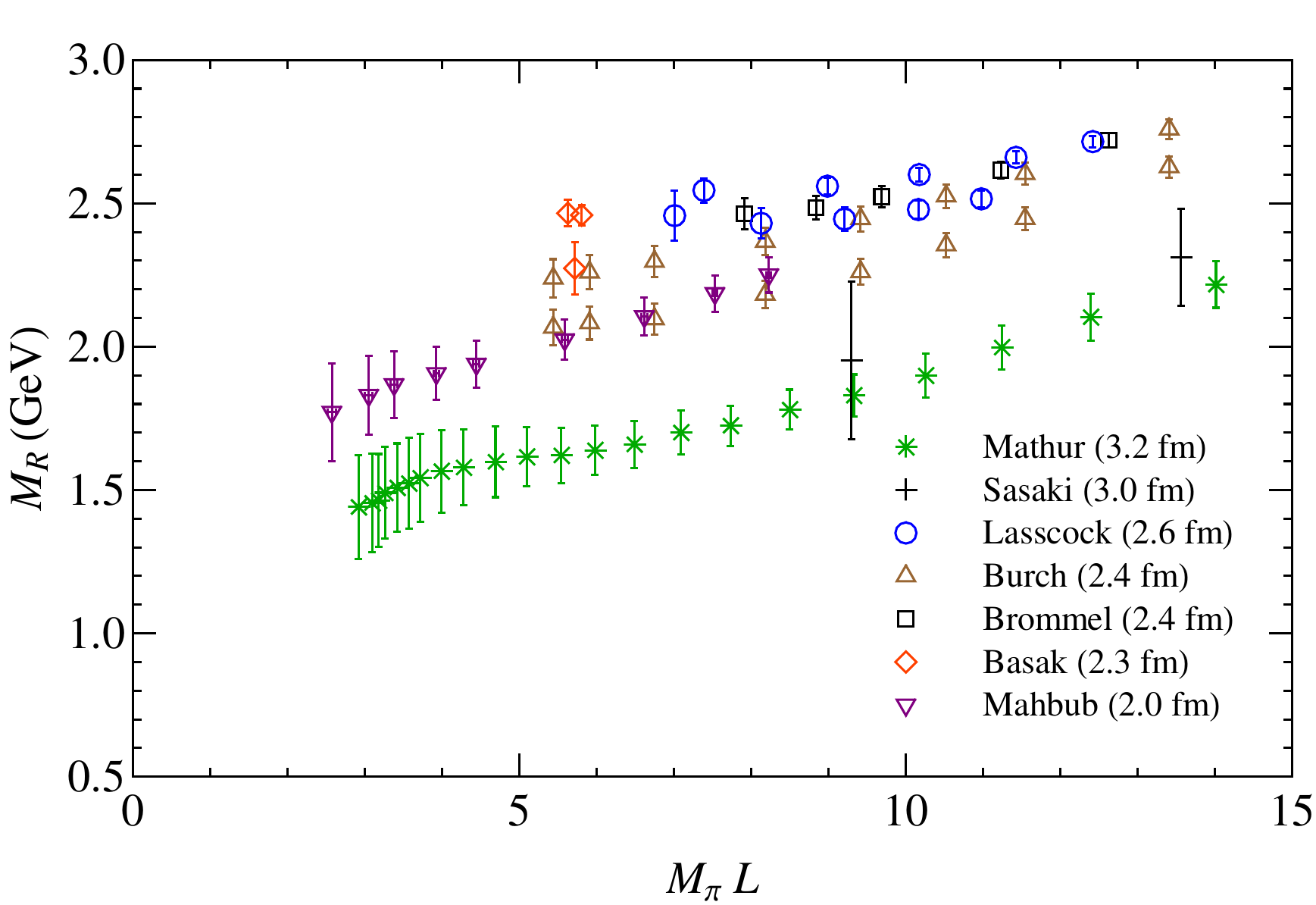}
\end{center}
\caption{\label{fig:Resonance-FV} Summary of published $N_f=0$ LQCD calculations of the nucleon and Roper masses in GeV as functions of $m_\pi^2$ (left) and Roper masses as a function of the dimensionless product of the pion mass and lattice size $L$ (right). Note that the errorbars are only statistical.
}
\end{figure}

We address the same issue on left-hand side of Fig.~\ref{fig:Resonance-FV}, sorting the results as functions of $m_\pi L$, which gives us a rough estimate of the expected finite-volume effects on these data. We find that certain discrepancies of the Roper masses seems to line up with $m_\pi L$. This suggests finite-volume effects can be more severe for excited states than ground states and that careful examination of such systematic errors is crucial. However, there are still two regions of Roper masses among different lattice groups for $m_\pi L > 5$ and further studies will be required to understand the difference between the different ways of extracting Roper masses and properly assign systematics, which are too often ignored.

\begin{table}
\begin{center}
\caption{\label{tab:SP_summary}Summary of existing published $S_{11}$ and $P_{11}$ calculations. Due to space limitations, we adopt these abbreviations for fermion actions: Domain-Wall Fermions~\cite{Kaplan:1992bt,Kaplan:1992sg,Shamir:1993zy,Furman:1994ky} (DWF), Chirally Improved Dirac Operator~\cite{Gattringer:2000js,Gattringer:2000qu} (CIDO), Fat-Link Irrelevant Clover~\cite{Zanotti:2001yb} (FLIC); and for the analysis methods: Variational Method~\cite{Michael:1985ne,Luscher:1990ck} (VM), Constrained Curve Fitting~\cite{Lepage:2001ym} (CCF), Maximum Entropy Method~\cite{Nakahara:1999vy,Asakawa:2000tr} (MEM), Black-Box Method~\cite{Fleming:2009wb,Lin:2007iq,Fleming:2004hs} (BBM). For those works which do not perform extrapolation, we use the lightest pion mass to represent their results.}
\footnotesize
\begin{tabular*}{160mm}{c@{\extracolsep{\fill}}|c|c|c|c|c|c|c|c|}
\hline\hline
Group & $N_{\rm f}$ & $S_{\rm f}$ & $a_t^{-1}$ (GeV)  & $m_\pi$ (GeV) & $L$ (fm) & Method & Extrapolation \\
\hline
Basak et~al.~\cite{Basak:2006ww}    & 0   & Wilson & 6.05   & 0.49 &  2.35  & VM & N/A \\
Burch et~al.~\cite{Burch:2006cc}    & 0   & CIDO & 1.68,1.35   & 0.35--1.1 &  2.4  & VM & $a+b m_\pi^2$ \\
Sasaki et~al.~\cite{Sasaki:2005ap}    & 0   & Wilson & 2.1   & 0.61--1.22 &  1.5,3.0  & MEM & $\sqrt{a+b m_\pi^2}$ \\
Guadagnoli et~al.~\cite{Guadagnoli:2004wm}    & 0   & Clover~\cite{Sheikholeslami:1985ij} & 2.55   & 0.51--1.08 &  1.85  & BBM & $a+b m_\pi^2 +c m_\pi^4$ \\
Leinweber et~al.~\cite{Leinweber:2004it}    & 0   & FLIC & 1.6   & 0.50--0.91 &  2.0  & VM  & N/A \\
Mathur et~al.~\cite{Mathur:2003zf}    & 0   & Overlap~\cite{Neuberger:1997fp} & 1.0   & 0.18--0.87 &  2.4,3.2  & CCF  & $a+b m_\pi +c m_\pi^2$ \\
Sasaki et~al.~\cite{Sasaki:2001nf}    & 0   & DWF & 2.1   & 0.56--1.43 &  1.5  & VM  & $a+b m_\pi^2$ \\
\hline\hline
\end{tabular*}
\end{center}
\end{table}

The examples given above are calculated in a vacuum with gluonic degrees of freedom only; that is, no sea fermion loops contribute to the ensembles. In recent years, newer calculations need not apply this approximation, and more results with $N_f=2+1$ (degenerate up and down plus strange in the sea) are available.

\begin{figure}
\includegraphics[width=.65\textwidth]{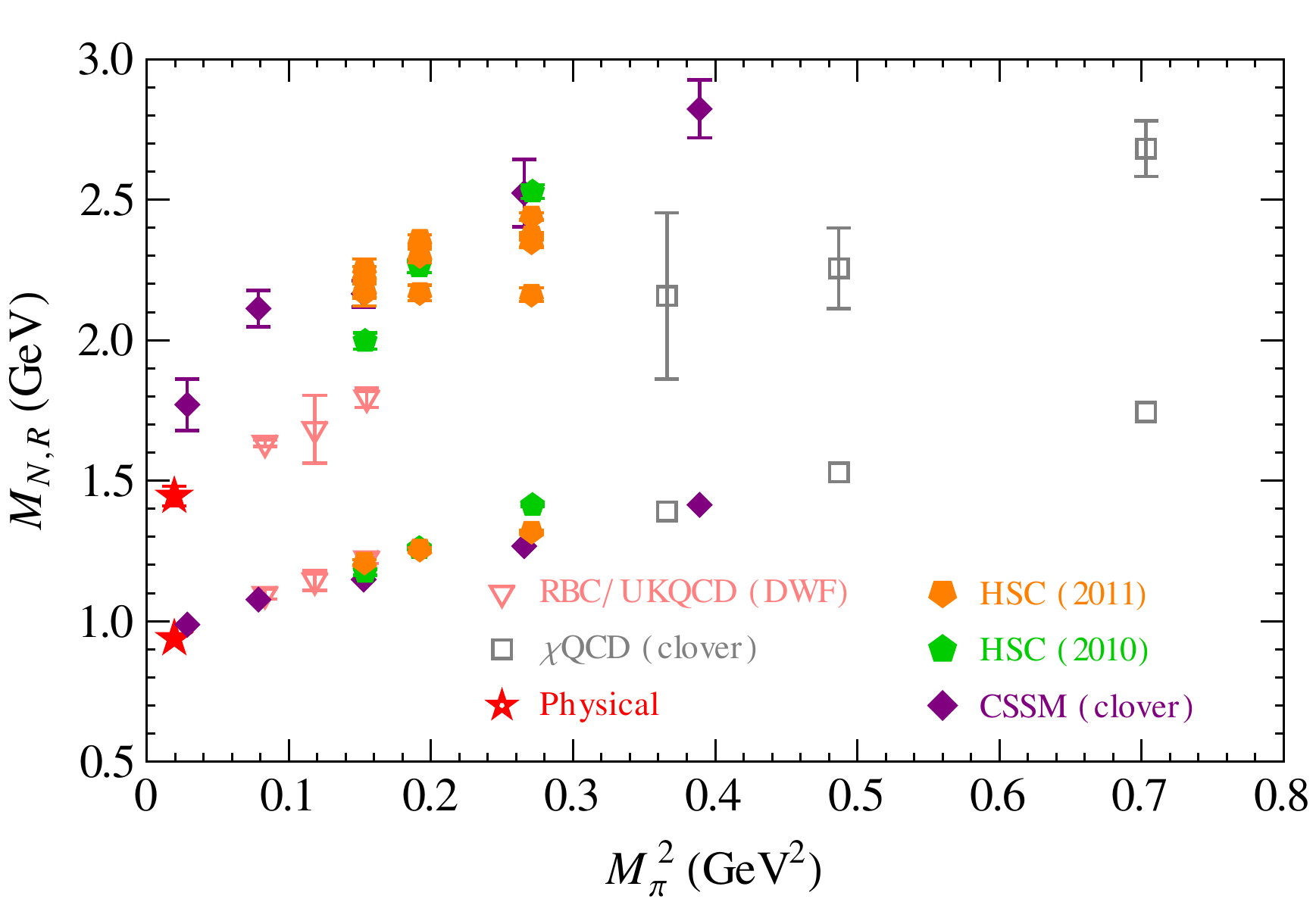}
\caption{\label{fig:nf21roper}
Summary of the latest $N_f=2+1$ Roper results as a function of pion mass squared.
Note that the filled symbols and solid errorbars (open symbols and dashed errorbars) indicate the results taken from published papers (the latest Lattice Conference proceedings).
}
\end{figure}

Two preliminary studies were reported in the annual lattice conference using $N_f=2+1$ DWF and clover lattices by RBC/UKQCD and $\chi$QCD; their results are shown as the open symbols in Fig.~\ref{fig:nf21roper}, but no refined numbers have been reported since. 
CSSM lattice group\cite{Mahbub:2010rm} uses local baryon operators with different smearing parameters to construct a large basis for the nucleon with spin 1/2 and analyze the data using the variational method. They use $N_f=2+1$ isotropic lattices at a fixed lattice spacing, $a=0.0907$~fm and fixed volume, $32^3\times 64$ with lightest pion mass 156~MeV. Their calculation of the Roper masses are shown in Fig.~\ref{fig:nf21roper} as 5 purple diamond points. However, the lightest two pion masses may suffer finite-volume effects (which have not currently been estimated), since their $m_\pi L$ values are less than 3. 
Hadron Spectroscopy Collaboration reported two studies using a large basis of independent operators organized according to representations of the cubic group, including operators where the three quarks inside the nucleon are located at different lattice sites\cite{Bulava:2010yg,Edwards:2011jj}. They have used $2+1$-flavor anisotropic clover lattices, with finer temporal lattice spacing to improve the excited-state signal in Euclidean time; that is, the lattices have better resolution in the direction from which the particle energies are extracted. The green pentagons in Fig.~\ref{fig:nf21roper} are taken from the lowest two states reported in their study last year using only operators in the $G_{1g}$ irrep. This year, a refined study mapped the cubic-irrep operators to definite spins using derivative operators constructed with continuum symmetries. They found four states closely aligned and roughly around the excited state found by the CSSM lattice group. 
A refined CSSM study using larger basis of their smeared operators reported in this year's Nstar Workshop a similar structure as Ref.~\cite{Edwards:2011jj}. 
This may suggest that multiple-particle operators need to be included in the correlator matrix to clearly identify the nature of these closely related states; both HSC and CSSM are currently investigating this.

The dynamical results in Fig.~\ref{fig:nf21roper} are impressive but the errorbars shown above are only statistical only; i.e., there is no estimation of the systematics due to the finite volume and nonzero lattice spacings in the calculations. This is sometimes difficult due to the limited amount of dynamical lattices available. 
Despite the difficulties, NPLQCD made such a calculation in Ref.~\cite{Beane:2011pc} with high statistics using $N_f=2+1$ anisotropic clover lattices at $m_\pi \approx 390$~MeV. Four volumes are used in this calculation, 2, 2.5, 3, and 4~fm with $m_\pi L \approx 3.9$, 4.8, 5.8, 7.7. The volume dependence is not noticeable in the earlier HSC study with fewer trajectories available at the time. To our surprise, the ground state with $m_\pi L$ around 4 (rule-of-thumb suggests that at this value finite-volume effects can be ignored) shows around 30~MeV discrepancy from its value in the infinite-volume limit. They also extract the axial decuplet-octet-meson couplings from volume-dependent heavy-baryon chiral perturbation theory, $|g_{\Delta N \pi }|=2.80(18)(21)$; similarly for other couplings such as $|g_{\Sigma^* \Lambda \pi }|=2.49(23)(35)$, $|g_{\Xi^* \Xi \pi }|=2.49(23)(35)$. 
The effect of the finite volume on the Roper mass is even more dramatic; one expects the central values to shift to by about 170~MeV. The multiple-volume study also allows the issue of the Roper's multiplicity to be addressed. Since the ratio of the ``weight'' or overlap factors associated with the Roper between the largest two volumes is $A_R(4~fm)/A_R(3~fm)=0.964(52)$ consistent with one, we see that these first-excited states are likely to be one-particle states instead of two. A scattering state would have overlap factors proportional to the lattice volume. 
More time should be devoted to studying volume dependence to get a better estimation on the finite-volume effects from these dynamical calculations.

We can also study the nature of the Roper resonance by examining its form factors, such as the transfer-momentum dependence of the Roper-nucleon transition form factors. A first exploratory study was done in 2008, using quenched anisotropic lattices at pion mass 720~MeV\cite{Lin:2008qv}. Additional pion masses (480 and 1100~MeV) were reported in Ref.~\cite{Lin:2008gv}. However, none the results yielded negative values for $F_2^{N*p}$ in the low-momentum region, as suggested by experimental studies. 
An updated study of the Roper-nucleon form factors using $2+1$-flavor anisotropic lattices~\cite{Edwards:2008ja,Lin:2008pr} seems to reproduce the behavior of CLAS's analysis, providing further strong evidence that the Roper is the first radially excited state of nucleon. Further study of the systematics and improved operators will be needed to strengthen this statement.

\subsection{Highly Excited States}
In the previous section, we found how difficult it is to extract the Roper resonance. What if we want to get to even higher excited states with LQCD? Unfortunately, in Euclidean space, excited-state contributions to correlation functions decay faster than the ground state. Therefore at large times, the signals for excited states are swamped by the signals for lower-energy states. One way resolve this issue is to improve resolution in the temporal direction. 
An anisotropic lattice where the temporal lattice spacing is finer than spatial spacings can provide better resolution while avoiding the increase in computational cost associated with a similar reduction of all spacings.

One also needs large basis of independent operators that allow the maximal overlap with higher-spin operators. The use of cubic-group irrep operators or operators constructed according to continuum symmetries is needed to expand the reach of the lattice calculation. The variational method becomes quite reliable when allowed to consider up to the 8th excited state. In this section, we will focus on the Hadron Spectroscopy Collaboration (HSC)'s concerted efforts and developments over the past few years.

\subsubsection{Dynamical Anisotropic Lattices}

Using a finer temporal lattice spacing helps to reduce the systematics associated with energy extraction for higher excited energies (even with the aid of the variational method). Such lattices provide many time slices of information to reconstruct more accurately the signal due to a particular state, which is crucial since the signal in Euclidean time declines exponentially. In principle, we could use very fine lattices while keeping the volume big enough to avoid the ``squeezing'' systematic effect; however, computational cost scales significantly (power of 5--6) with the inverse lattice spacing. 
Tuning the parameters for dynamical anisotropic lattices is extremely difficult. Not only do we have to set the usual coupling constants and quark masses, but also the anisotropy for both gauge and fermion fields. The task is further complicated by the feedback among the parameters. 
HSC adopted anisotropic clover lattices to keep the spatial lattice spacing coarse, avoiding finite-volume systematic error, and to make the temporal lattice spacing fine enough to extract towers of nucleon excited states. However, tuning the $O(a)$-improved parameters for fermion actions in the dynamical gauge generation correctly is a much more difficult task for anisotropic lattices. 
They use Symanzik-improved gauge action and clover fermion action with 3-dimensional stout-link smeared gauge fields; the gauge ensemble is generated using the (R)HMC algorithm. The spatial lattice spacing is $a_s=0.1227(8)$~fm (determined using $m_\Omega$), and the renormalized anisotropy $\xi_R=a_s/a_t$ is $3.5$. 
Ref.~\cite{Edwards:2008ja} gives a detailed study of how to set the dynamical anisotropic lattice parameters, and Ref.~\cite{Lin:2008pr} reports basic lattice properties along with the ground-state hadron spectrum.

\begin{figure}[t]
\begin{center}
\includegraphics[width=0.32\textwidth]{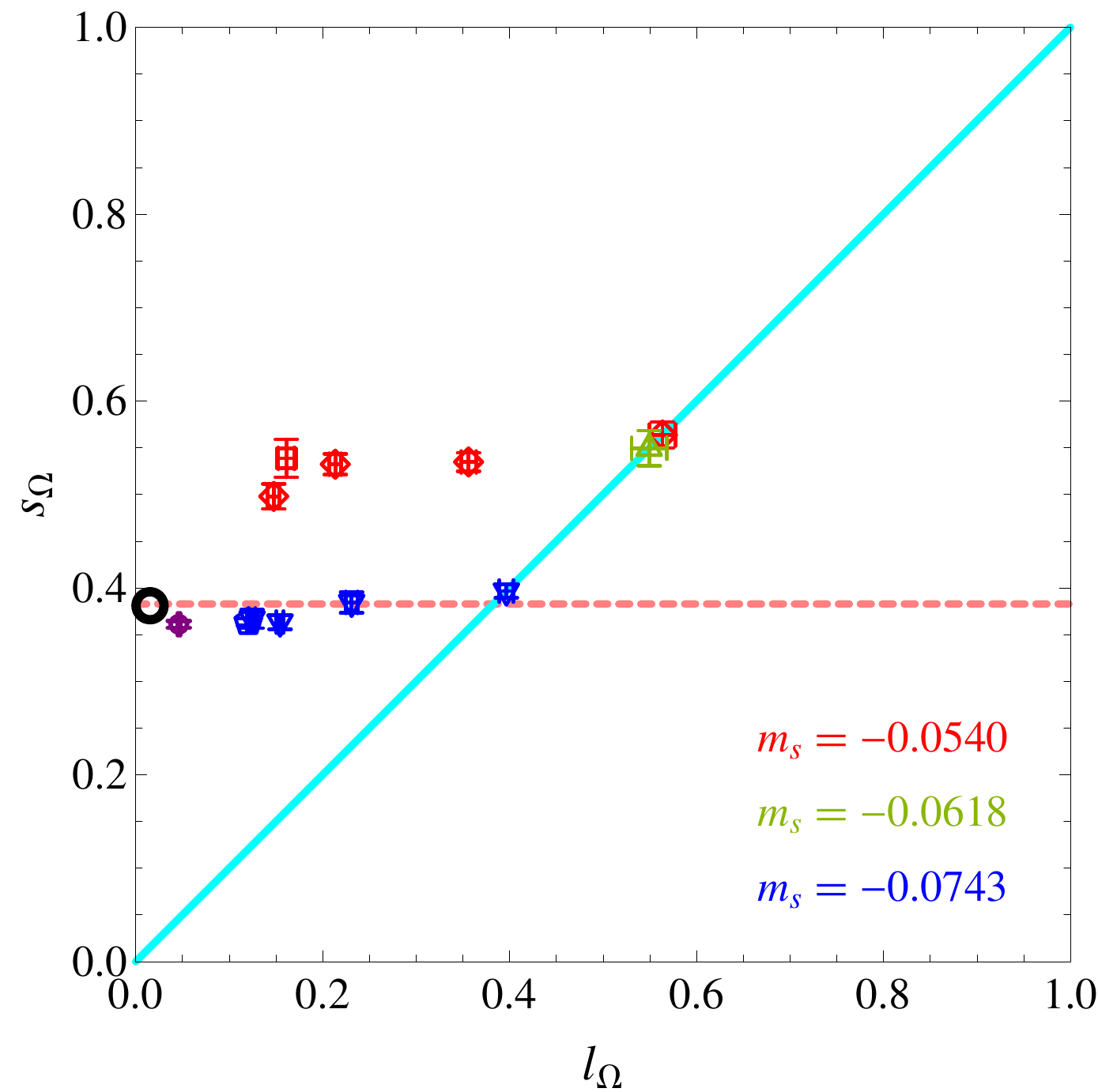}
\includegraphics[width=0.32\textwidth]{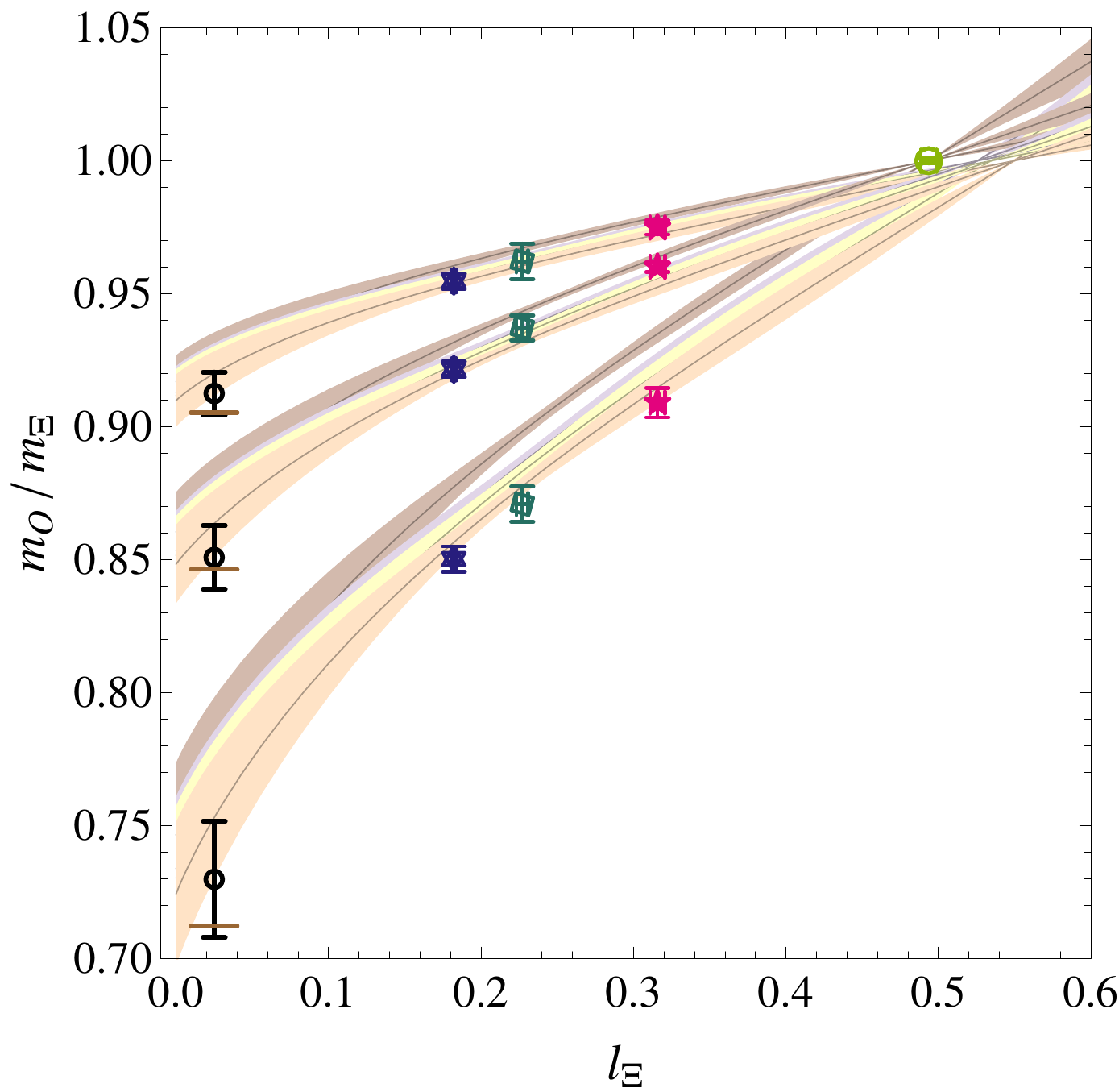}
\includegraphics[width=0.32\textwidth]{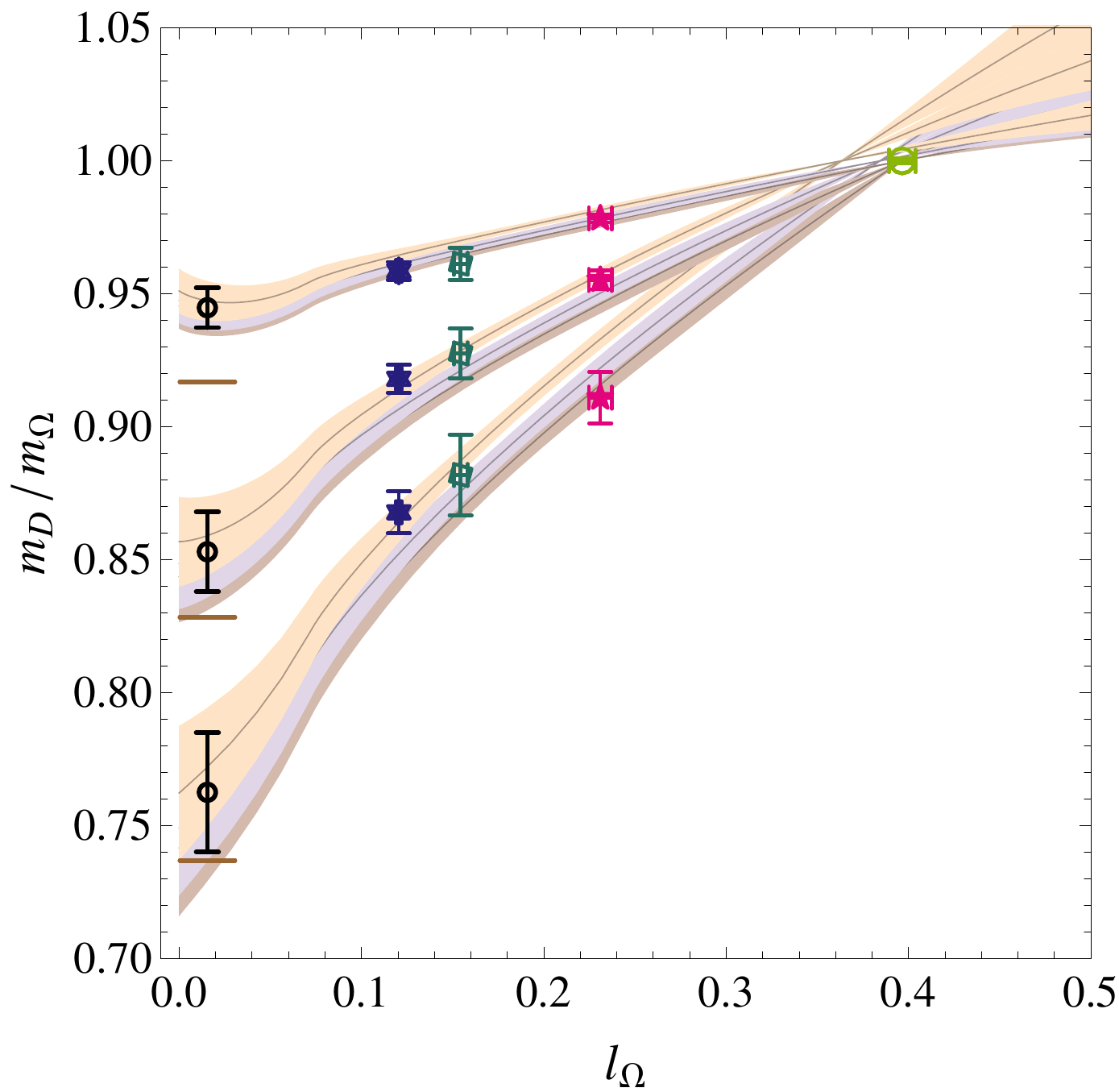}
\end{center}
\vspace{-0.3in}
\caption{\label{fig:ChPTMassRatio}
(Left panel) The location of the dynamical ensembles used in Ref.~\cite{Cohen:2009zk} in the $s_\Omega$-$l_\Omega$ plane. The leftmost circle (black) indicates the physical point \{$l_\Omega^{\rm phys}$, $s_\Omega^{\rm phys}$\}, while the red, green and blue points are the lattices with $a_tm_s=-0.0540, -0.0618, -0.0743$ respectively in Ref.~\cite{Lin:2008pr}. The horizontal dashed (pink) line indicates constant $s_\Omega$ at the physical point, and the diagonal line indicates three-flavor degenerate theories. The purple points are from recent measurements of the 230~MeV ensemble, showing that it does not deviate away from the $N_f=3$ value.
Mass-ratio chiral extrapolations as functions of the $l_\Xi$ and $s_\Xi$ for octets (center panel) and $l_\Omega$ and $s_\Omega$ for decuplets (right panel).
The lines indicate the ``projected'' leading chiral extrapolation fits in $l_\Xi$ and $s_\Xi$ ($l_\Omega$ and $s_\Omega$) while keeping the other one fixed. The black (circular) point is the extrapolated point at physical $l_\Xi$ and $s_\Xi$ ($l_\Omega$ and $s_\Omega$).
The masses for $N$, $\Sigma$ and $\Lambda$ are 0.962(29), 1.203(11), 1.122(16)~GeV, respectively, and the masses for $\Delta$, $\Sigma^*$ and $\Xi^*$ are 1.275(38), 1.426(25) and 1.580(13)~GeV, respectively.
}
\end{figure}

HSC tunes the strange-quark mass using the dimensionless parameter $s_\Omega=\frac{9(2m_K^2-m_\pi^2)}{4m_\Omega^2}$. The parameter $s_\Omega$ is set as close as possible to the corresponding experimental value at the $SU(3)_f$-symmetric point (875-MeV pion mass). When the sea-quark mass is reduced, the $s_\Omega$ parameter is seen to remain roughly constant for pion masses as low as 383~MeV in Ref.~\cite{Lin:2008pr}. (Furthermore, the $s_\Omega$ parameter is more sensitive to the strange-quark mass than other methods, such as those using the $J$-parameter. Since it is a ratio, there is no need to worry about estimating the shift in the lattice spacing when decreasing the sea-quark mass or generating more statistics.) Ref.~\cite{Cohen:2009zk} reported an updated measurement of the same quantity on a 230-MeV $24^3$ ensemble, and $s_\Omega$ does not deviate from the expected value. The $s_\Omega$ parameter is a stable and useful observable for setting the strange-quark mass, as shown in the leftmost of Fig.~\ref{fig:ChPTMassRatio}.

Ref.~\cite{Cohen:2009zk} also employs a refined baryon mass extrapolation using $m_\Omega$ or $m_\Xi$ as a reference mass. 
Using mass ratios instead of masses decreases the statistical (and possibly systematic) errors due to explicit cancellation of systematics and gauge-field noise and removes the ambiguity associated with setting the lattice spacing. Using the same data, they update the mass-ratio chiral extrapolations by modifying the next-to-leading-order heavy-baryon chiral perturbation theory. They find the finite-volume corrections are negligible according to ChPT estimates, and the extrapolated baryon masses agree with experimental values significantly better than with a linear extrapolation. The results are shown in middle and right-hand side of Fig.~\ref{fig:ChPTMassRatio}.

\subsubsection{Cubic-Group Irrep Operators}

Another important tool for extracting highly excited states is the variational method~\cite{Michael:1985ne,Luscher:1990ck}, which uses a matrix of different source and sink operators to project more exactly onto the eigenstates of the Hamiltonian and make the analysis more reliable. To use the method effectively, we need a large number of independent operators that overlap well with excited states with desired quantum numbers and linearly independent of with each other. 
Unfortunately, the single-site baryon operators of form given in Eq.~\ref{eq:proton-op} can only give us 3 different operators. We need to expand our consideration to a bigger operator space to extract highly excited states reliably.

\begin{figure}[t]
\begin{center}
\includegraphics[width=.85\textwidth]{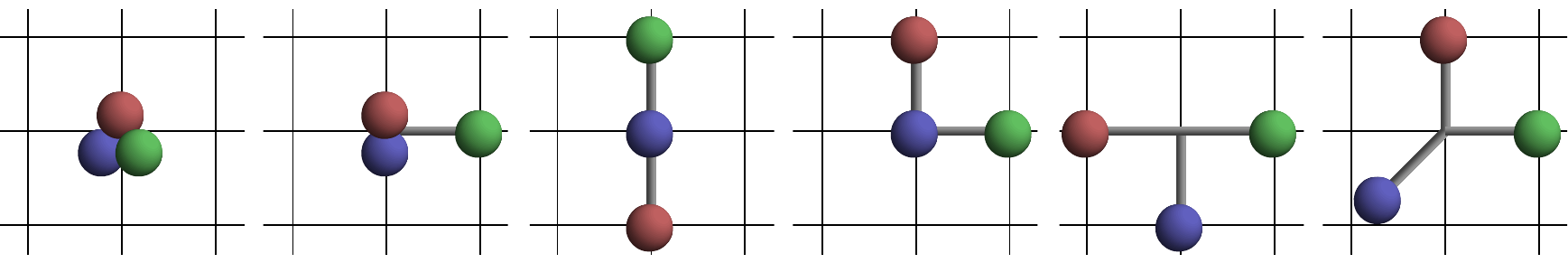}
\caption{\label{fig:baryon-ops} The orientations of 3 quarks within a single baryon. The right-most one involves all 3 spatial axes. The quark displacements shown above use only quarks separated from the central core by one unit; however, larger displacements can be used to improve overlap with higher orbitally excited states.
}
\end{center}
\end{figure}

The Hadron Spectrum Collaboration (HSC) has been investigating interpolating operators projected into irreducible representations (irreps) of the cubic group~\cite{Basak:2005aq,Basak:2005ir} in order to better calculate two-point correlators for nucleon spectroscopy. 
First, we construct various gauge-invariant baryonic operators by displacing (with appropriate gauge parallel transport) the component quarks in spatial directions as shown in Fig.~\ref{fig:baryon-ops} and compose them with the appropriate flavor structure, isospin and strangeness. This large basis of operators does not respect rotational symmetry nor the subgroup of rotations that are also respected by the lattice. We apply group-theoretical projections such that operators will transform according to an irreducible representation under lattice rotations and reflections. Since the action of the lattice theory respects these symmetries, the representation of a state is conserved as it propagates through the lattice gauge field; operators belonging to different representations will not mix even as they move from source to sink. Using such a design for our operators, we can create large correlator matrices, and the technique can be extended to meson and multi-hadron computations.

In the cubic group $O_h$, for baryons, there are four two-dimensional irreps $G_{1g}, G_{1u}, G_{2g}$, $G_{2u}$ and two four-dimensional irreps $H_g$ and $H_u$. (The subscripts ``$g$'' and ``$u$'' indicate positive and negative parity, respectively.) Each lattice irrep contains parts of many continuum states. The $G_1$ irrep contains $J\in\{\frac{1}{2},\frac{7}{2},\frac{9}{2},\frac{11}{2},\dots\}$ states, the $H$ irrep contains $J\in\{\frac{3}{2},\frac{5}{2},\frac{7}{2}, \frac{9}{2},\dots\}$ states, and the $G_2$ irrep contains $J\in\{\frac{5}{2},\frac{7}{2},\frac{11}{2},\dots\}$ states. 
The continuum-limit spins $J$ of lattice states must be deduced by examining patterns of degeneracy between the different $O_h$ irreps. For example, $G_1$ will be dominated by spin-1/2 baryons with some contribution from highly excited spin-7/2 states. The $H$ irrep will contain the spin-3/2 baryons (such as the ground-state Delta), but it will also contain many closely spaced states from all higher spins. $G_2$ will contain spin-5/2 and 7/2 states and serves as an important check of these spins in the $H$ and $G_1$ channels. 
See Table~\ref{fig:baryon-ops} for further details about spin and irreps.

Using these operators, we construct an $r \times r$ correlator matrix and extract individual excited-state energies by applying the variational method, as described above. 
More details on the baryon correlation functions evaluated using the displaced-quark operators are described in Refs.~\cite{Basak:2005aq,Basak:2005ir}. 
Demonstrations of how these operators work using purely gluonic vacuum with light valence quarks for nucleon and delta spectroscopy are reported in Refs.~\cite{Lichtl:2006dt,Basak:2006ww,Basak:2007kj}. Further calculations of isospin-$\frac{1}{2}$ excited nucleons in two-flavor QCD, using $u$ and $d$ quarks that have the same mass, are reported in Ref.~\cite{Bulava:2009jb} with 2 pion masses: 416(36) and 578(29)~MeV.

\begin{table}
\begin{center}
\begin{tabular}{|c|c|c|c|c|c|c|c|c|c|c|c|c|c|}
\hline\hline
\text{\textit{$j$}} & $\frac{1}{2}$ & $\frac{3}{2}$ &
$\frac{5}{2}$ & $\frac{7}{2}$ & $\frac{9}{2}$ & $\frac{11}{2}$
& $\frac{13}{2}$ & $\frac{15}{2}$ & $\frac{17}{2}$ &
$\frac{19}{2}$ & $...$
\\\hline
\text{Irreps} & $G_1$ & $H$ & $G_2\oplus H$ & $G_1\oplus
G_2\oplus H$ & $G_1\oplus 2 H$ & $G_1\oplus G_2\oplus 2 H$
& $G_1\oplus 2 G_2\oplus 2 H$ & $G_1\oplus G_2\oplus 3 H$
& $2 G_1\oplus G_2\oplus 3 H$ & $2 G_1\oplus 2 G_2\oplus
3 H$
& $...$
\\
\hline
\hline
\end{tabular}
\end{center}
\caption{\label{tab:irrep-spin} Subduction of continuum baryon spin irreps into representations of the cubic group}
\end{table}

One drawback of calculations using these cubic-group operators is that they require multiple orientations in order to maximally overlap with a wide range of quantum numbers, and generating so many correlators is quite expensive. Furthermore, as we go to lighter and lighter pion masses, there will be increasingly many decay modes open, even for the lowest energy at a specific quantum number. We need to extend the matrix to include multiple-particle operators (so that we can further understand the nature of the ``resonance'' in our calculation) and ``disconnected'' operators. Further, we need to achieve better precision for each state to distinguish among them.

A new way to calculate timeslice-to-all propagators, ``distillation'', has been proposed in Ref.~\cite{Peardon:2009gh}. The method is useful for creating complex operators, such as those used in the variational method, since it reduces the amount of time needed for contractions and allows the structure of the operators to be decided after performing the Dirac inversions. Rather than using pure noise in its estimators, distillation uses sources derived from the eigenvectors of the gauge-Laplacian, giving better coverage of relevant degrees of freedom. Increasing the number of sources in this scheme improves statistics faster than 1/$\sqrt{N}$. Distillation can combined with stochastic methods, which might be desirable if the number of sources needed to cover the volume becomes too large.

The distillation operator on time-slice $t$ can be written as
\begin{equation}
 \Box(t) = V(t) V^\dagger(t)
 \rightarrow
\Box_{xy}(t) = \sum_{k=1}^N v_x^{(k)} (t) v_y^{(k)\dag} (t),
    \label{eqn:box}
\end{equation}
where the $V(t)$ is a matrix containing the first through $k^{\rm th}$ eigenvectors of the lattice spatial Laplacian. The baryon operators involve displacements (${\cal D}_i$) as well as coefficients ($S_{\alpha_1\alpha_2\alpha_3}$) in spin space:
\begin{equation}
\chi_B(t) = \epsilon^{abc} S_{\alpha_1\alpha_2\alpha_3}
({\cal D}_1\Box d)^a_{\alpha_1}
({\cal D}_2\Box u)^b_{\alpha_2}
({\cal D}_3\Box u)^c_{\alpha_3}(t),
\end{equation}
where the color indices of the quark fields acted upon by the displacement operators are contracted with the antisymmetric tensor, and sum over spin indices. Then one can construct the two-point correlator; for example, in the case of proton,
\begin{align}
C^{(2)}_B[\tau_d,\tau_u,\tau_u](t',t) &=
  \Phi^{(i,j,k)}(t')
  \tau_d^{(i,\bar{i})}(t',t)
  \tau_u^{(j,\bar{j})}(t',t)
  \tau_u^{(k,\bar{k})}(t',t)
  \Phi^{(\bar{i},\bar{j},\bar{k})*}(t)
  \nonumber\\
 &\quad- \Phi^{(i,j,k)}(t')
  \tau_d^{(i,\bar{i})}(t',t)
  \tau_u^{(j,\bar{k})}(t',t)
  \tau_u^{(k,\bar{j})}(t',t)
  \Phi^{(\bar{i},\bar{j},\bar{k})*}(t),
\end{align}
where the ``baryon elemental''
\begin{equation}
 \Phi^{(i,j,k)}_{\alpha_1\alpha_2\alpha_3}(t) = \epsilon^{abc}
\left({\cal D}_1 v^{(i)}\right)^{a}
\left({\cal D}_2 v^{(j)}\right)^{b}
\left({\cal D}_3 v^{(k)}\right)^{c}(t)\;
S_{\alpha_1\alpha_2\alpha_3}
\label{eqn:baryon_op}
\end{equation}
can be used for all flavors of baryon and quark masses with the same displacements on the same ensemble, and the ``perambulator''
\begin{equation}
  \tau_{\alpha\beta }(t',t) = V^\dagger(t') M^{-1}_{\alpha\beta}(t',t) V(t)
\label{eqn:peram}
\end{equation}
can be reused for different baryon (and meson) operators after a single inversion of the $M$ matrix. There is a large factor of computational power saved by ``factorizing'' correlators in terms of elementals and perambulators, and the same elementals and perambulators can be used to contract various different correlators.

A variations of the distillation Ref.~\cite{Morningstar:2010kb,Bulava:2010em,Morningstar:2011ka}, called
stochastic LapH method, has been developed to improve the efficiency when
work on even larger volumes; a demonstration on the excited-meson and -baryon, and
pion-pion scattering phase shift are shown in these references.

A procedure called ``pruning'' can be used to reduce the number of operators when the size of a correlator matrix begins to get out of hand. A practical procedure is to calculate the correlators with the same source and sink operator (i.e., the diagonal elements of the full correlator matrix) and sort them according to some metric of their ``individuality''. One way to systematically prune is to take the matrix of inner products of the effective masses of each correlator across all time slices and sort them from there. This measures a sense in which the effective masses of different correlators have the same shape, implying that they also have the same excited-state content, and takes advantage of the black-box nature of the effective mass to eliminate ambiguity.

The middle panel of Fig.~\ref{fig:spec-840} shows a subset of various $G_{1g}$ displacement-operator correlators sorted by the values of their inner products with respect to one another. Any overlap larger than 70\% is marked by yellow, while the remaining values are depicted as magenta to blue colors. ($G_{1g}$ has the largest overlap amongst its operators of all the baryon irreps.) Notice that the inner-product matrix has a clear block structure when organized this way. Each yellow block consists of a set of operators that yield nearly identical effective masses as functions of time. HSC exclude many operators by such a selection process and pick a couple dozen operators among different quark orientations for each irrep. This is then refined to a smaller sub-matrix by filtering the matrix by condition numbers; a variational-method analysis is performed on the resulting matrix, retaining the lowest eigenstates.

\begin{figure}[t]
\begin{center}
\includegraphics[width=.45\textwidth]{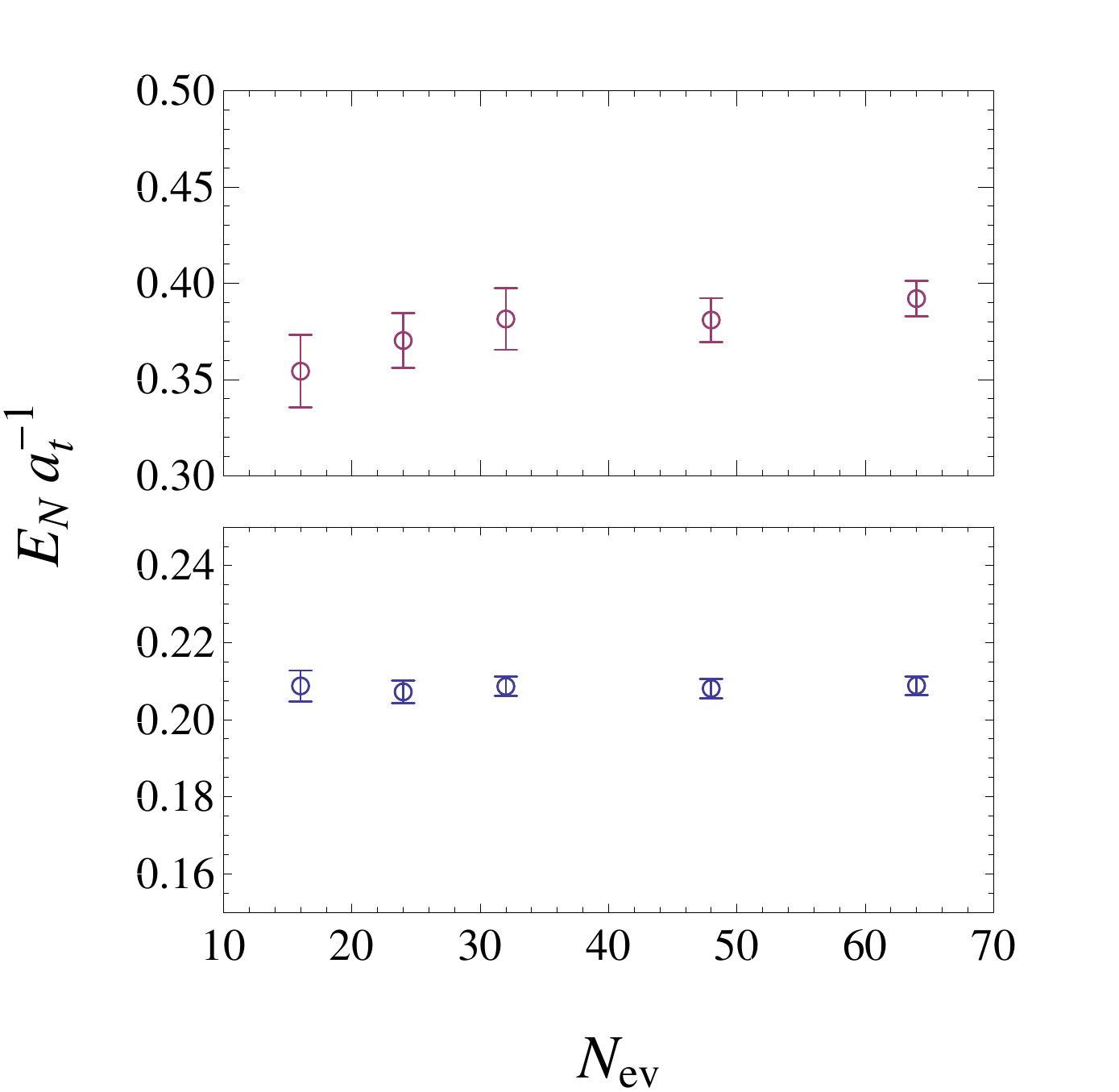}
\includegraphics[width=.4\textwidth]{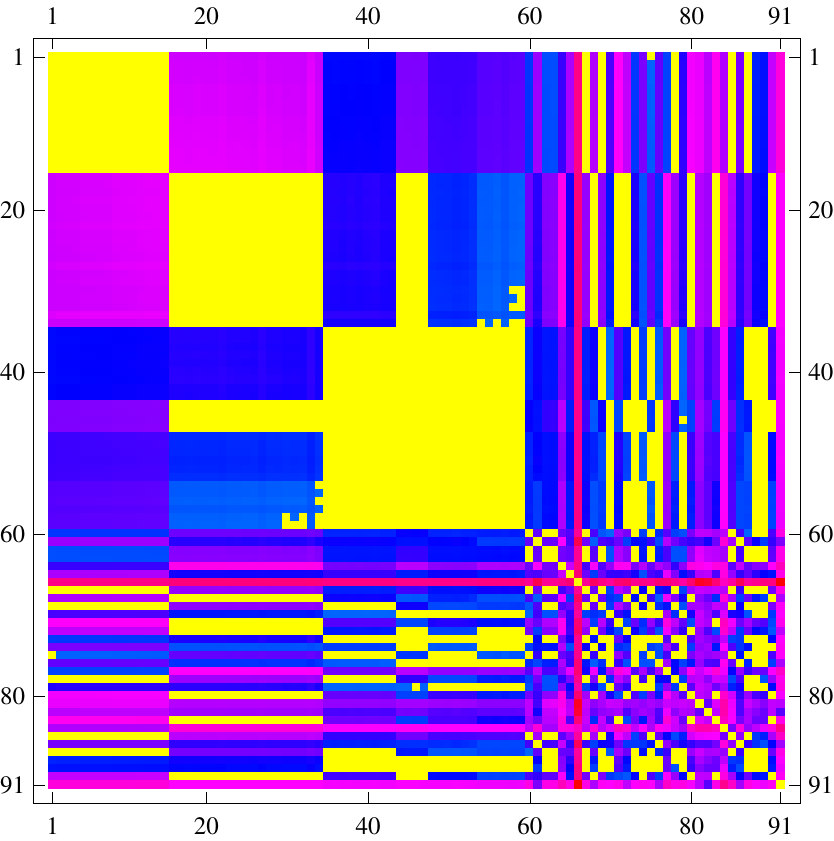}
\caption{\label{fig:dist-prunning}(Left) The fitted energies for the first-excited (top) and ground (bottom) states as functions of the number of distillation eigenvectors $N$.
(Right) A subset of $G_{1g}$ irrep operator correlators, grouped by their inner product. The yellow blocks indicate overlap ${}>70$\%. $G_{1g}$ is the worst case among all the irreps.
}
\end{center}
\end{figure}

Using the ``distillation''~\cite{Peardon:2009gh} technique, HSC has been able to produce highly excited states faster than previous two-flavor and quenched studies. The results also has better signal-to-noise ratios than the conventional approach. In Ref.~\cite{Bulava:2010yg}, HSC uses $N_f=2+1$ anisotropic clover lattices with pion masses of 392(4), 438(3) and 521(3)~MeV at a fixed volume, $16^3\times 128$ (which gives roughly 2-fm spatial volume).

Fig.~\ref{fig:spec-840} shows the results for nucleon and Delta spectroscopy with lightest pion mass around 390~MeV, classified by the cubic-group irreps. To improve the clarity of the graphic, we also regroup the experimental 3- and 4-star states according to their subduced lattice irreps. Note that multiple cubic-group states will merge into the same continuum state; thus, we see a few repeated states on the experimental summary plot. Out of hundreds operators, only the best 7 and 11 operators are chosen to calculate the full correlator matrix for the nucleon and Delta. Each principal correlator is fitted with a two-exponential form to get rid of possible contamination from excited states that have not been fully projected out by the variational method. In the end, the lowest 6 states are reported.

In the nucleon spectrum, the first excited state from the $G_{1g}$ irrep is the potential Roper state and appears at roughly 1.9~GeV, which is heavier than its negative-parity partner, which is only around 1.7~GeV. Fig.~4 in Ref.~\cite{Bulava:2010yg} shows the pion mass dependence from 390 to 520~MeV, and the Roper mass decreases much faster than the $S_{11}$; it is likely at the physical limit, the Roper becomes lighter than its negative-parity partner but studies at lighter pion masses need to be performed. The higher states in $G_{1g,u}$ may correspond to spin-7/2 states.

Overall, the states extracted from the lattice calculations are clustered; these states maybe degenerate in the continuum limit. For example, the $G_2$ irrep corresponds to states with at least spin 5/2, which has six linearly independent components in the continuum limit. Each $G_2$ state must have a partner $H$ state with the same parity in order to have an interpretation as a physical state. However, since unlike irrep identity, spin is not conserved on the lattice, lattice discretization effects can cause the $H$ and $G_2$ partner states to have energy difference of $O(a^2)$.

The experimental $H_g$ spectrum has a wide spread of states ranging from 1.6 to 3~GeV, while lattice data for pion mass around 400~MeV forms 2 clusters and ranging from 1.9 to 2~GeV; however, Fig.~4 in Ref.~\cite{Bulava:2010yg} shows that as pion mass decreases, the spectrum spreads out for this irrep. The $H_g$ states correspond to five experimental resonances: $N(1720; 3/2^+)$ and $N(1900; 3/2^+)$, $N(1680; 5/2^+)$, $N(2000; 5/2^+)$ and $N(1990; 7/2^+)$. 
In the $H_u$ irrep, the lattice results have better determined spectrum. The four low-lying lattice states spread within 1.7--1.9~GeV may correspond to the experimental resonances $N(1530; 3/2^-)$, $N(1650; 3/2^-)$ and $N(1675; 5/2^-)$. However, the threshold for scattering states is near the same energy as this group of lattice states. We cannot distinguish clearly between single- and multi-particle contributions to the spectrum in this calculation.

The $\Delta$ results from HSC at the lowest pion mass (400~MeV) is shown in the lower row of Fig.~\ref{fig:spec-840}. As in the nucleon case, overall the lattice data points are higher due to the heavier pion mass used in the calculation. The extracted energies all decrease as the pion mass is reduced closer to the physical pion mass. In the experimental irrep figure, there are more degenerate states due to the cubic symmetry.

The lowest $H_g$ state corresponds to the ground state of $\Delta$; the nearby excited state is likely to be the spin-3/2 state $\Delta(1600)$. The higher states are likely to be some combination of spin-$5/2^+$ and $7/2^+$ resonances and scattering states. 
The negative-parity irrep, $H_u$, forms two clusters of states. The lowest two states are likely to be $\Delta(1700; 3/2^-)$ and $\Delta(1930; 5/2^-)$, while some of the higher states could correspond to spin-9/2 states, possibly mixing with the $G_{2u}$ irrep.

The lowest two $G_{1g}$ states seem to correspond with the experimental resonances $\Delta(1910; 1/2^+)$ and $\Delta(1950; 7/2^+)$. However, the same spin-$7/2^+$ could show up the spectrum of irreps $H_g$ and $G_{2g}$ due to the degeneracy caused by the broken rotational symmetry. 
Similarly for the spin-9/2 state and the parity-negative irrep $G_{1u}$.

Finally, the $G_2$ irrep starts with spin-5/2 states and then 7/2. The 3 lowest lattice states of the $G_{2g}$ spectrum might correspond to $\Delta(1905; 5/2^+)$ and scattering states. The $G_{2u}$ is are less well determined, with states collapsing together; no strong identification with the experimental resonances can be made.

\begin{figure}[t]
\begin{center}
\includegraphics[width=.45\textwidth]{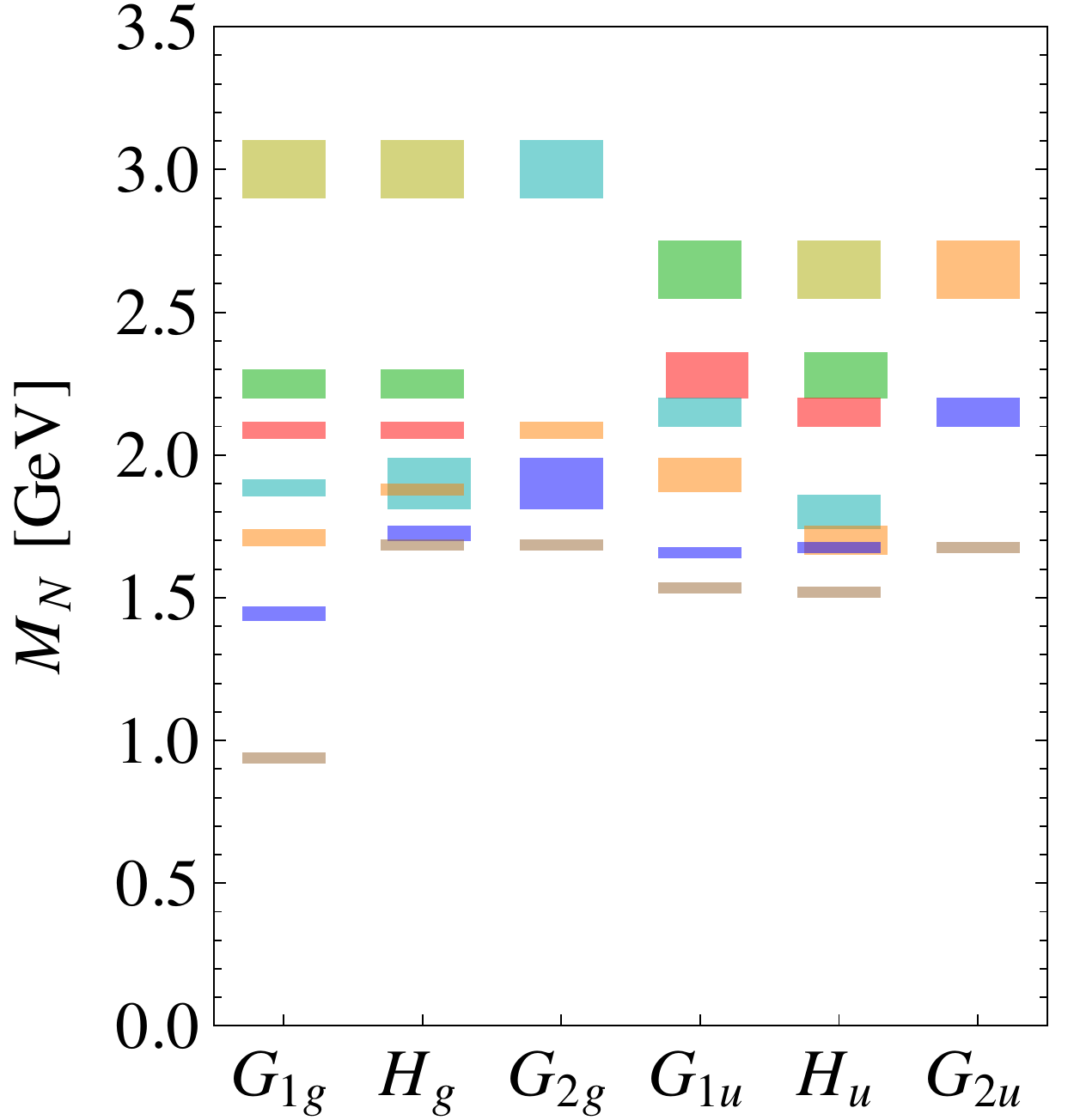}
\includegraphics[width=.45\textwidth]{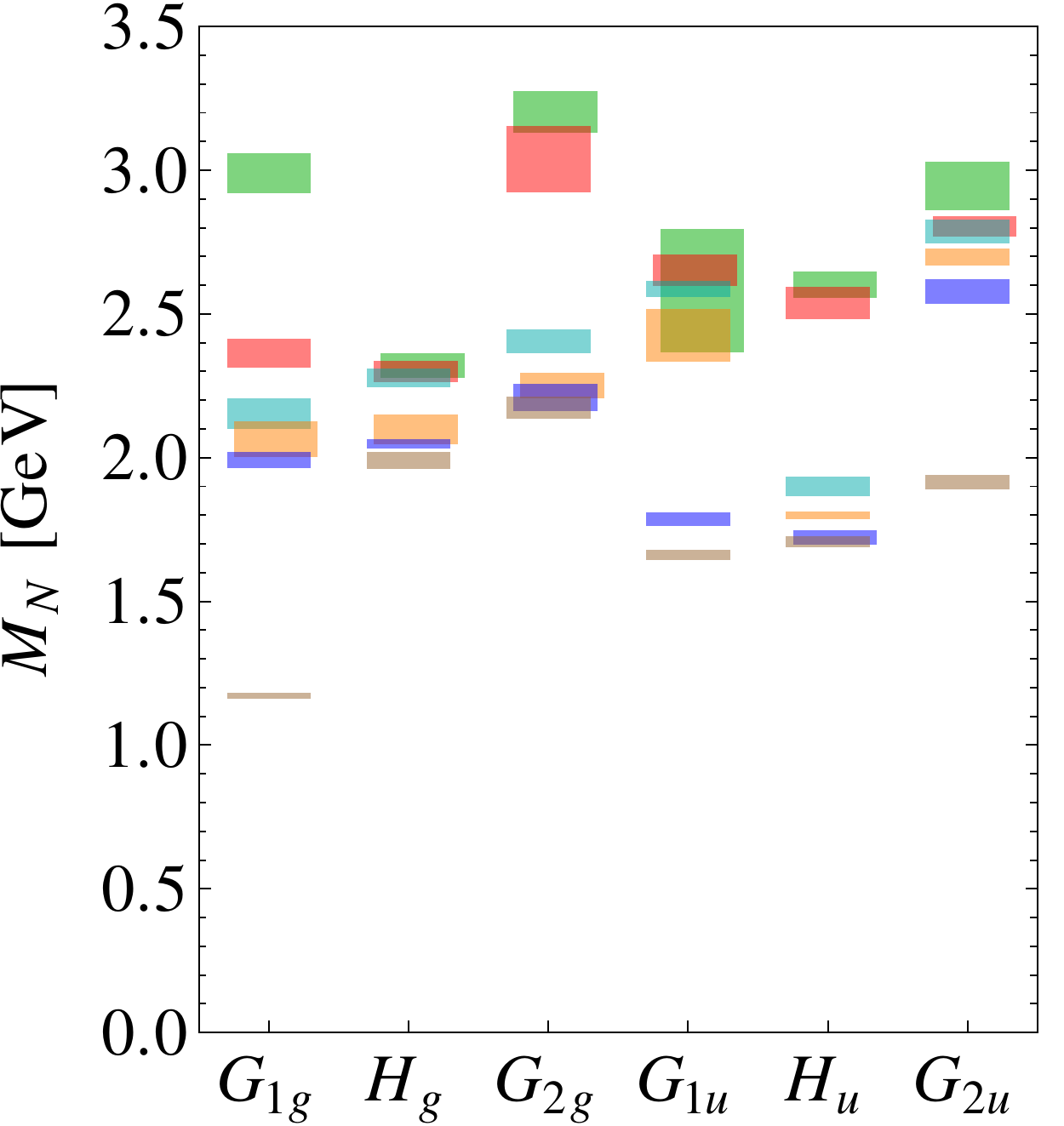}
\includegraphics[width=.45\textwidth]{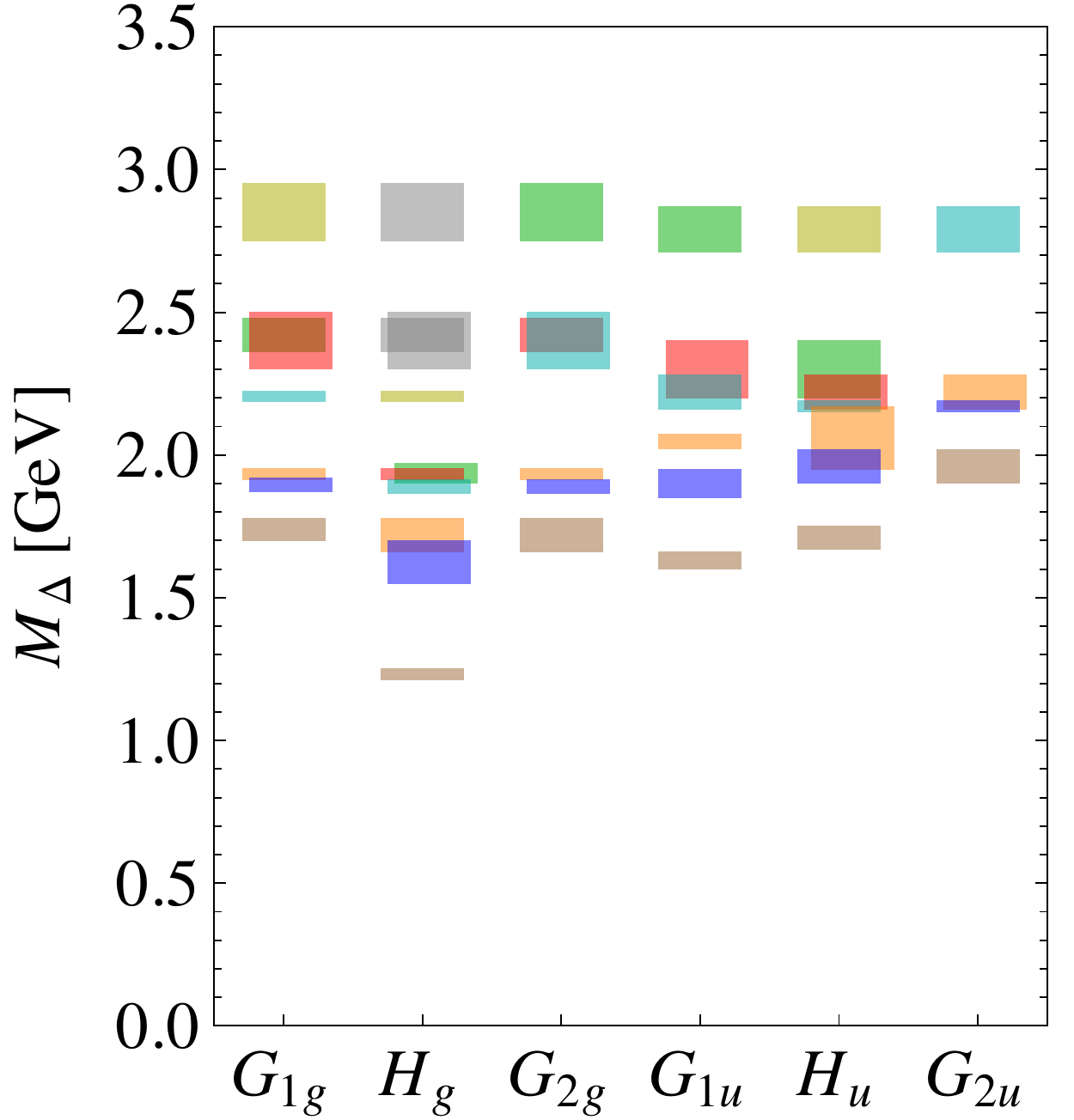}
\includegraphics[width=.45\textwidth]{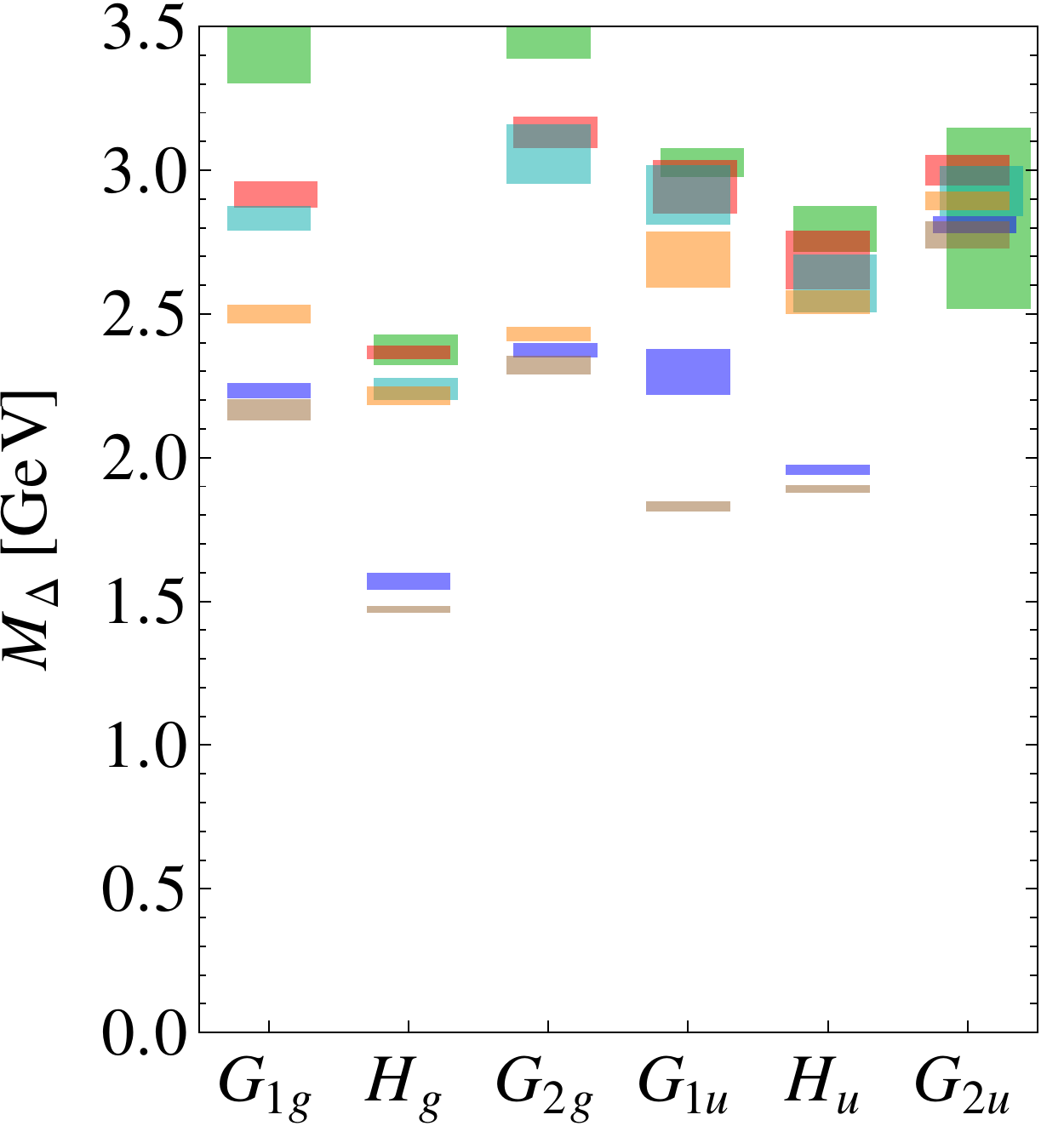}
\caption{\label{fig:spec-840} Nucleon (upper row, left: experiment and right: lattice with pion mass 400~MeV) and Delta (lower row, left: experiment and right: lattice) excited spectrum sorted according to cubic-group irrep.
}
\end{center}
\end{figure}

As mentioned in the discussion of the results from the nucleon and Delta spectra, the resonance states mix with scattering states; thus, it is important to include multi-particle states with in the correlator matrix. Then processing by the variational method will help us to correctly identify the nature of the extracted excited states.
Such an effort has been undertaken for the simplest possible scattering system, $\pi\pi (I=2)$. In Ref.~\cite{Dudek:2010ew}, HSC used the distillation technique to study the $\pi\pi$ scattering process elastic phase-shift using Luscher's method.  They report both the phase shift for isospin-2 scattering S-wave and also the D-wave channels using the same $N_f=2+1$ anisotropic clover lattices with pion masses between 390 and 520~MeV. 
Using pairs of pions with five relative momenta (summing to zero)
and 2 smearing parameters, a $10\times 10$ correlator matrix is formed and analyzed with the variational method to extract a spectrum of states. 

However, the above study only included the ``connected'' diagram, while many baryon states could mix with scattering states, such as $N\pi$ and $N\pi\pi$, which include ``disconnected'' diagrams. In such diagrams, quarks and antiquarks contract within the initial state or final state. Such diagrams are notoriously difficult to calculate; not only do they require more sources to be calculated 
but also the resulting signals are much noisier. 
A new method for distillation which uses stochastic estimation of the low-lying effects of quark propagation was investigated in Ref.~\cite{Morningstar:2011ka}. Such an improved method resolves the problem of rapidly increasing numbers of eigenvectors as one increases the spatial volume, and also allows accurate determinations of disconnected diagrams. 
The method exploits Laplacian Heaviside (LapH) quark-field smearing and variance reduction is achieved using judiciously-chosen noise dilution projectors.
They study $I=0,1$ S-wave $\pi\pi$ scattering using the $N_f=2+1$ anisotropic clover lattices with pion mass as low as 240~MeV.

\begin{figure}[t]
\begin{center}
\includegraphics[width=.75\textwidth]{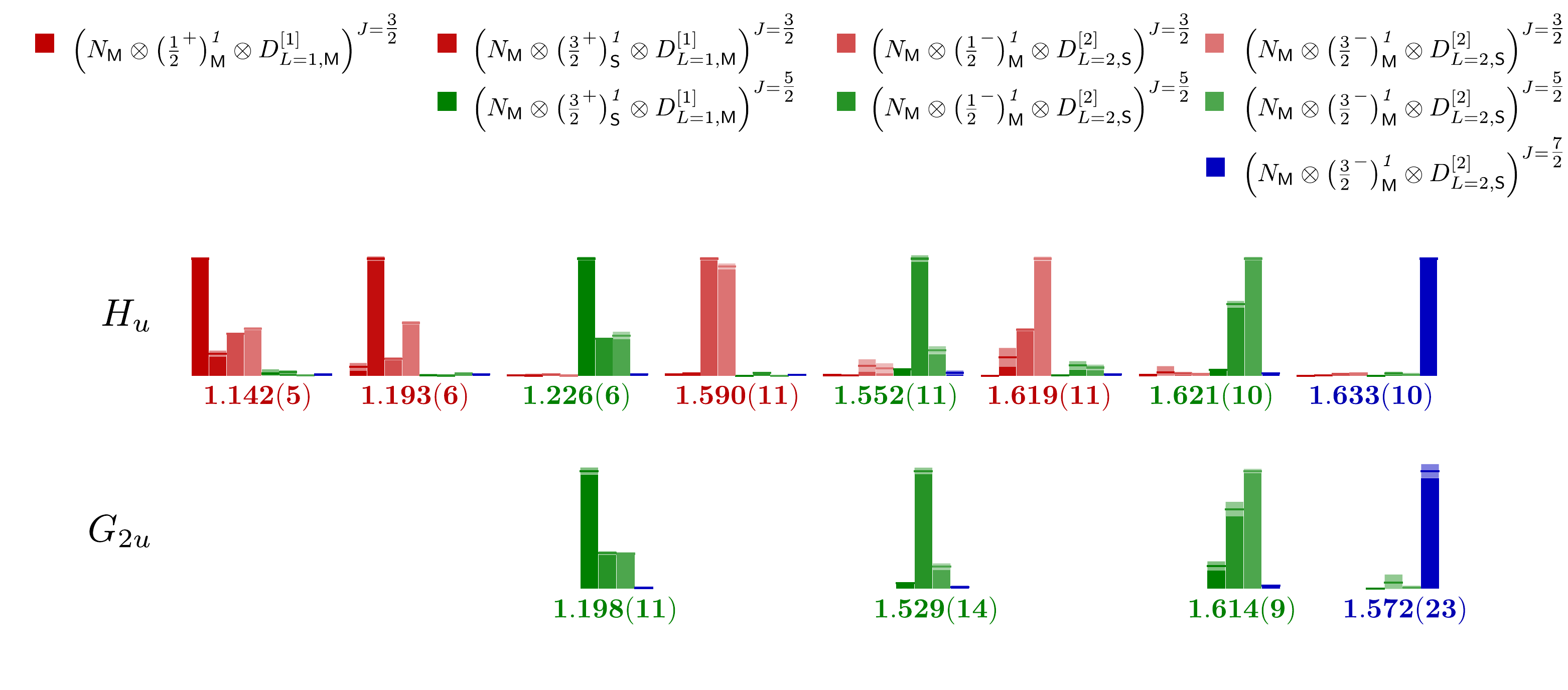}
\includegraphics[width=.75\textwidth]{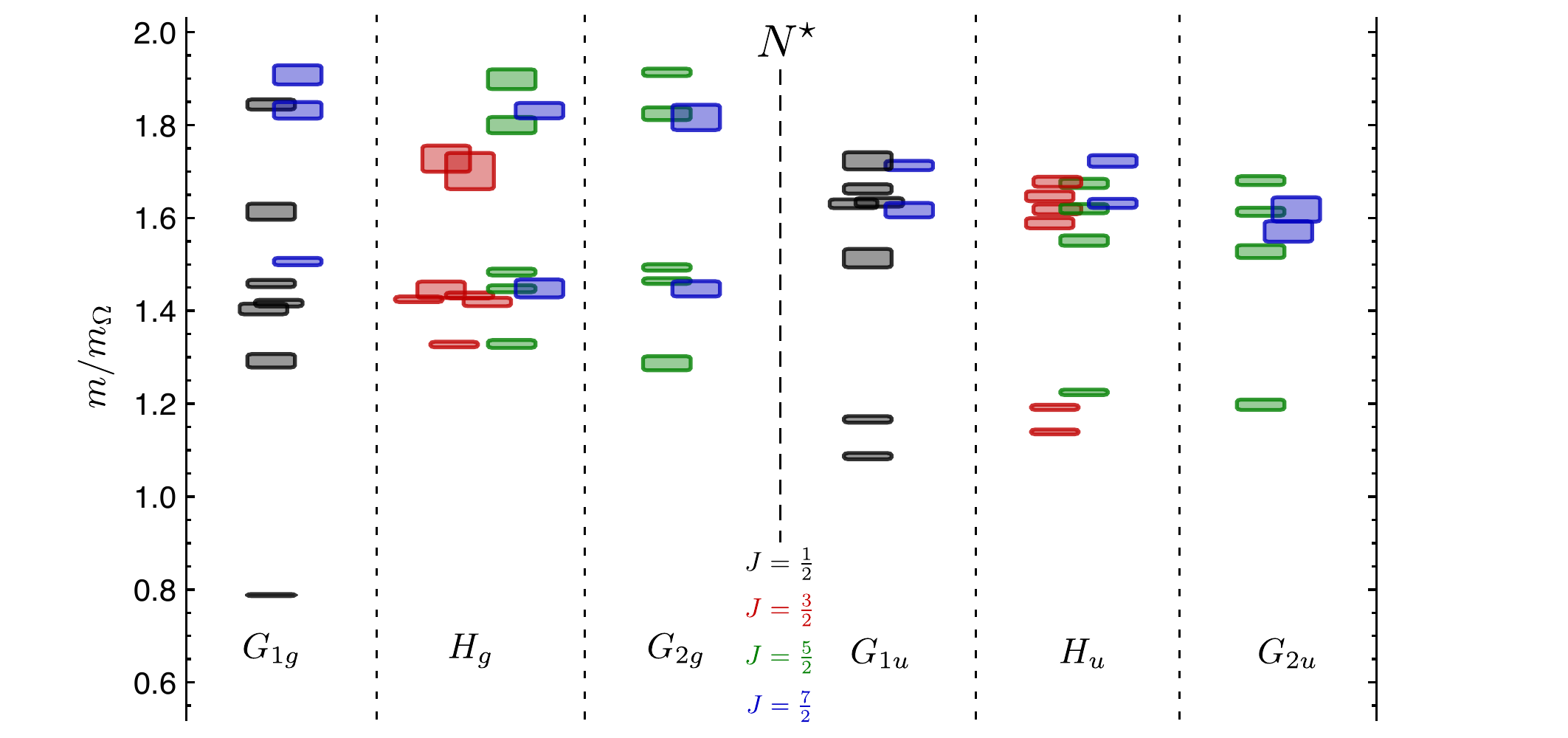}
\caption{\label{fig:Jlab-overlap} The nucleon overlap factors for a subset of operators in $H_g$ and $G_{2g}$ irreps and their eigenstates (up) and extracted nucleon excited states using full set of operators (bottom) from Ref.~\cite{Edwards:2011jj}. The calculation was done on the $N_f=2+1$ anisotropic clover lattice at pion mass around 520~MeV. }
\end{center}
\end{figure}

The cubic-irrep operators provide many independent operators to work with on the lattice; however, we have already seen in the above discussion of the nucleon and Delta spectra that the multiplicity of states across different irreps can make them difficult to clearly identify the spin for comparison with experiments. 
In Ref.~\cite{Edwards:2011jj}, HSC overcomes the difficulty caused by the reduced symmetry of the cubic lattice by introducing a method for operator construction that allows for reliable identification of continuum spins; for details, see Sec.~IV in Ref.~\cite{Edwards:2011jj}. 
Basically, we examine the overlap between the operator and the eigenstate in question obtained from the variational method: $Z_n^i = \langle n| O_i| 0\rangle$. The relative magnitude tells us how each operator $O_i$ overlaps with the eigenstate $n$. 
For example, Fig.~\ref{fig:Jlab-overlap} shows the overlaps for a set of low-lying states in the nucleon $H_u$ and $G_{2u}$ irreps at pion mass 520~MeV and box size 2~fm. The overlaps for a given operator show a clear preference for overlap onto only operators of a definite spin. The assignment of spin must hold for states with continuum spin $J$ subduced across multiple irreps. As can be seen, there is good agreement between $Z$ values in the different irreps that are subduced from spin $J$ after accounting for Clebsch-Gordan coefficients, with only small deviations. 
These results demonstrate that the $Z$ values of carefully constructed subduced operators can be used to identify the continuum spin of states extracted for the lattices and operators shown. The mass values determined from fits to principal correlators in each irrep differ slightly due to the discretization effects and fitting variations (such as selection of the fitting intervals). 
Using this method, HSC determine the spectrum of single-particle states for spins up to 7/2, as shown in the bottom part of Fig.~\ref{fig:Jlab-overlap}. 
Lighter pion masses around 400~MeV at 2~fm spatial volume are also studied in Ref.~\cite{Edwards:2011jj}, as well as a comparison of the spectrum with the non-relativistic quark model.

\subsubsection{Hyperon Spectra}

\begin{table}
\centering
\begin{tabular}{cccc}
Flavor & $G_{1g,u}$ & $H_{g,u}$ \\ \hline
$\Lambda$ & 4 & 1 \\
$\Sigma$ & 4 & 3  \\
$\Xi$ & 4 & 3  \\
$\Omega$ & 1 & 2
\end{tabular}
\caption{\label{tag:hyperon-op}The number of linearly independent interpolating operators in each irreducible representation of the cubic group from local-site operators}
\label{tab:operators}
\end{table}

\begin{figure}
\includegraphics[width=0.5\textwidth]{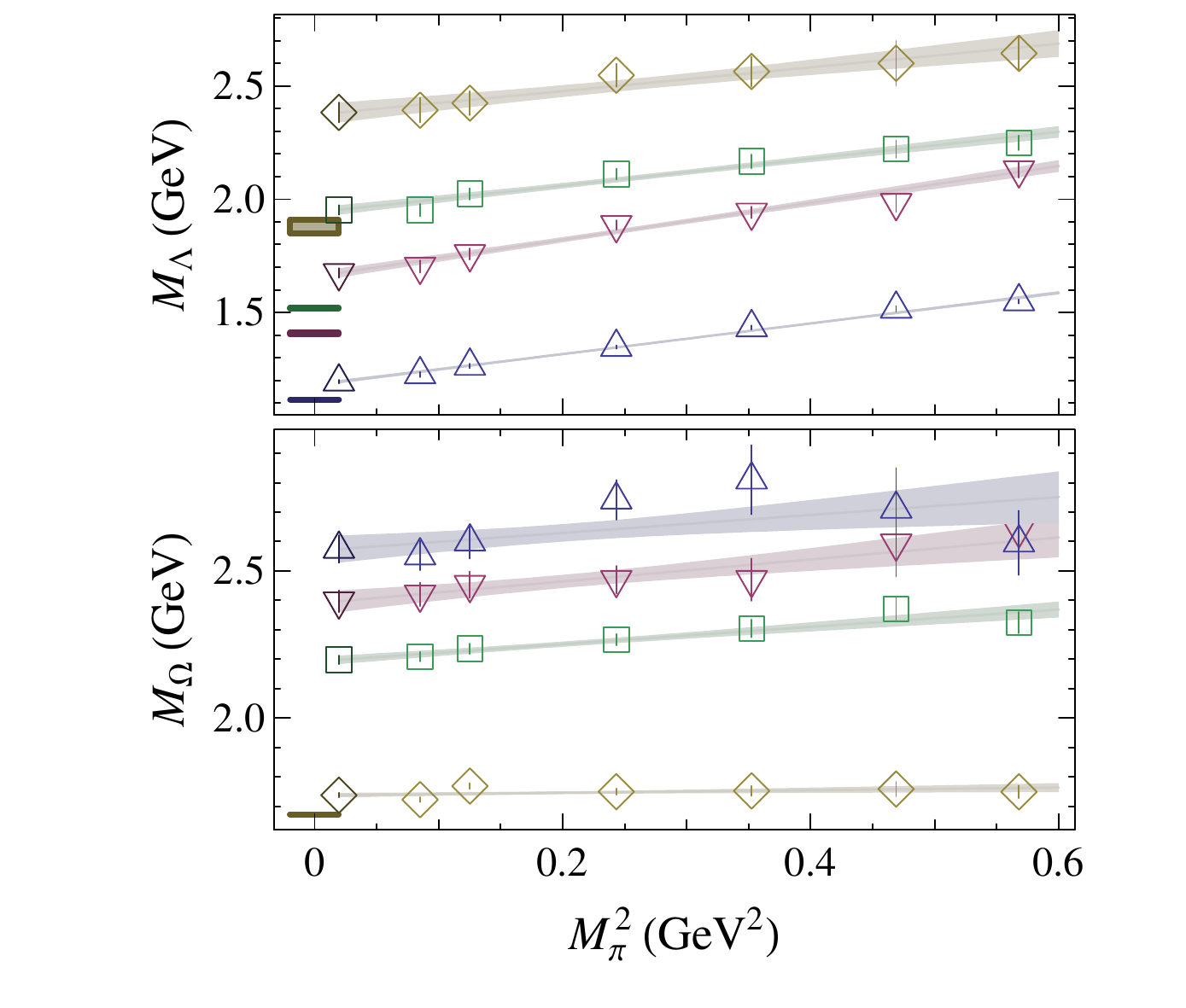}
\includegraphics[width=0.45\textwidth]{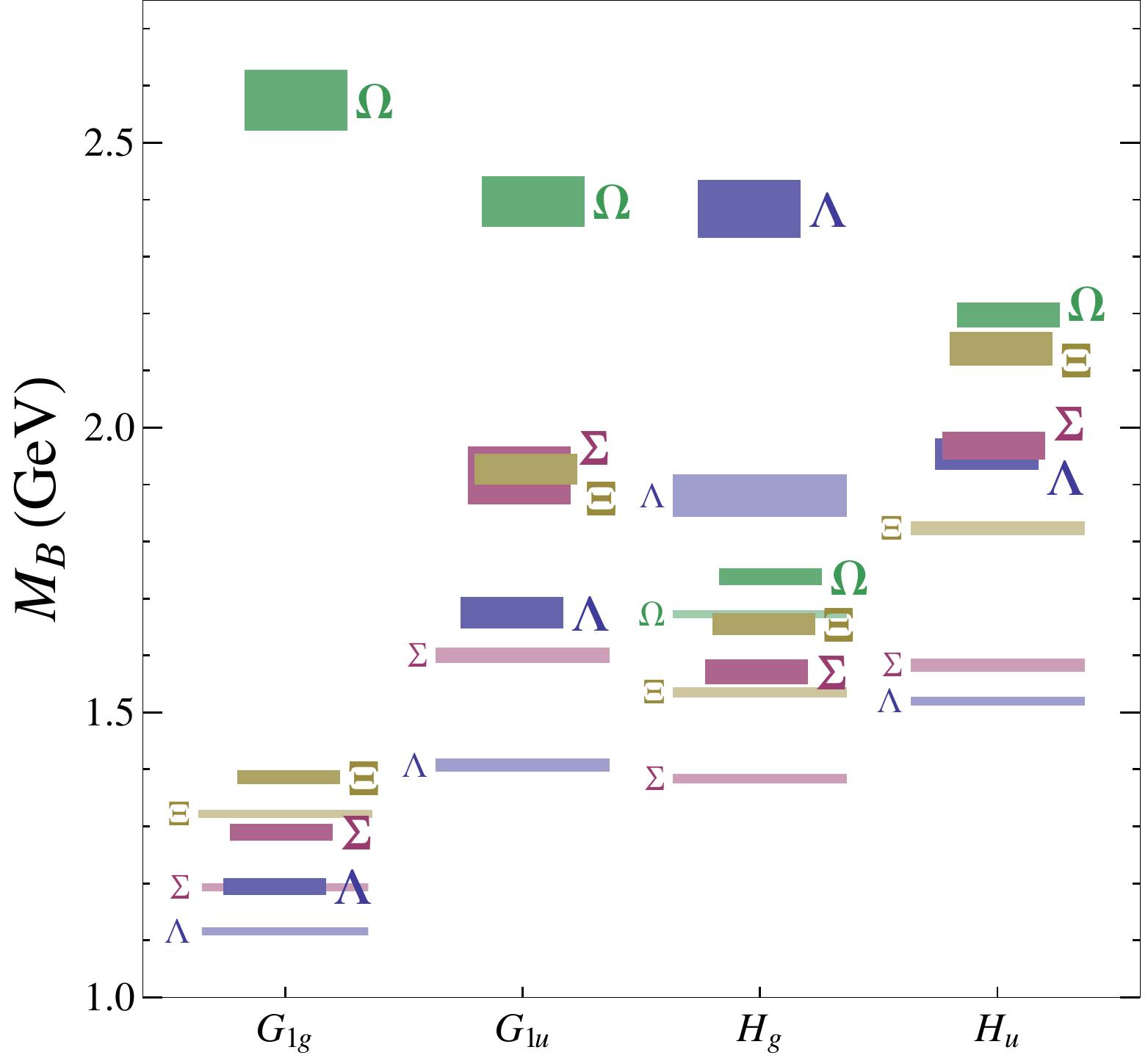}
\caption{(left) The squared pion-mass dependence of $\Lambda$ and $\Omega$-flavor baryons with their extrapolated values. The bar at the left indicates the experimental values for the corresponding spin.
(right) A summary of our hyperon spectra compared to experiment. The lattice data from Ref.~\cite{Lin:2008rb} are the short bars with large labels on the right. The experimental states are long bars with small labels on the left.
}\label{fig:spec-mpi2}
\end{figure}

Using the techniques discussed in previous section, we can construct all the possible baryon interpolating operators that can be formed from local or quasi-local $u/d$ and $s$ quark fields. 
In Ref.~\cite{Lin:2008rb}, the masses of the lowest-lying states in the $G_{1g,u}$ and $H_{g,u}$ representations are calculated using domain-wall valence fermions on 2+1 flavors of asqtad sea quarks (generated by the MILC collaboration\cite{Bernard:2001av}). The pion masses range from 300 to 700~MeV in a lattice box of size 2.6~fm at lattice spacing 0.12~fm. The gauge fields are hypercubic-smeared and the source field is Gaussian-smeared to improve the signal. Details on the configurations can be found in Ref.~\cite{WalkerLoud:2008bp}. 
The ground-state masses of the octet and decuplet and the SU(3) Gell-Mann--Okubo mass relation can be found in Ref.~\cite{WalkerLoud:2008bp}. This reference has an extensive description of baryon-mass extrapolation (to the physical pion mass) using continuum and mixed-action heavy-baryon chiral perturbation theory for two and three dynamical flavors. However, no chiral perturbation theory results have been published for orbitally excited hyperon resonances; in such cases, naive linear extrapolations in terms of $m_\pi^2$ are used. 
Only the local-site operators are constructed, as shown in Table~\ref{tag:hyperon-op}; the $G_{2g,u}$ irreps are excluded, since they require non-(quasi-)local operators. 
Figure~\ref{fig:spec-mpi2} summarizes our results for the lowest-lying $\Lambda$ and $\Omega$ $G_{1g,u}$ (upward/downward-pointing triangles) and $H_{g,u}$ (diamonds and squares) (since these have less overlap with the data in in Ref.~\cite{WalkerLoud:2008bp}) and their chiral extrapolations. The leftmost points are extrapolated masses at the physical pion mass, and the horizontal bars are the experimental masses (if they are known).

The lattice hyperon-mass calculations for the $\Sigma$, $\Lambda$, $\Xi$ and $\Omega$ are summarized on the right-hand side of Figure~\ref{fig:spec-mpi2}, divided into vertical columns according to their discrete lattice spin-parity irreps ($G_{1g,u}$ and $H_{g,u}$), along with experimental results obtained by subduction of continuum $J^P$ quantum numbers onto lattice irreps. $G_1$ ground states only overlap with spin-$1/2$. The spin identification for $H$ can be a bit trickier, since this irrep could match either to spin-$3/2$ or $5/2$ ground states. We simply select the lowest-lying of $3/2$ or $5/2$ indicated in the PDG, so it could be either depending on which one is the ground state for a particular baryon flavor. (We note below if the lowest $H$ is not spin-$3/2$.)

Although our naive extrapolation neglects contributions at next-to-leading order in chiral perturbation theory, several interesting patterns are seen. 
The better known $\Lambda$ (with $H_g$ being spin-$5/2$) and $\Sigma$ spectra match up with our calculations well. $G_{1u}$ $\Xi$ lines up well with the $\Xi(1690)$, indicating that the spin-parity of this resonance could be $1/2^-$; this agrees with SLAC's spin measurement. 
The spin assignments for the $\Omega$ channel are the least known; from our mass pattern, we predict that $\Omega(2250)$ is likely to be $3/2^-$ (although we cannot rule out the possibility of $5/2^-$), and $\Omega(2380)$ and $\Omega(2470)$ are likely spin $1/2^-$ and $1/2^+$, respectively.

To successfully extract even higher excited hyperon states, we would need finer lattice spacing, which would require massive computational resources to achieve. In Euclidean space, the excited signals exponentially decay faster than the ground state. One possible solution is to use an anisotropic lattice, where the temporal lattice spacing is made finer than the spatial ones to reduce overall costs. 
Such an effort have been taken by Hadron Spectroscopy Collocation and there are a few results reported by the HSC. 
Ref.~\cite{Mathur:2008gw} uses $N_f = 2$ anisotropic Wilson (non-$O(a)$-improved) lattices ($a_s/a_t = 3$) with inverse temporal lattice spacing 5.56~GeV at a single pion mass at 416~MeV. 
However, only the lowest states of the $G_{1g,u}$, $G_{2g,u}$ and $H_{g,u}$ of the cascade are reported.

\begin{figure}[t]
\begin{center}
\includegraphics[width=.45\textwidth]{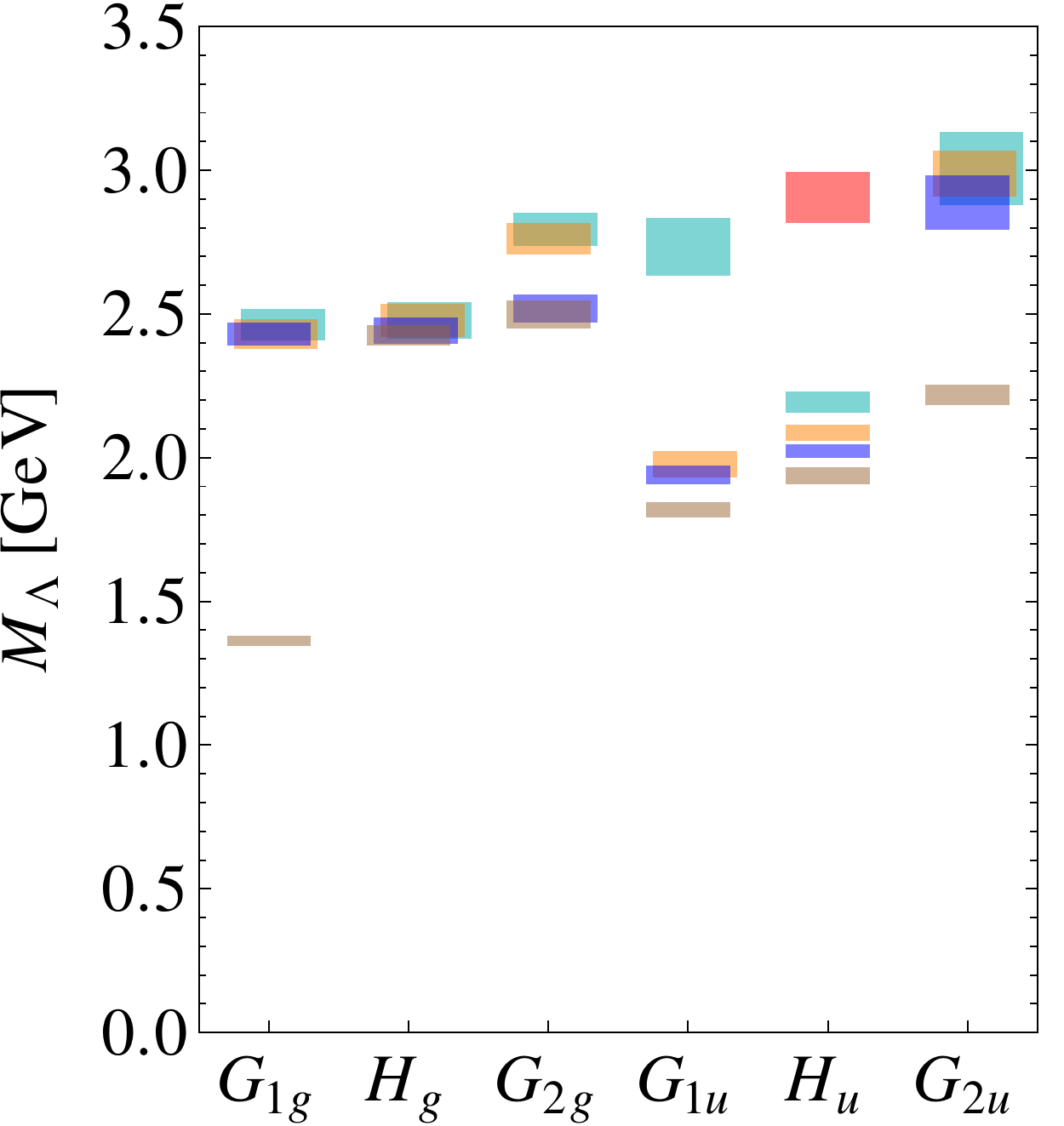}
\includegraphics[width=.45\textwidth]{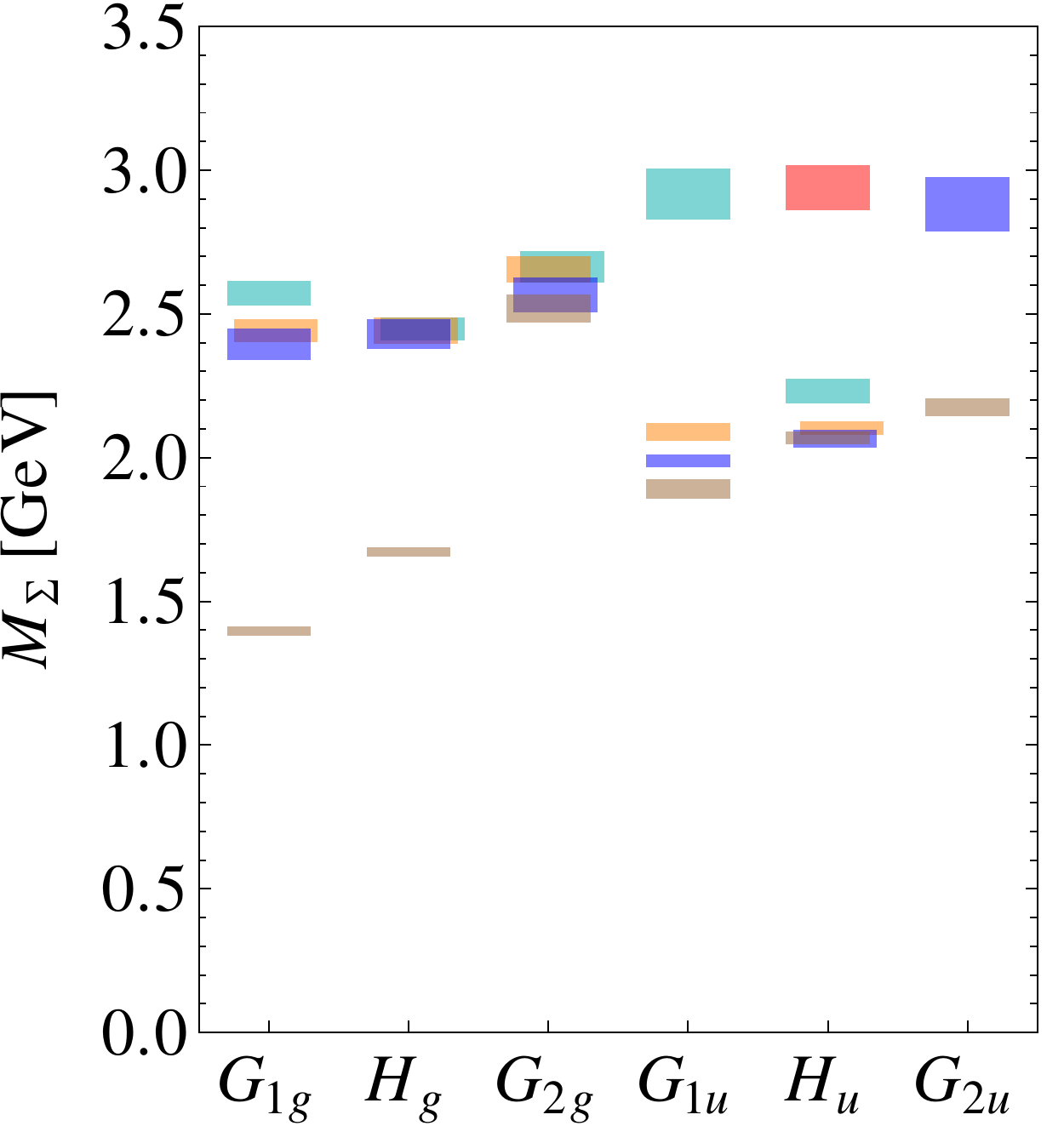}
\includegraphics[width=.45\textwidth]{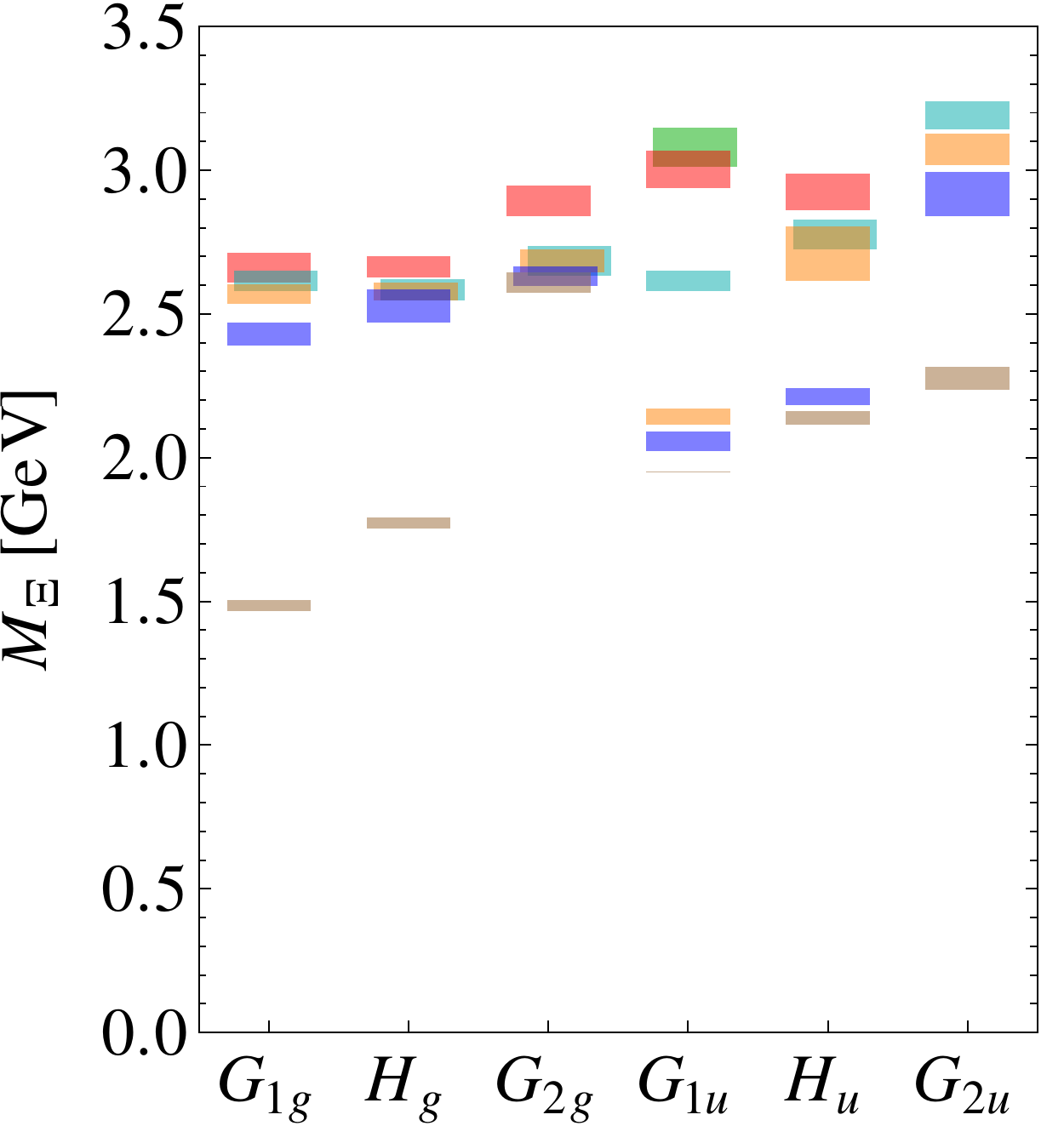}
\includegraphics[width=.45\textwidth]{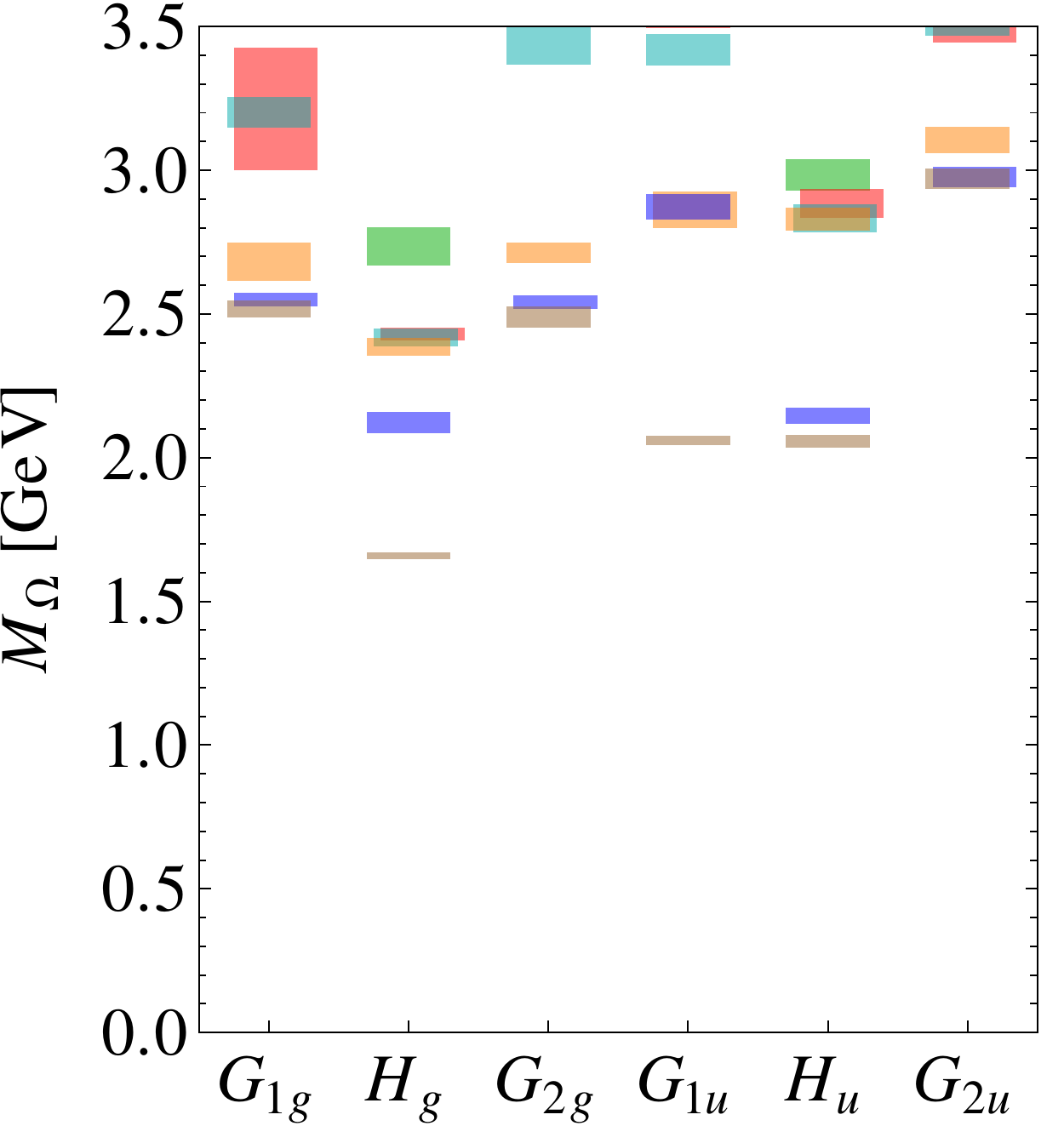}
\caption{\label{fig:HSC-hyperon-840} Lambda (upper left), Sigma (upper right), cascade (lower left) and Omega (lower right) spectra sorted according to cubic-group irrep from Refs.~\cite{Morningstar:2010kb,Bulava:2010yg}.
}
\end{center}
\end{figure}

Higher excited states of other hyperon ($\Sigma$, $\Lambda$ and $\Xi$ flavors) states are followed up at low-statistics (50 to 100 configurations) on the $N_f = 2+1$ anisotropic clover lattice in Ref.~\cite{Morningstar:2010kb}.
Fig.~\ref{fig:HSC-hyperon-840} shows results obtained from 2~fm box at pion masses around 390~MeV with eight optimal operators (from hundreds) in each symmetry channel. 
A detailed study of the Omega was reported in Ref.~\cite{Bulava:2010yg}. 
Similar to the cases of the nucleon and Delta, the higher excited states may be subject to decay modes; thus, we need to add multiple-particle operators to the correlator matrix to separate them.

\section{Heavy Flavor}\label{sec:heavy}

Lattice QCD is now a mature field capable of providing accurate results that can be directly compared to experiment. In the previous sections, we showed that calculations in the light-quark sector are well established and on-going. However, the study of heavy quarks requires careful treatment of discretization errors. Lattice heavy quarks have $O((m_Q\,a)^n)$ errors, where $m_Q$ is the mass of the heavy quark and $a$ is the lattice spacing. Lattice spacings for most currently accessible dynamical ensembles are still too coarse ($a^{-1} \approx 2\mbox{ GeV}$) to make such systematic errors small. To assert better control over the discretization errors for heavy quarks on the lattice, a few specialized heavy-quark approaches have proven useful. 
For example, non-relativistic QCD (NRQCD)~\cite{Lepage:1992tx}, which is an expansion of the lattice quark action in powers of $\frac{1}{am_Q}$, is commonly applied to bottom quarks. However, the charm-quark mass is not heavy enough to justify the use of NRQCD in the charm sector. 
Relativistic heavy-quark (RHQ) actions~\cite{ElKhadra:1996mp,Aoki:2001ra,Christ:2006us,Lin:2006ur} systematically remove $O((m_Qa)^n)$ terms and are better suited to charm-quark calculations. Recent updates on the state of heavy-quark physics on the lattice can be found in several reviews~\cite{Kronfeld:2003sd,Wingate:2004xa,Okamoto:2005zg,Onogi:2006km,DellaMorte:2007ny,Gamiz:2008iv,Aubin:2009yh,Wingate:2011fb} and references therein.

Measurements of mesons involving heavy quarks have been carried out successfully for some time and are now entering an era of precision. Lattice-QCD calculations of decay constants, semileptonic form factors, and neutral-meson mixing parameters that are important to CKM matrix elements are often quoted and considered as the Standard Model inputs. These inputs can then be combined with experimental measurements to probe physics beyond the Standard Model. However, knowledge about heavy baryons is less extensive, and experimentally, there are still missing doubly and triply heavy baryons yet to be discovered. In this section, we emphasize recent progress in heavy-baryon spectroscopy from lattice QCD.

\subsubsection{Charmed Baryons}

The construction of operators in the charm sector proceeds similarly to the light-flavor ones. The interpolating operators used for $J=1/2$ singly and doubly charmed baryons, for example, are 
\begin{eqnarray}
& \mathcal{O}_{\Lambda_{c}}  :   \epsilon^{ijk}(q_u^{iT} C\gamma_5 q_d^j)Q_c^k,  \,\, 
	\mathcal{O}_{\Sigma_{c}}  : \epsilon^{ijk}(q_u^{iT} C\gamma_5 Q_c^j)q_u^k,& \,\,\nonumber\\
& \mathcal{O}_{\Xi_{c}}  :   \epsilon^{ijk}(q_u^{iT} C\gamma_5 q_s^j)Q_c^k,  \,\, 
    \mathcal{O}_{\Omega_{c}}  :  \epsilon^{ijk}(q_s^{iT} C\gamma_5 Q_c^j)q_s^k, & \,\,\nonumber\\ 	
& \mathcal{O}_{\Xi_{cc}}  :  \epsilon^{ijk}(Q_c^{iT} C\gamma_5 q_u^j)Q_c^k,  \,\, 
	\mathcal{O}_{\Omega_{cc}}  :  \epsilon^{ijk}(Q_c^{iT} C\gamma_5 q_s^j)Q_c^k,  \,\, \nonumber &\\
& \mathcal{O}_{\Xi^\prime_{c}}  : \frac{1}{\sqrt{2}} \epsilon^{ijk}\left[(q_u^{iT} C\gamma_5 Q_c^j)q_s^k
	+ (q_s^{iT} C\gamma_5 Q_c^j)q_u^k\right], & \,\,
\end{eqnarray}
where $q_{u,d}$ are the up- and down-quark fields, $q_s$ is strange-quark field and $Q_c$ is charm-quark field.

As in the light-flavor case, people began lattice charmed-baryon calculations using the quenched approximation. In the early days, there were a few calculations using an $O(a)$-improved light-quark action on either isotropic or anisotropic lattices with a single lattice spacing: 
Bowler et~al.~\cite{Bowler:1996ws} used tree-level clover action for both light and heavy quarks to calculate the singly charmed baryon spectrum for spin 1/2 and 3/2. 
Later, Flynn et~al.~\cite{Flynn:2003vz} updated this project with nonperturbative clover action and extended the calculation to doubly charmed baryons. 
Chiu et~al.~\cite{Chiu:2005zc} used improved chiral fermion action for charm quarks and calculated both positive and negative parity for singly and doubly charmed baryons. 
Such calculations using light-quark actions to simulate heavy quarks could introduce large systematic errors proportional to $(am_Q)^2$, which must be carefully addressed.

One calculation has used a higher-order improved fermion action: 
Lewis et~al.~\cite{Lewis:2001iz} did a calculation on both doubly and singly charmed baryons using D234-type fermion action (which would leave a leading error of $O(a^3)$) for both light and heavy quarks but on a coarse anisotropic ensemble (with anisotropy $\xi=2$). 
Finally, heavy-quark effective theory was applied to charm calculation: 
Mathur et~al.~\cite{Mathur:2002ce} continued to use anisotropic lattices, adding two more lattice spacings, but changed their heavy-quark action to NRQCD, which reduces the lattice-spacing discretization effects. 
For all of these calculations, the quenched approximation remains a significant source of a systematic error that is difficult to estimate.

Since new charmed baryons have been discovered in experiments, more interest has been raised to revisit charmed baryon calculations on the lattice. With the increasing computational resources and improved algorithms, one can use $N_f=2+1$ dynamical gauge ensembles to remove the uncontrollable systematics due to the quenched approximation. 
Although more dynamical ensembles are available these days, not many charmed baryon calculations have been published so far. 
Refs.~\cite{Liu:2008rza,Liu:2009jc,Lin:2010wb} compute the ground-state spectrum of the spin-1/2 singly and doubly charmed baryons. The RHQ action is used for the charm quarks and domain-wall fermions for the light valence quarks on gauge configurations with 2+1-flavor asqtad staggered fermions and a range of quark masses resulting in pion masses as light as 290~MeV at fixed lattice spacing and volume. The fermion anisotropy is nonperturbatively tuned; temporal and spatial clover coefficients are set separately through the tree-level tadpole improved coefficients, and two input bare masses for charm quarks. The results are extrapolated to the physical light-quark masses using both heavy-hadron chiral perturbation theory (HHChPT) as well as HHChPT-inspired polynomial extrapolations. 
Refs.~\cite{Na:2007pv,Na:2008hz} also use 2+1-flavor asqtad staggered lattices but using four lattice spacings. The light-quark valence action is the same as the sea and the spin-3/2 charm baryons are also studied.

\begin{table}
\begin{center}
\footnotesize
\begin{ruledtabular}
\begin{tabular}{c|ccccccc}
Group & $N_{\rm f}$ & $S_{\rm H}$ & $a_t^{-1}$ (GeV)   & $L$ (fm)\\
\hline
Bowler et~al.~\cite{Bowler:1996ws}    & 0   & tree clover~\cite{Sheikholeslami:1985ij} & 2.9   &  1.63 \\
Lewis et~al.~\cite{Lewis:2001iz}  & 0   & D234~\cite{Alford:1995dm} & 1.8, 2.2, 2.6   &  1.97\\
Mathur et~al.~\cite{Mathur:2002ce} & 0 & NRQCD~\cite{Sheikholeslami:1985ij} & 1.8, 2.2 & 2.64,2.1\\
Flynn et~al.~\cite{Flynn:2003vz} & 0 & NP clover  & 2.6 & 1.82\\
Chiu et~al.\cite{Chiu:2005zc} & 0 & ODWF~\cite{Chiu:2002ir} & 2.23 &1.77\\
Na et~al.\cite{Na:2007pv,Na:2008hz} & $2+1$ & Fermilab & 2.2, 1.6, 1.3& 2.5\\
Liu et~al.\cite{Liu:2008rza,Liu:2009jc,Lin:2010wb} & $2+1$ & RHQ~\cite{ElKhadra:1996mp} & 1.6 & 2.5
\end{tabular}
\end{ruledtabular}
\caption{\label{tab:charmedB_summary}Summary of published charmed baryon calculations from lattice QCD. Please refer to the above references and references within for more details.}
\end{center}
\end{table}

\begin{figure}
\includegraphics[width=0.65\textwidth]{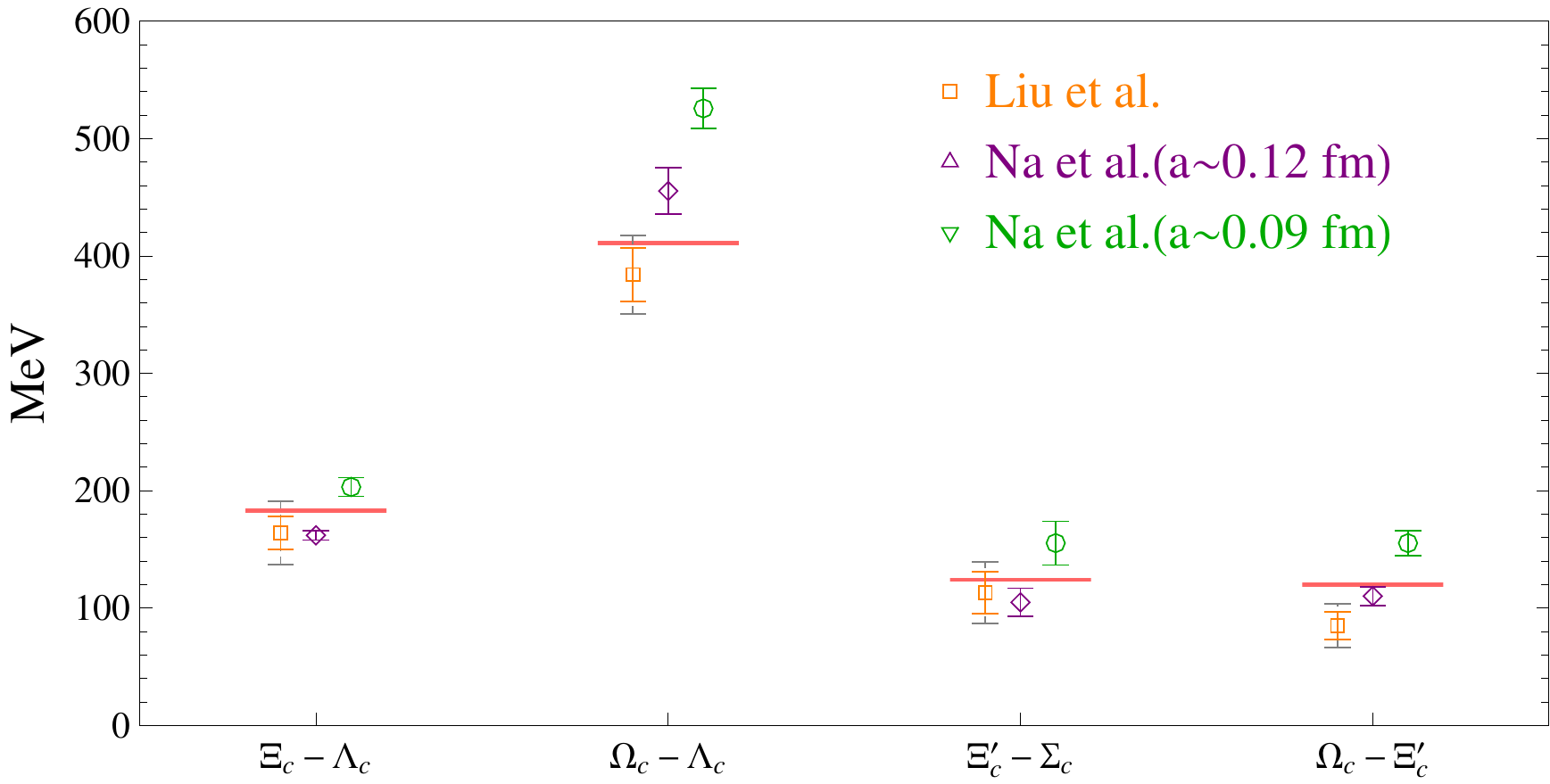}
\includegraphics[width=0.65\textwidth]{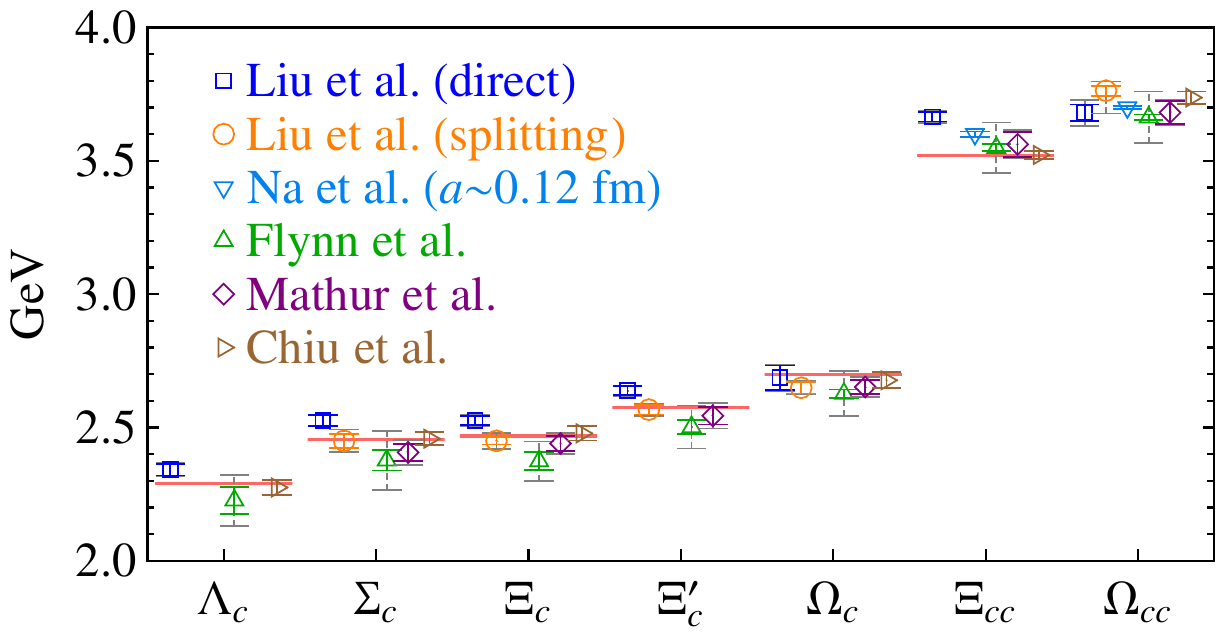}
\caption{\label{fig:all2p1-lat}
(Left) Comparison among charmed-baryon mass splittings of dynamical lattice calculations~\cite{Na:2007pv,Na:2008hz,Liu:2008rza,Liu:2009jc,Lin:2010wb}.
(Right) A summary of charmed-baryon masses (in GeV) calculated using LQCD.
Two mass extractions are taken from Ref.~\cite{Liu:2009jc}: the lighter (orange) points are taken from a splitting extrapolation and the darker (blue) points are taken from a direct mass extrapolation.
}
\end{figure}

All published lattice calculations on the charm baryon spectrum are summarized in Table~\ref{tab:charmedB_summary}.
Most of the calculations were done in the quenched approximation with relatively heavy pion masses. 
A few selected mass splittings from dynamical charmed baryon calculations~\cite{Na:2007pv,Na:2008hz,Liu:2008rza,Liu:2009jc,Lin:2010wb} are compared in Fig.~\ref{fig:all2p1-lat}. 
In Ref.~\cite{Na:2007pv,Na:2008hz}, a calculation with staggered light quarks, an interpretation of the Fermilab action was used, defining the charm mass with the kinetic mass instead of the rest mass. They observe a significant lattice-spacing dependence in their calculation, which is very different from previous calculations~\cite{Mathur:2002ce} with quenched approximation. Their calculation covers ensembles at more than one lattice spacing, but their results are not as precisely determined statistically. Their uncertainties given here are merely the statistical ones; no systematic uncertainties are given yet. 
We selected a few to compare with the numbers we obtained in this work, as shown in Fig.~\ref{fig:all2p1-lat}. (For those calculations with more than one lattice spacing, we show only the results from the ensemble with lattice spacing closest the commonly used $a\approx0.12$~fm among all the works.) Note that the quenched systematic is not included in the given errorbars.

\begin{figure}
\includegraphics[width=0.65\textwidth]{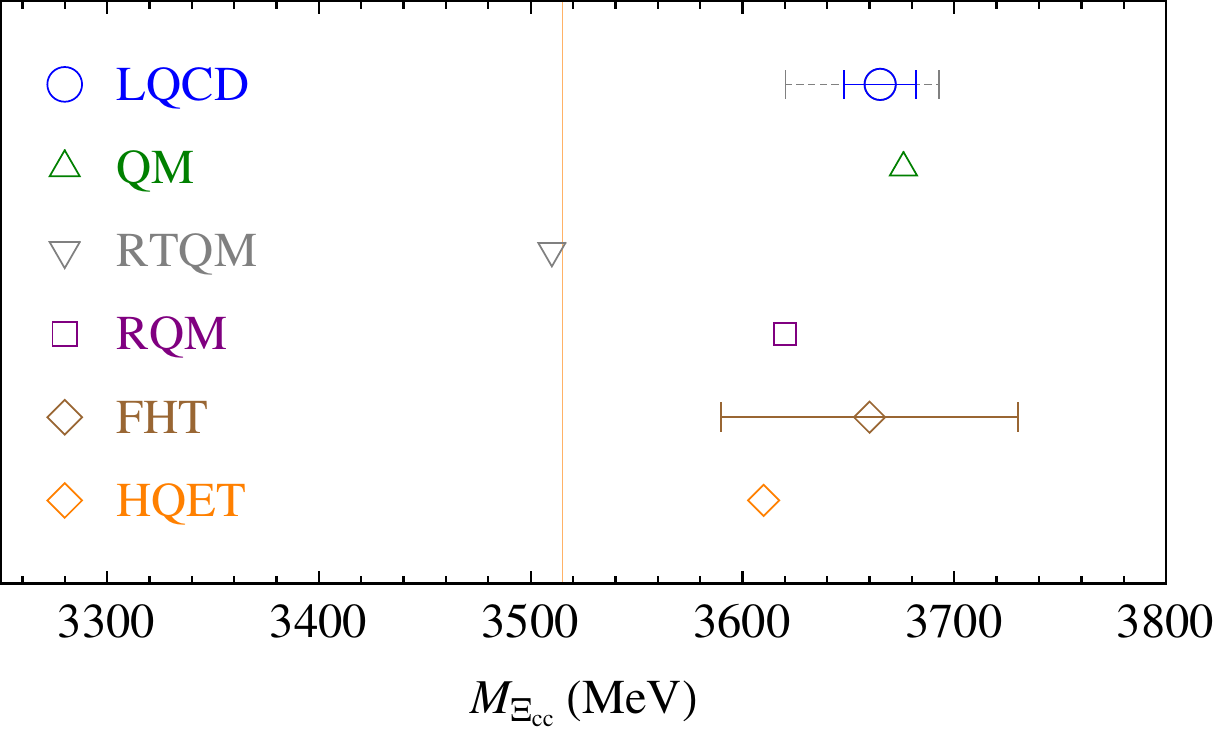}
\includegraphics[width=0.65\textwidth]{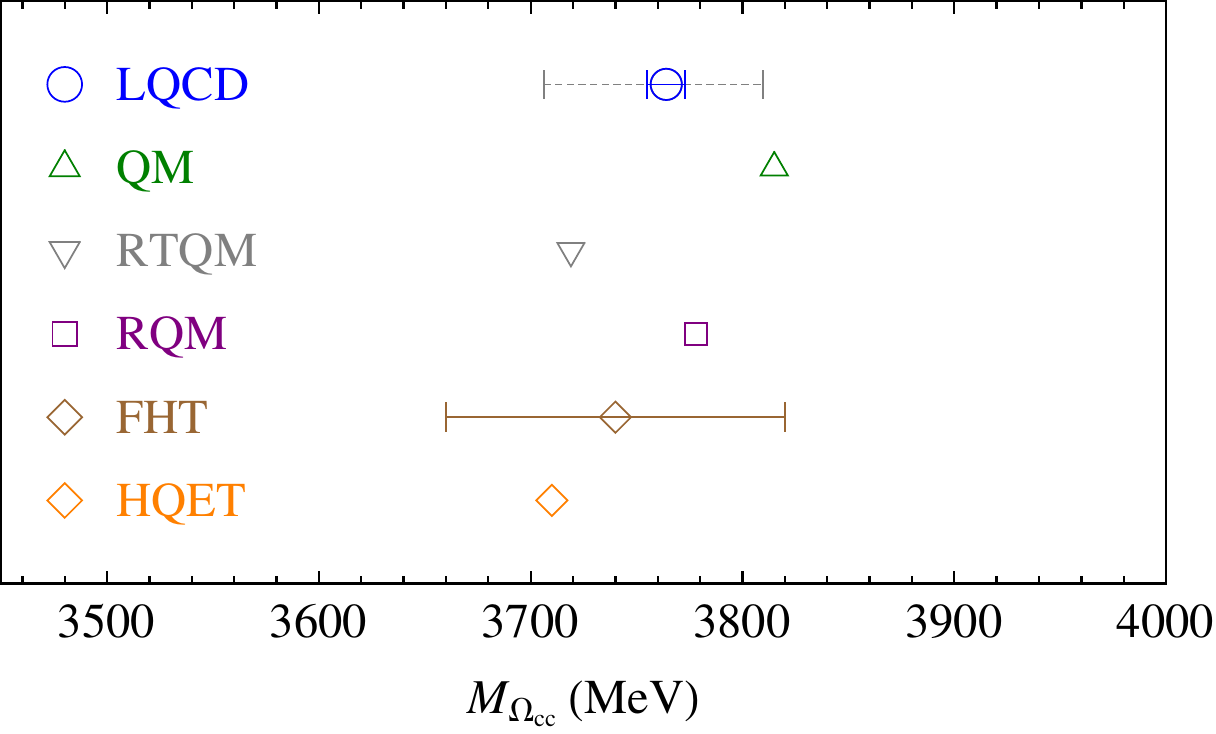}
\caption{\label{fig:DoublyCharmedB} Comparison of theoretical predictions for doubly charmed baryons of spin 1/2. ``LQCD'' is the lattice QCD calculation taken from more precisely determined values in Refs.~\cite{Liu:2009jc,Lin:2010wb} with solid error bars for the statistical error and dashed bars for the total error including the estimated systematic;
``QM'' is taken from a recent quark-model calculation~\cite{Roberts:2007ni};
``RTQM'' is the result of relativistic three-quark model~\cite{Martynenko:2007je};
``RQM'' and ``HQET'' are from the relativistic quark model~\cite{Ebert:2002ig} and the heavy-quark effective theory~\cite{Korner:1994nh} respectively;
note that there is no error estimation done in these calculations.
``FHT'' is based on the Feynman-Hellmann theorem~\cite{Roncaglia:1995az}, where rough uncertainties are estimated.}
\end{figure}

Finally, let us compare the doubly charmed baryons with the predictions of theoretical models, as shown in Fig.~\ref{fig:DoublyCharmedB}. Although the SELEX Collaboration has reported a first observation of doubly charmed baryons, searches by the BaBar~\cite{Aubert:2006qw}, Belle~\cite{Chistov:2006zj} and Focus~\cite{Ratti:2003ez} Collaborations have not confirmed their results. This makes it interesting to look back to the theory to see where the various predictions lie. 
We compare with a selection of other theoretical results, such as a recent quark-model calculation~\cite{Roberts:2007ni}, relativistic three-quark model~\cite{Martynenko:2007je}, the relativistic quark model~\cite{Ebert:2002ig}, the heavy quark effective theory~\cite{Korner:1994nh} and the Feynman-Hellmann theorem~\cite{Roncaglia:1995az}. 
The lattice-computed mass of $\Xi_{cc}$ is $3665\pm17\pm14\,{}^{+0}_{-35}$~MeV, which is higher than what SELEX observed; most theoretical results suggest that the $\Xi_{cc}$ is about 100--200~MeV higher. 
The $\Omega_{cc}$ mass prediction is $3763\pm9\pm44\,{}^{+0}_{-35}$~MeV, and the overall theoretical expectation is for the $\Omega_{cc}$ to be 3650--3850~MeV. We hope that upcoming experiments will be able to resolve these mysteries of doubly charmed baryons.

\subsubsection{Bottom Baryons}

The bottom quark is relatively heavy; thus a first-order approximation of $b$-quark would be to imagine a single static infinitely massive quark. Such an approximation introduces corrections suppressed by powers of $\Lambda_{\rm QCD}/m_b$, which can be included systematically in heavy-quark effective field theory. In this limit, the quark mass is removed, leaving only the forward hopping term in the action: 
\begin{equation}
G_Q({\bf x},t;t_0) = \frac{1+\gamma_4}{2}\prod^{t}_{t^\prime=t_0}U_4({\bf x},t^\prime).
\end{equation} 
The links $U_4$ that enter this propagator can be any set of gauge covariant paths that connect neighboring points in time. Different choices of paths result in actions that are equivalent up to discretization errors. All of them have the same continuum limit. It is often accompanied by the usage of HYP-smeared links, which give significantly better signals for the lattice correlators.

In the static-quark limit, the heavy-quark propagator is reduced to a Wilson line. Thus the hadron correlator is quite different from the light-flavor case. For example, the static-light baryon correlation functions have the form 
\begin{equation}
G_{\Gamma}(\vec{x},t;\vec{x}_0,t_0) =
  \langle
    q_f^a(\vec{x},t) \Gamma q_{f^\prime}^b(\vec{x},t)
    \epsilon_{cab}
    P^{cc^\prime}(\vec{x},t;\vec{x}_0,t_0)
    \bar{q}_{f^\prime}^{a^\prime}(\vec{x}_0,t_0) \Gamma \bar{q}_f^{b^\prime}(\vec{x}_0,t_0)
    \epsilon_{c^\prime a^\prime b^\prime}
  \rangle,
\end{equation} 
and the static-light meson correlator is 
\begin{equation}
M(\vec{x},t;\vec{x}_0,t_0) =
  \langle
    q_f^a(\vec{x},t)
    P^{aa^\prime}(\vec{x},t;\vec{x}_0,t_0)
    \bar{q}_{f^\prime}^{a^\prime}(\vec{x}_0,t_0)
  \rangle,
\end{equation} 
where $f \in \{u,d,s\}$ is the light-quark flavor index and $P^{cc^\prime}$ is the Wilson line connecting the source and the sink which are separated by time $t$. The spin matrix $\Gamma$ is either $C\gamma_5$ or $C\gamma_\mu$ for scalar or vector diquarks, respectively.

Many bottom spectroscopy lattice calculations have been performed in the meson sector. 
Ref.~\cite{Li:2008kb} calculated the bottomonium and bottom-light meson spectra using the relativistic heavy-quark action\cite{Christ:2006us,Lin:2006ur} with fermion action parameters determined through nonperturbative tuning to eliminate systematic uncertainties; their ensembles used 2+1 flavors of domain-wall fermions (DWF)\cite{Kaplan:1992bt,Furman:1994ky} with lightest pion mass 275~MeV.
Ref.~\cite{Meinel:2009rd} also used DWF gauge ensembles but with NRQCD\cite{Lepage:1992tx} to calculate bottom-quark quantities. 
A more impressive work\cite{Foley:2007ui} done by the TriLat Collaboration used anisotropic 2-flavor Wilson-type lattices with renormalized anisotropy as high as 6 and operators projected into irreducible representations of the lattice cubic group. These techniques allowed them to obtain very clean signals including multiple static-light excited states.

Unlike bottomonium and $B$ mesons, bottom baryons have not received as much attention from lattice QCD. Some pioneering works\cite{Bowler:1996ws,AliKhan:1999yb,Mathur:2002ce} were done using extrapolations with light-fermion actions or the NRQCD action in the quenched approximation, where the fermion loop degrees of freedom are absent. Since it is difficult to estimate the systematic error due to the quenched approximation, high-precision calculations cannot be achieved. 
Recently, more dynamical ensembles have become available due to the increase of the computer resources available for numerical research, and since the recent discoveries of the $\Sigma_b^*$, $\Xi_b$ and $\Omega_b$ baryons at CDF and D0, more lattice calculations have emerged. 
Ref.~\cite{Lewis:2008fu} calculated single- and double-$b$ baryons with NRQCD action for the bottom quark on isotropic 2+1-flavor clover ensembles, having lightest pion mass around 600~MeV. 
Ref.~\cite{Meinel:2009vv} also used NRQCD for bottom baryons but used 2+1 flavors of domain-wall fermions for valence/sea fermions at a single lattice spacing for spin-1/2 and 3/2 singly to triply charmed baryons.
A selection of single-$b$ baryons were calculated using static-quark (infinitely massive) action on 2-flavor chirally improved lattice Dirac operator at pion masses as light as 350~MeV in Ref.~\cite{Burch:2008qx}. 
Ref.~\cite{Detmold:2008ww} went to a lighter pion mass (275~MeV) using 2+1 flavors of domain-wall fermions but used static-quark action to simulate the bottom quarks. Since the static-quark action tends to result in noisy signals, one needs high statistics to yield numbers comparable with experimental results. 
Ref.~\cite{Wagner:2010hj}
reports static-light spin-1/2 and 3/2 baryons on $N_f = 2$ twisted-Wilson fermions at a fixed lattice spacing of 0.079~fm. The lightest pion mass is 340~MeV, and they predict masses of negative parity states. 
Ref.~\cite{Lin:2009rx} also uses static-quark action for the bottom quarks, domain-wall fermion for the valence and 2+1 asqtad staggered fermion ensembles with  lattice spacing $a=0.124$~fm and volume about $(2.5\mbox{ fm})^3$ with lightest pion masses of 290~MeV. The number of available configurations for allowed this work to achieve high statistics and thus better determined resonance. 
An ongoing work using staggered and Fermilab fermion actions\cite{ElKhadra:1996mp} on 2+1 asqtad staggered sea and valence fermion with multiple lattice spacings was presented in Ref.~\cite{Na:2008hz}.

A summary of results for the splittings in the bottom-hadron spectrum from Ref.~\cite{Lin:2009rx} is shown in Table~\ref{tab:splittings-final}. The leading-order HChPT extrapolation in terms of the $m_\Omega$ scale is adopted as the central values, and we explicitly give three sources of error: statistical, extrapolation and scale. The extrapolation systematic error is estimated by the discrepancy between the leading-order extrapolation and the result using a form including the next-order $m_\pi^3$ term. These errors are generally about the same size as the statistical error. The error due to scale setting and discretization is estimated using the spread amongst the three scale-setting methods: $f_\pi$, $r_1$ and $m_\Omega$. This systematic is quite large, up to four times the statistical error, and could be resolved by repeating the calculation on finer lattices. The finite-volume corrections are negligible for a 2.5-fm box, since no such correction is observed in the light-hadron masses\cite{WalkerLoud:2008bp} either. The remaining systematics, such as the effects of the $\Lambda_{\rm QCD}/m_b$ corrections are omitted, since they are substantially smaller than the main systematics (less than one percent).

\begin{table}
\begin{tabular}{|c|cc|}
\hline
$ $ & Splitting(Stat)(Extrap)(Scale) & Experiment \\
\hline \hline
$B_s-B_d$ &  71.2(2.2)(1.2)(4.2) & 87.1(0.6) \\
$\Lambda_b-B_d$ &  340(11)(8.1)(24) & 341.0(1.6) \\
$\Xi_b-B_d$ &  484.7(7.7)(6.2)(32) & 513(3) \\
$\Sigma_b-B_d$ &  615(15)(12)(40) & 554(3) \\
$\Xi_b^\prime-B_d$ &  672(10)(9.1)(44) & -- \\
$\Omega_b-B_d$ &  749.2(9.8)(9.0)(49) & 786(7)/886(16) \\
\hline
$\Lambda_b-B_s$ &  261(10)(8.5)(19) & 253.9(1.7) \\
$\Xi_b-\Lambda_b$ &  157.2(5.2)(3.4)(9.0) & 172(3) \\
$\Sigma_b-\Lambda_b$ &  274(13)(13)(17) & 213(3) \\
$\Xi_b^\prime-\Lambda_b$ &  335(15)(10)(20) & -- \\
$\Omega_b-\Lambda_b$ &  414(12)(9.5)(25) & 445(7)/545(16) \\
\hline
$\Xi_b^\prime-\Sigma_b$ &  62.6(2.6)(2.0)(3.9) & -- \\
$\Omega_b-\Xi_b^\prime$ &  81.5(2.7)(3.4)(4.9) & -- \\
\hline
\end{tabular}
\caption{\label{tab:splittings-final}Our consensus heavy-hadron splittings in units of MeV. The first error is statistical, the second is a systematic associated with the mass extrapolation and the third is a systematic associated with the setting of the lattice scale.}
\end{table}

\begin{figure}[ht]
\includegraphics[width=0.65\textwidth]{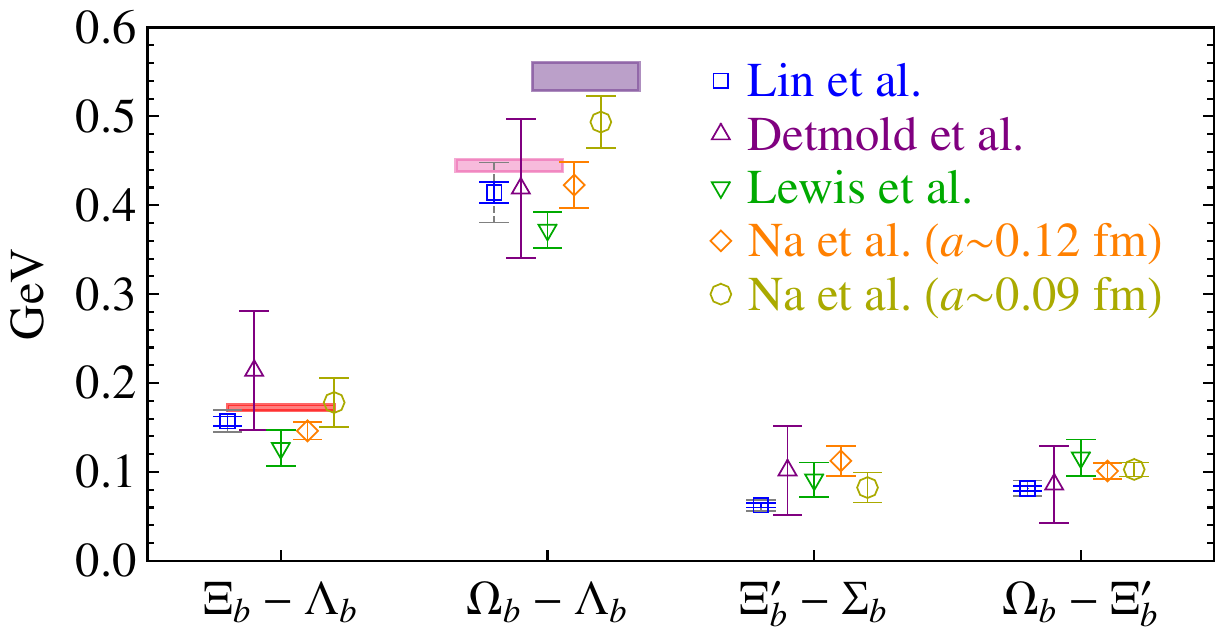}
\caption{\label{fig:all2p1-mixed-comp}
Comparison of mass splittings with all available 2+1-flavor lattice calculations of bottom baryons. The square (blue) points are the points extrapolated using the $\Omega$-mass reference scale; the solid error bars indicate the statistical error, and the dashed bars indicate the total errors (including the estimated systematic ones). The solid (red) bars indicate the experimental values given in the PDG, where available. For the $\Omega_b$, we show both the D0 result\cite{Abazov:2008qm} (upper-right, purple) and the CDF result\cite{Aaltonen:2009ny} (lower-left, magenta).
}
\end{figure}

To reduce the systematic error and because static quarks cannot produce sensible bare masses, many works present their results in terms of mass splittings with other nearby bottom baryons. However, not all the 2+1-flavor works\cite{Detmold:2008ww,Lin:2009rx,Na:2008hz,Lewis:2008fu} compute the same mass splittings, so we cannot have a comprehensive comparison. Lewis~et~al.\cite{Lewis:2008fu} used the NRQCD action to simulate their bottom quarks on CP-PACS clover lattices; Na~et~al.\cite{Na:2008hz} calculated using the same MILC asqtad lattices as Ref.~\cite{Lin:2009rx} but with Fermilab heavy-quark action and staggered light-quark action; they also included another two lattice spacings. All four calculations include the splittings $\Xi_b-\Lambda_b$ and $\Sigma_b-\Lambda_b$; we calculate the additional splittings $\Xi_b^\prime-\Sigma_b$ and $\Omega_b-\Xi_b^\prime$ in order to make direct comparisons. The results are shown in Fig.~\ref{fig:all2p1-mixed-comp}. We again see good agreement amongst all lattice calculations, and mild scaling in the Na~et~al. results. Agreement with experimental values as given by the PDG is fairly good. The discrepancy in the $\Sigma_b-\Lambda_b$ splitting may be due to discretization effects, as suggested by the trend visible in the Na~et~al. results.

These results are in fairly good agreement with experimental results. The largest discrepancies appear for the $\Sigma_b$, although it is only about one sigma away, given the estimated systematics. We give both the D0\cite{Abazov:2008qm} and new CDF\cite{Aaltonen:2009ny} results for the $\Omega_b$ mass. Our result (like most theory predictions) is more consistent with the lower CDF value.

\begin{figure}[ht]
\includegraphics[width=0.65\textwidth]{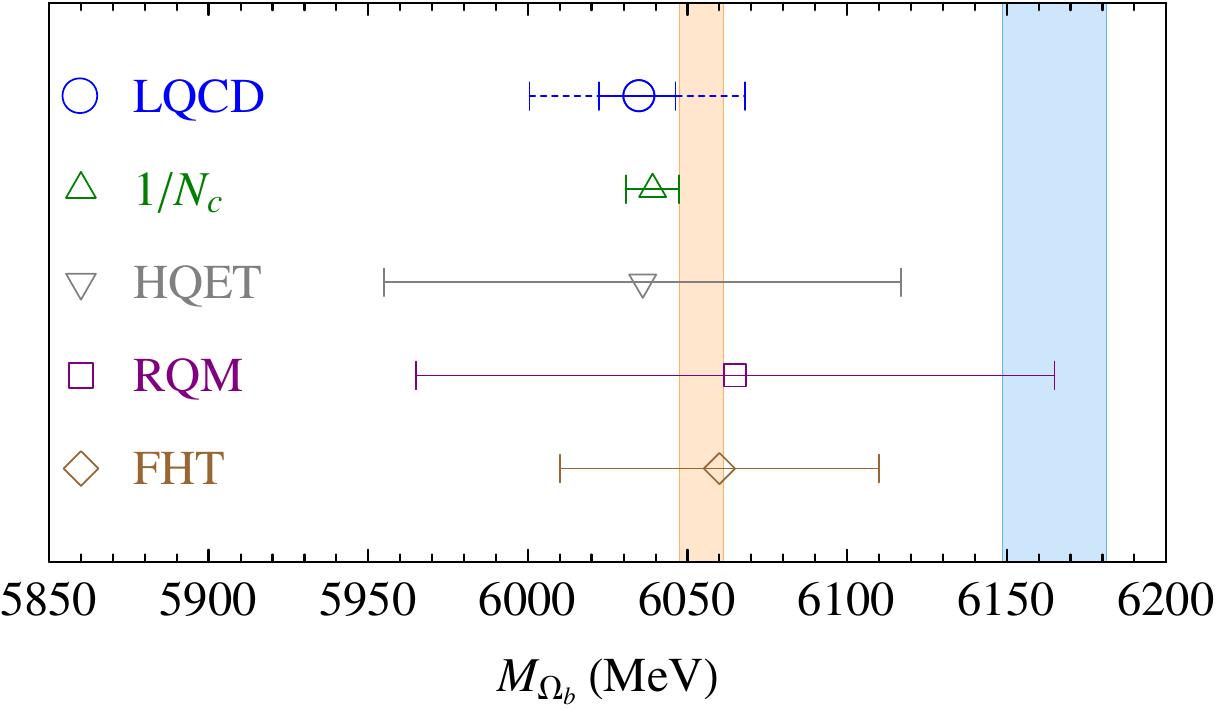}
\includegraphics[width=0.65\textwidth]{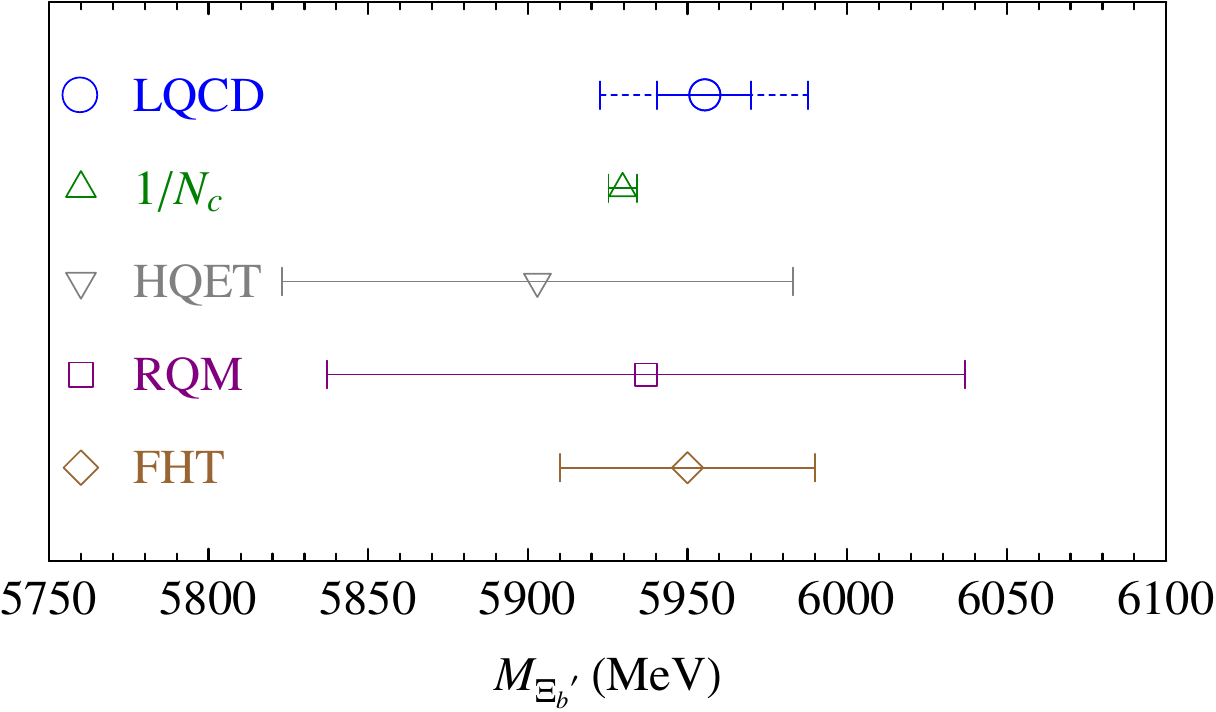}
\caption{\label{fig:theory-comp}
Comparison of various theoretical estimates of the $\Omega_b$ (top) and $\Xi^\prime_b$ (bottom) masses (and experimental values for the $\Omega_b$).
``LQCD'' is the lattice QCD calculation done in this work with solid error bars for the statistical error and the dashed bars for the total error including the estimated systematic;
``$1/N_c$'' is taken from a combined expansion in $1/N_c$, $1/m_Q$ and SU(3) flavor symmetry breaking\cite{Jenkins:2007dm}, where the errorbar is derived solely from experimental inputs;
``HQET'' is the result of heavy-quark effective field theory (HQET) using QCD sum rules\cite{Liu:2007fg};
``RQM'' is from the relativistic quark model\cite{Ebert:2005xj};
and ``FHT'' is based on the Feynman-Hellmann theorem\cite{Roncaglia:1995az}.
}
\end{figure}

The $\Omega_b$ and the yet-to-be-observed $\Xi^\prime_b$, masses are summarized along with those of other theoretical approaches in Fig.~\ref{fig:theory-comp}.

Finally, Meinel et al.~\cite{Meinel:2010pw} use NPQCD to simulate the bottom quark on $2+1$ DWF and MILC asqtad ensembles. On each ensembles, 2 lattice spacings around 0.012 and 0.08~fm are used. The lightest sea pion ensemble is around 300~MeV and predicts the $\Omega_{bbb}$ mass to be $14.371(4)_{\rm stat}(11)_{\rm sys}$~GeV.

\section{Summary and Outlook}\label{sec:conclusion}

Lattice QCD has shown great potential and progress in solving the mysteries of the QCD and its resonances. Many experimentally well-measured quantities have been successfully reproduced. In some cases, lattice QCD can provide higher-precision numbers than experiments, such as the scattering length of $\pi\pi(I=2)$ and B-meson decay constants, and provide the fundamental parameters of the Standard Model, such as the strong coupling constant and quark masses. 
In nuclear physics, the H-dibaryon, a theoretical $\Lambda\Lambda$ bound state predicted by Jaffe in the late 1970s but yet to be observed in experiment, has recently been reported with significant signal by two independent lattice groups. 
In addition, there are cases where experiments are impossible or extremely difficult, for example, the hyperon axial couplings and form factors involving initial- and final-state weak decays. 
Lattice QCD can provide all the necessary information on these quantities if provided sufficient computational resources.

In this article, we have reviewed a few selected topics in baryon spectroscopy from lattice QCD. 
The mysterious nature of Roper state is studied through direct calculation of its mass and electromagnetic properties of its transition to or from the nucleon. Early studies on the Roper were mainly done in the quenched approximation, where the systematics are uncontrolled. In the recent years, more dynamical calculations have been reported by various groups (for example, HSC and CSSM) with different choices of lattice fermion actions and operator bases. The variational method is used to reliably extracted the states that are nearby the Roper. Unfortunately, the latest dynamical calculation still have not yet observed a Roper mass smaller than its negative-parity partner. 
The lightest pion mass used so far is 156~MeV but the systematics due to finite-volume effects could be significant and need to be properly taken into account. At lighter pion mass and large volume, we also need to include the multiple-particle operators along to distinguish resonances from scattering states. Collaborations like HSC, CSSM are currently investigating this area, and we expects tremendous progress to be made in the near future.

In the realm of highly radially and orbitally excited baryon states, HSC has been taking a leading role, developing large operators bases, special lattices (with finer temporal resolution), distillation (a novel method to speed up quark contractions and improve signal) and spin identification techniques (to map between cubic-group and SO(4) irreps). 
The nucleon and Delta spectra at the lightest pion mass around 390~MeV first shown in terms of cubic irreps classification (where one sees similar experimental results with multiple degeneracy states among irreps) and connected to the continuum world by examining the overlap factors with the eigenstate from diagonalization. As in the case of the Roper, we need to be able to treat all the excited sates, and being able to distinguish the non-resonance ones (multiple particles) is an important task. HSC has two independent studies for the mesonic case with high angular momentum and notoriously difficult channels involving disconnected diagrams with significant signals. 
Channels like $\pi N$ and $\pi\pi N$ are important to baryon resonance programs and we expect more progress will soon be made, given the promising techniques developed by HSC.

Hyperon excited states are also interesting, but there have been fewer calculations of them in lattice QCD. 
Like Roper, the nature of $\Lambda (1405)$ (though not reviewed in this article) could be the first radially excited state of the $\Lambda$ baryon or a five-quark $\overline{K}N$ bound state (among other possibilities). In cascade spectroscopy, there are resonances being observed experimentally whose quantum numbers remain unknown; the spin-identification carried out by the HSC would surely be very helpful to fill in this missing information. 
New experimental facilities are starting strange production and we expect more experimental data will be available soon to map out the unknown territory of the strange-baryon spectroscopy. For example, the non-zero strangeness photoproduction by CLAS at Jefferson lab is starting to collect high-statistics on hyperon resonances. A coherent effort by both experiment and lattice QCD would greatly enhance our knowledge of the hyperon spectrum.

In the heavy-flavor sector, the recent experimental discoveries of new charmed and bottom baryons have re-energized the field. They have inspired more lattice-QCD studies using variety of heavy-quark formulations, such as static approximation, NRQCD and relativistic heavy-quark action. These new calculations have been performed on dynamical sea ensembles thanks to increases in the available computational resources and improved algorithms for QCD-vacuum generation. Many calculations are using multiple lattice-spacing ensembles to remove discretization systematic errors. High-precision heavy-baryon masses from lattice QCD can be achieved within the next couple years; these would provide useful input for experimental programs to look for the predicted resonances at specific energies.

Much progress has been made in lattice QCD improving first-principles calculations of baryon spectroscopy in both light and heavy sectors. Recent developments have been discussed in this article, and we have seen the maturity of various techniques that will begin to resolve one of the remaining challenges: to identify the unstable resonances. This will involve firstly expanding the single-baryon operator basis to include multiparticle operators, sorting them according to their lattice symmetry and corresponding continuum spin assignment. 
Currently, scattering phase shifts for resonances and their decay modes can be computed via Luscher's method by varying the spatial volume in the lattice calculation. 
Process involving ``disconnected contributions'', such as $\pi\pi(I=0)$, has shown improvement with the introduction of improved stochastic estimators.
Process such as $\pi N$ and $\pi\pi N$ which are more directly relevant to excited baryon spectrum would be the next challenge, since the noise-to-signal on lattice QCD calculation increases greatly with the number of nucleons involved in the process.

As the era of extreme-scale computing approaches, it allows lattice QCD to be able to greatly improve precision and attack problems that were previously considered computationally infeasible. 
There are multiple collaborations starting the gauge generation at the physical pion mass; thus, the uncertainty due to chiral extrapolation from around 300-MeV pions to the physical mass will be nearly eliminated. This is extremely helpful to highly excited baryon spectroscopy, since many of the forms and low-energy constants needed for chiral extrapolation are unknown. 
One challenge is to extend the current methods for analyzing elastic processes to the treatment of inelastic decays involving multiple final states. This would also open avenues to investigate gluonic degrees of freedom, such as in exotic mesons or glueballs. There will be an intense experimental effort, such as in the GlueX experiment at the 12-GeV upgrade at the Jefferson Laboratory, to look for evidence of gluonic degrees of freedom in the spectrum. Calculations so far are limited to significantly higher pion mass than the currently available pion-mass range. In addition, due to the uncontrolled mixing with mesonic states, signal-to-noise is generally orders of magnitude worse than fermionic observables, necessitating use of the cheap quenched approximation.

With the increasing computational resources made available to lattice QCD calculations, many obstacles have been overcome to reveal more of the nature QCD. 
Extreme-scale computing for physics has caught the attention of major funding agencies, and it is likely that we will see an exaflops system by 2017. For lattice QCD, this means we can look forward to greater things!

\section*{Acknowledgments}
The author is supported in part by the U.S. Dept. of Energy under grant No. DE-FG03-97ER4014.

\end{document}